\RequirePackage{fix-cm}
\documentclass[superscriptaddress,secnumarabic,amssymb,amsmath,nobibnotes,aps,prd,showkeys,showpacs,nofootinbib,twocolumn,a4paper,epjc3]{svjour3}
\smartqed  
\RequirePackage{graphicx}
 \RequirePackage{mathptmx}      
\RequirePackage{latexsym}
\RequirePackage[numbers,sort&compress]{natbib}
\RequirePackage[colorlinks,citecolor=blue,urlcolor=blue,linkcolor=blue]{hyperref}
\journalname{Eur. Phys. J. C}

\usepackage{graphicx}
\usepackage{dcolumn}
\usepackage{bm}

\usepackage[
scale=0.7, marginratio={1:1, 2:3}, ignoreall,
text={7in,10in},centering,
margin=1.5in,
total={6.5in,8.75in}, top=1.2in, left=0.9in, includefoot,
height=10in,a5paper,hmargin={3cm,0.8in},
]{geometry}
\RequirePackage{graphicx}
 \RequirePackage{mathptmx}      
\RequirePackage{latexsym}
\RequirePackage[colorlinks,citecolor=blue,urlcolor=blue,linkcolor=blue]{hyperref}
\usepackage[english]{babel}
\usepackage{amssymb,amsmath,amsfonts,graphicx, graphics, setspace}
\usepackage{bigints}
\usepackage{anysize}
\usepackage{float}
\usepackage{fancyhdr}
\usepackage{subfigure}
\usepackage{pdflscape}
\usepackage{epsfig}
\usepackage{epsf}
\usepackage{epstopdf}
\usepackage{hyperref} 
\newtheorem{thm}{Theorem}

\newtheorem{lem}[thm]{Lemma}

\usepackage{amsmath}
\usepackage{amsfonts}
\usepackage{amssymb}
\usepackage{subfigure}
\usepackage{epstopdf}
\usepackage{epsf}
\usepackage{url} 
\usepackage{pdflscape} 
\usepackage{amsmath}
\usepackage{amssymb} 
\usepackage{epsfig} 
\usepackage{amsfonts}
\usepackage{graphicx}
\usepackage{subfigure}
\usepackage{times,fancyhdr}
\usepackage{enumitem}
\setlist[enumerate,2]{label=\roman*)}

\def\case#1/#2{\textstyle\frac{#1}{#2}}

\newcommand{\be}{\begin{equation}}
\newcommand{\ee}{\end{equation}}
\newcommand{\ben}{\begin{eqnarray}}
\newcommand{\een}{\end{eqnarray}}

\usepackage{amsmath}
\usepackage{amsfonts}
\usepackage{amssymb}
\usepackage{widetext}
\usepackage{subfigure}
\usepackage{epstopdf}
\usepackage{epsf}
\def\ex{e_1{}^1}
\def\ey{e_2{}^2}

\def\R{{}^3\!R}

\def\ex{e_1{}^1}
\def\ey{e_2{}^2}

\def\R{{}^3\!R}

\def\y{\vartheta}
\def\z{\zeta}

\begin{document}

\title{Averaging Generalized Scalar Field Cosmologies I: Locally Rotationally Symmetric Bianchi III and open Friedmann-Lemaître-Robertson-Walker models}

\author{Genly Leon \thanksref{e1,addr1} \and Esteban Gonz\'alez \thanksref{e2,addr2} \and Samuel Lepe \thanksref{e3,addr3}  \and Claudio Michea \thanksref{e4,addr1} \and  Alfredo D. Millano \thanksref{e5,addr1}}
\thankstext{e1}{genly.leon@ucn.cl}
\thankstext{e2}{esteban.gonzalezb@usach.cl}
\thankstext{e3}{samuel.lepe@pucv.cl}
\thankstext{e4}{claudio.ramirez@ce.ucn.cl}
\thankstext{e5}{alfredo.millano@alumnos.ucn.cl}
\institute{Departamento  de  Matem\'aticas,  Universidad  Cat\'olica  del  Norte, Avda. Angamos  0610,  Casilla  1280  Antofagasta,  Chile\label{addr1} \and Universidad de Santiago de Chile (USACH), Facultad de Ciencia, Departamento de F\'isica, Chile \label{addr2}
\and Instituto de F\'isica, Facultad de Ciencias, Pontificia Universidad Cat\'olica de Valpara\'iso, 
Av. Brasil 2950, Valpara\'iso, Chile \label{addr3}
}
\date{\today}

\maketitle

\begin{abstract}
Scalar field cosmologies with a generalized harmonic potential and a  matter fluid with a barotropic Equation of State (EoS) with barotropic index $\gamma$ for Locally Rotationally Symmetric (LRS) Bianchi III metric and open Friedmann-Lemaître-Robertson-Walker (FLRW) metric are investigated. Methods from the theory of averaging of nonlinear dynamical systems are used to prove that time-dependent systems and their corresponding time-averaged versions have the same late-time dynamics. Therefore,  simple time-averaged systems determine the future asymptotic behavior. Depending on  values of  barotropic index $\gamma$  late-time attractors  of physical interests for LRS Bianchi III metric  are Bianchi III flat spacetime, matter dominated FLRW universe (mimicking de Sitter, quintessence or zero acceleration solutions) and matter-curvature scaling solution.  For open FLRW metric late-time attractors are a matter dominated FLRW universe and Milne solution. With this approach, oscillations entering nonlinear system through Klein-Gordon (KG) equation can be controlled and  smoothed out as the Hubble factor $H$ - acting as a time-dependent perturbation parameter - tends monotonically to zero. Numerical simulations are presented  as evidence of such behaviour.
\end{abstract}


\section{Introduction}
Scalar fields have played important roles in the physical description of the universe  in inflationary scenario \cite{Guth:1980zm}  as well as an explanation of late time acceleration of the universe. Examples of the latter are a quintessence scalar field  \cite{Ar1,qqq01,qqq02,qqq03} (generalizing the cosmological constant), a phantom scalar field (which, however, suffers ghosts instabilities \cite{lurena}),  a quintom scalar field model  \cite{quin00,Guo:2004fq,Feng:2004ff,Wei:2005nw,Zhang:2005eg,Zhang:2005kj,Lazkoz:2006pa,Lazkoz:2007mx,Setare:2008pz,Setare:2008pc,Leon:2008aq,Leon:2012vt,Leon:2014bta,Leon:2018lnd,Mishra:2018dzq,Marciu:2019cpb,Marciu:2020vve,Dimakis:2020tzc},   a chiral cosmology \cite{atr6,atr7,Dimakis:2020tzc,Paliathanasis:2020wjl}, or   multi-scalar field models (which describe various epochs of the cosmological history \cite%
{Paliathanasis:2018vru,Elizalde:2004mq,Elizalde:2008yf}).  Scalar field theories like scalar tensor theories, and many others have been exhaustively studied for example in \cite{Jordan,Brans:1961sx,Horndeski:1974wa,Copeland:1993jj,Lidsey:1995np,Ibanez:1995zs,Coley:1997nk,Copeland:1998fz,Coley:1999mj,Billyard:2000bh,Coley:2000zw,Coley:2000yc,Coley:2003tf,Curbelo:2005dh,Gonzalez:2005ie,Capozziello:2005tf,Gonzalez:2006cj,Gonzalez:2007ht,Hrycyna:2007gd,Leon:2009dt,Leon:2009rc,Leon:2009ce,Leon:2010pu,Miritzis:2011zz,Basilakos:2011rx,Xu:2012jf,Jamil:2012vb,Leon:2012mt,Leon:2013qh,Skugoreva:2013ooa,Fadragas:2013ina,Minazzoli:2014xua,Kofinas:2014aka,Paliathanasis:2014yfa,Leon:2014yua,Paliathanasis:2015gga,Leon:2015via,Harko:2015pma,Solomon:2015hja,DeArcia:2015ztd,Barrow:2016qkh,Barrow:2016wiy,Dimakis:2017kwx,Cruz:2017ecg,Matsumoto:2017gnx,Giacomini:2017yuk,Alhulaimi:2017ocb,Karpathopoulos:2017arc,Paliathanasis:2017ocj,DeArcia:2018pjp,Tsamparlis:2018nyo,Barrow:2018zav,VanDenHoogen:2018anx,Leon:2018skk,Humieja:2019ywy,Quiros:2019ktw,Leon:2019mbo,Paliathanasis:2019qch,Basilakos:2019dof,Shahalam:2019jgs,Paliathanasis:2019pcl,Leon:2019jnu,Nojiri:2019riz,Foster:1998sk,Miritzis:2003ym,Dania&Yunelsy,Leon:2008de,Giambo:2009byn,Giambo:2008ck,Leon:2010ai,Leon:2014rra,Fadragas:2014mra,vandenHoogen:1999qq,Copeland:1997et,Tzanni:2014eja,Giambo:2019ymx,Cid:2017wtf,Alho:2014fha}  by means of qualitative techniques of dynamical systems from \cite{Coddington55,Hale69,AP,wiggins,perko,160,Hirsch,165,LaSalle,aulbach,TWE,coleybook,Coley:94,Coley:1999uh,bassemah,LeBlanc:1994qm,Heinzle:2009zb}. Some works related to  Einstein-Klein-Gordon, Maxwell,  Yang-Mills,   Einstein-Vlasov systems, etc., are also studied in  \cite{Rendall:2003ks,TchapndaN.:2003bv,Alcubierre:2003sx,Rendall:2006cq,Liebscher:2012xt,Reiris:2015zaa,Lozanov:2017hjm,Wang:2018fay,Klainerman:2018mge,Ionescu:2019spj,Fajman:2019vma,Alho:2019pku,Siemonsen:2020hcg,Chatzikaleas:2020zmq,Chatzikaleas:2020twz,Barzegar:2020pna,Barzegar:2019nue,Barzegar:2020vzk,Alho:2015cza}. Perturbation methods and averaging methods were used in  \cite{Rendall:2006cq,Alho:2015cza} with interest in early and late-time dynamics. Slow-fast methods were used for example in theories based on a  Generalized
Uncertainty Principle (GUP), say in \cite{Paliathanasis:2015cza,Paliathanasis:2021egx}.
The amplitude-angle
transformation  was used in \cite{Leon:2019iwj,Leon:2020ovw} to study scalar field's oscillations driven by generalized harmonic potentials.  In reference \cite{Leon:2020pvt} interacting scalar field cosmologies with generalized harmonic potentials for flat and negatively curved  FLRW  metrics, and  for Bianchi I metrics were studied. Asymptotic and averaging methods were used to obtain stability conditions for several solutions of interest as $ H \rightarrow 0 $, where $H$ is the Hubble parameter. These results suggest that the asymptotic behavior of time-averaged model is independent of  coupling function and  geometry.
Averaging theory was used in reference \cite{Llibre:2012zz}  to study periodic orbits of Hamiltonian systems describing a universe filled with a scalar field; and in reference \cite{Fajman:2020yjb} to study future asymptotics of LRS Bianchi type III cosmologies with a massive scalar field. In reference \cite{Fajman:2021cli}  a theorem about  large-time behaviour of solutions of Spatially Homogeneous (SH) cosmology with oscillatory behaviour (when $H$ is non negative and monotonic decreasing to zero) was presented.  In references \cite{Leon:2020pfy,Leon:2020ovw}  scalar field cosmologies with arbitrary potential and with arbitrary coupling to matter were studied. In particular, generalized harmonic potentials and exponential couplings to matter in the sense of \cite{Dania&Yunelsy,Leon:2008de,Giambo:2009byn,Tzanni:2014eja} were examined.  This paper is a sequel of \cite{Leon:2020pfy,Leon:2020ovw}, where  asymptotic methods and averaging theory  \cite{dumortier,fenichel,Fusco,Berglund,holmes,Kevorkian1,Verhulst} were used to obtain relevant information about   solution's space of scalar field cosmologies with generalized harmonic potential: (i) in vacuum, (ii) in  presence of matter. 
As in \cite{Fajman:2020yjb}, we construct  averaged versions of  original systems where  oscillations of solutions  are smoothed out. Then, the analysis is reduced to study late-time dynamics of a simpler averaged system where oscillations entering the full system through KG equation can be controlled.  

This research program -- named ``Averaging Generalized Scalar Field Cosmologies''-- has three steps according to three cases of study:  (I)  Bianchi III and open FLRW model, (II)  Bianchi I and flat  FLRW model and (III) Kantowski-Sachs (KS) and closed FLRW. This paper is devoted to case I, and cases II and III will be studied in two companion papers \cite{Leon:2021rcx,Leon:2021hxc}. The main aspect in the present work is the interaction of KG fields and  field equations.  The paper is organized as follows. In section \ref{section-potential-choice} we discuss the class of generalized harmonic potentials in which we are interested.  In section \ref{Sect2} we introduce the models. 
 In section  \ref{SECT:II} we apply  averaging methods to analyze periodic solutions of a scalar field  with self-interacting potentials within the class of generalized harmonic potentials \cite{Leon:2019iwj}. In particular, in section \ref{SECT3.4}  LRS Bianchi III metric is studied. In section \ref{SECT:IIIA} is investigated  FLRW metrics with negative  curvature. 
 In section  \ref{SECT:III} we study averaged systems. Bianchi III metric is studied  in section \ref{LRSBIII}, and section \ref{FLRWflatopen} is devoted to FLRW metric with negative curvature (open FLRW).  Finally, in section \ref{Conclusions} our main results are discussed. 
In  \ref{appendixeq28} is proved  our main Theorem. In   \ref{AppCenter}  center manifold calculations for nonhyperbolic equilibrium points are presented. In  \ref{numerics} numerical evidences supporting the results of section \ref{SECT:II} are presented.

\section{Generalized harmonic potential}
\label{section-potential-choice}

Chaotic inflation is a model of cosmic inflation which takes the potential term $V$ of a hypothetical inflaton field $\phi$ is $V(\phi)= \frac{m_\phi^2 \phi^2}{2}$, the so-called harmonic potential ($\phi^2$-interaction) \cite{Linde:1983gd,Linde:1986fd,Linde:2002ws,Guth:2007ng}.  Whereas other inflationary models assume a monotonic decreasing potential with $\phi$; assuming in an ad hoc way that inflaton field has a large amplitude ``at Big Bang'', then slowly ``roll down''  the potential. The idea of \cite{Linde:1983gd} is that instead of inflaton rolls down and sits on its potential minimum at equilibrium,  quantum fluctuations stochastically (``chaotically'') drive it out of its minimum back and forward. Wherever this happens cosmic inflation sets in and blows up the region of ambient spacetime in which inflaton happened to  fluctuate out of its equilibrium. Relevant experimental results disfavoring  $\phi^2$-interaction are due to \cite{Planck:2013jfk,Ade:2015tva}. These results state that chaotic inflation generically predicts large values of tensor-to-scalar ratio $r$. In  contrast to recent measurements which show low upper bounds on $r$. Notwithstanding, we investigate variations of $\phi^2$- potential and we do not refer to tensor-to-scalar ratio issue for the potential \eqref{pot_v2}. 

The action integral of interest is
\begin{align}&
\int  d{ }^4 x \sqrt{|g|}\left[\frac{1}{2} R  -\frac{1}{2} g^{\mu
\nu}\nabla_\mu\phi\nabla_\nu\phi-V(\phi)+ \mathcal{L}_{m}\right].  \label{eq1}
\end{align} 
It is expressed in a system of units in which $8\pi G=c=\hslash=1$ where $\mathcal{L}_{m}$ is the Lagrangian density of matter, $R$ is the curvature scalar, $\phi$ is the scalar field,  $\nabla_\alpha$ is the covariant derivative  and the scalar field potential $V(\phi)$ of interest in this research is given by  \begin{equation}\label{pot}
 V(\phi)= \mu ^3 \left[b f \left(1-\cos \left(\frac{\phi }{f}\right)\right)+\frac{\phi ^2}{\mu}\right], \;    b> 0.
\end{equation} 
It is related but not equal to monodromy potential of  \cite{Sharma:2018vnv} used in the context of loop-quantum gravity, which is a particular case of general monodromy potential  \cite{McAllister:2014mpa}. 
In references  \cite{Leon:2019iwj,Leon:2020ovw,Leon:2020pvt} were proved that  potential of \cite{Sharma:2018vnv,McAllister:2014mpa} for $p=2$, say 
$V(\phi)= \mu^3 \left[\frac{\phi^2}{\mu} + b f \cos\left(\frac{\phi}{f}\right)\right]$, $b\neq 0$ is not good to describe the late-time FLRW universe driven by a scalar field  because it has two symmetric local negative minimums  which are related to  Anti-de-Sitter solutions. 
\begin{figure*}
    \centering
    \subfigure[$V(\phi)$ defined by \eqref{pot_v2} for $\mu=\frac{\sqrt{2}}{2}, \quad \omega = \sqrt{\Big{|}\frac{f-1}{f}\Big{|}} i, \quad f=\frac{1}{10}$. ]{\includegraphics[scale=0.8]{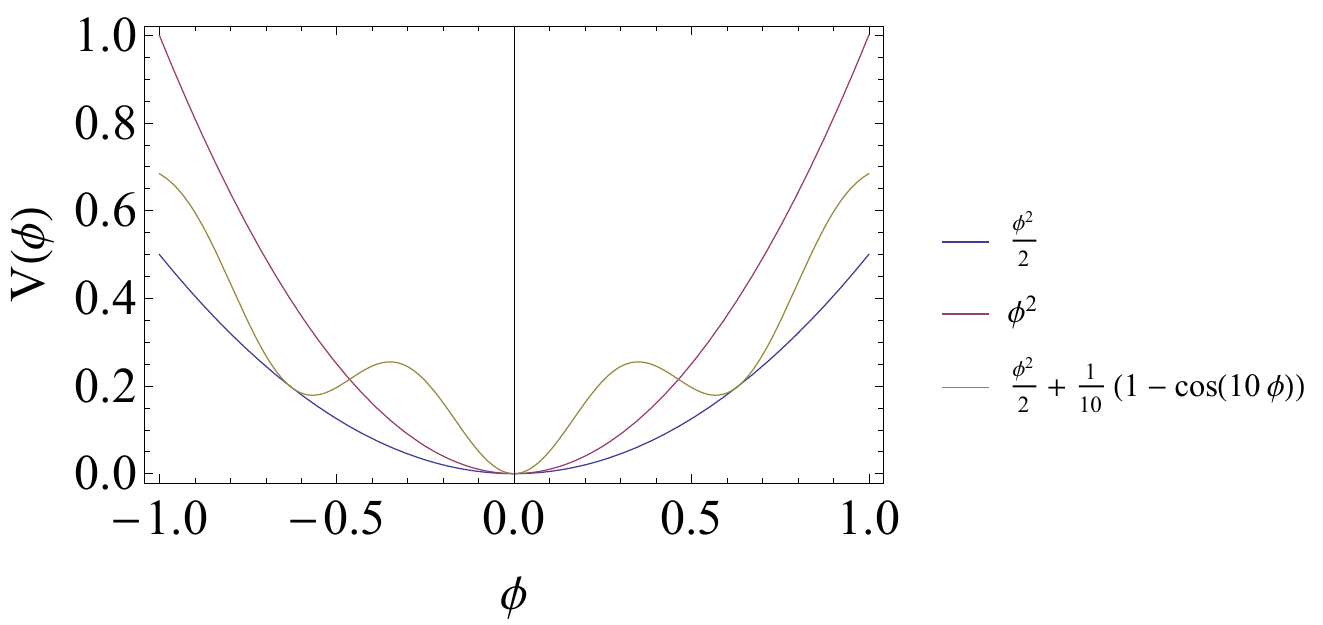}}
    \subfigure[$V(\phi)$ defined by \eqref{pot_v2} for $\mu=\frac{\sqrt{2}}{2}, \quad \omega=\sqrt{2}, \quad f=\frac{1}{10}$]{\includegraphics[scale=0.8]{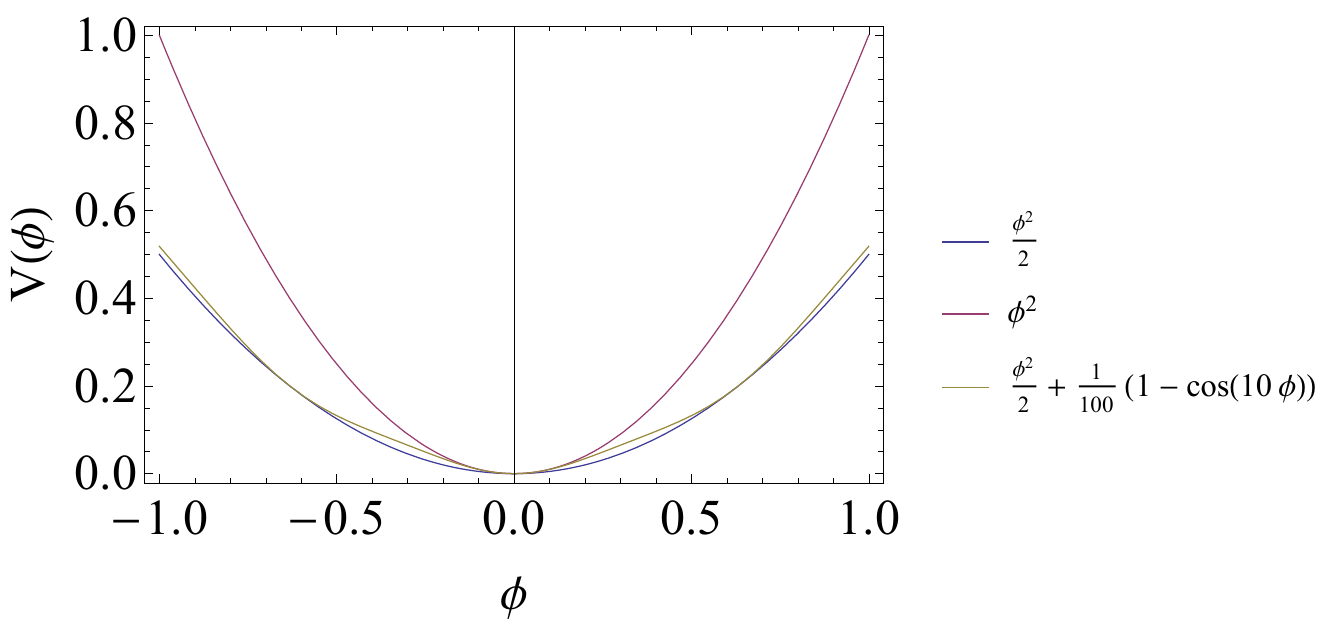}}
    \caption{Generalized harmonic potentials. Comparison with $\phi$-squared potentials.}
    \label{genharmpot}
\end{figure*}
Therefore, in \cite{Leon:2019iwj,Leon:2020ovw} we have studied the potential
\begin{equation}
\label{EQ:23}
    V(\phi)= \frac{\phi^2}{2}+ f\left[1- \cos\left(\frac{\phi}{f}\right)\right] 
\end{equation}
obtained by setting  $\mu=\frac{\sqrt{2}}{2}$ and $b \mu=2$ in equation \eqref{pot}. The potential \eqref{EQ:23} provides non-negative local minimums  which can be related to a late-time accelerated universe.  In section 2.4 of \cite{Leon:2020ovw}  a scalar field cosmology with  potential \eqref{EQ:23} non--minimally coupled to matter with coupling function  $\chi=\chi_0 e^{\frac{\lambda \phi}{4-3\gamma}}$ was studied, where  $\lambda$ is a constant and the barotropic index satisfies $0\leq \gamma \leq 2, \quad \gamma \neq \frac{4}{3}$ for FLRW metrics with $k=-1, 0$ and  Bianchi I metric. The late time attractors are associated to equilibrium points  with $\phi=\phi^*$   whenever  $\phi^*$ is a local non zero minimum of $V(\phi)$.  For FLRW metrics, global minimum is unstable to curvature perturbations for $\gamma>\frac{2}{3}$. Therefore, the result in \cite{Giambo:2019ymx} is confirmed, that for $\gamma> 2/3$  the curvature has a dominant effect on late evolution of the universe and it will eventually dominate both perfect fluid and scalar field energy densities. For Bianchi I model, the global minimum with $V(0)=0$   is unstable to shear perturbations.
\newline 
Additionally, potentials like  
$V(\phi)=\Lambda^4 \left[1- \cos \left(\frac{\phi}{f}\right)\right]$ are of interest in context of  axion models \cite{DAmico:2016jbm}. In  \cite{Balakin:2020coe} axionic dark matter model with a modified periodic potential for pseudoscalar
field  $V(\phi, \Phi_*)= \frac{m_A^2 {\Phi_*}^2}{2 \pi^2}\left[1- \cos \left(\frac{2 \pi \phi}{\Phi_*}\right)\right]$ in the framework of  axionic extension of Einstein-aether theory was studied. This periodic potential has minima at $\phi =n \Phi_*, n \in \mathbb{Z}$, whereas maxima when $n \rightarrow m +
\frac{1}{2}$ are found. Near the minimum when $\phi =n \Phi_* + \psi$ and $|\psi|$ is small, $V \rightarrow \frac{m_A^2 \psi^2}{2}$ where $m_A$ the axion rests mass.
\newline 
The previous statements justify the study of potential \eqref{pot}, which can be expressed as \begin{equation}
\label{pot_v2}
    V(\phi)=\mu ^2 \phi ^2 + f^2 \left(\omega ^2-2 \mu ^2\right) \left(1-\cos \left(\frac{\phi
   }{f}\right)\right), 
\end{equation}
by introducing an angular frequency $\omega \in\mathbb{R}$ through  conditions $b \mu ^3+2 f \mu ^2-f \omega ^2=0$ and $\omega^2-2 \mu^2>0$. Their applicability will be revealed in section \ref{SECT:II}. 

The generalized harmonic potentials \eqref{EQ:23} and \eqref{pot_v2} 
belong to the class of potentials studied by  \cite{Rendall:2006cq}. Potential \eqref{pot_v2}  has the following generic features:
\begin{enumerate} 
    \item$V$ is a real-valued smooth function  $V\in C^{\infty} (\mathbb{R})$  with  $\lim_{\phi \rightarrow \pm \infty} V(\phi)=+\infty$. 
        \item $V$ is an even function  $V(\phi)=V(-\phi)$.
    \item  $V(\phi)$ has always a local minimum at $\phi=0$;  $V(0)=0, V'(0)=0, V''(0)= \omega^2> 0$.
    \item There is a finite number of values $\phi_c \neq 0$ satisfying $2 \mu ^2 \phi_c +f \left(\omega ^2-2 \mu ^2\right) \sin \left(\frac{\phi_c
   }{f}\right)=0$, which are extreme points of $V(\phi)$. They are local maximums or local minimums depending on whether  $V''(\phi_c):= 2 \mu ^2+\left(\omega ^2- 2 \mu ^2\right) \cos \left(\frac{\phi_c }{f}\right)<0$ or $V''(\phi_c)>0$. For $\left|\phi_c\right| >\frac{f(\omega^2-2 \mu^2)}{2 \mu^2}= \phi_*$ this set is empty. 
    \item There exist 
    $V_{\max}= \max_{\phi\in [-\phi_*,\phi_*]} V(\phi)$   and $V_{\min}= \min_{\phi\in [-\phi_*,\phi_*]} V(\phi)=0$. The function $V$ has no upper bound  but it has a lower bound equal to zero.
  \end{enumerate}
The asymptotic features of  potential \eqref{pot_v2} are the following.  Near global minimum $\phi=0$, we have 
$V(\phi) \sim \frac{\omega ^2 \phi ^2}{2}+\mathcal{O}\left(\phi ^3\right), \quad \text{as} \; \phi\rightarrow 0$. 
That is, $\omega^2$ can be related to the mass of the scalar field near its global minimum.  As $\phi\rightarrow \pm \infty$  cosine- correction is bounded, then, $V(\phi) \sim \mu ^2 \phi ^2+\mathcal{O}\left(1\right) \quad \text{as} \; \phi\rightarrow \pm \infty$.  This makes it suitable to describe oscillatory behavior in cosmology. 

Setting
$\mu=\frac{\sqrt{2}}{2}$,  $\omega=\sqrt{2}$, we have 
\begin{equation}
\label{pot28}
 V(\phi)= \frac{\phi ^2}{2} + f^2 \left[1-\cos \left(\frac{\phi }{f}\right)\right]. 
\end{equation}
Setting 
$\mu=\frac{\sqrt{2}}{2},  \;  \omega = \sqrt{\frac{f-1}{f}}$, 
the potential  of \cite{Leon:2019iwj,Leon:2020ovw,Leon:2020pvt} \eqref{EQ:23} is recovered.  Although \eqref{EQ:23} can be derived as a particular case of our study, cases of interest with $f\ll 1$ leads to complex frequency $\omega = \sqrt{\Big{|}\frac{f-1}{f}\Big{|}} i$, in contradiction to $\omega \in \mathbb{R}$.  Therefore, potential \eqref{pot_v2} will not contain potential studied in  \cite{Leon:2019iwj,Leon:2020ovw,Leon:2020pvt}, unless we set $f>1$ and $\omega = \sqrt{\frac{f-1}{f}}$.
In Figure \ref{genharmpot} potentials \eqref{pot28} and \eqref{EQ:23} are depicted. 

\section{Spatially homogeneous scalar field cosmologies}
\label{Sect2}
In General Relativity (GR) the SH but anisotropic spacetimes are known as either  Bianchi or KS metrics.  
In Bianchi models,
the spacetime manifold is foliated along the time axis with three dimensional
homogeneous hypersurfaces.  Bianchi spacetimes contain many important cosmological models that have been used to study  anisotropies of
primordial universe and its evolution towards current observed isotropy \cite{jacobs2,collins,JB1,JB2}. The list  includes standard FLRW model in the limit of isotropization;  Bianchi III isotropizing to open FLRW models and Bianchi I isotropizing to flat FLRW models.  Hubble parameter $H$ is always monotonic for Bianchi I and Bianchi III. For Bianchi I anisotropy decays on time for $H>0$ and isotropization occurs \cite{nns1}. 

There is an interesting hierarchy in Bianchi models \cite{WE,Ryan2016,coleybook,Plebanski2006}. In  particular,  LRS Bianchi I model naturally appears as a boundary subset of  LRS Bianchi III model. The last one is an invariant boundary of the LRS Bianchi type VIII model as well. Additionally, LRS Bianchi type VIII can be viewed as an invariant boundary of  LRS Bianchi type IX model \cite{Byland:1998gx,BC1,BC2,BC3,BC4,BC5,BC6}. 
Bianchi spacetimes in presence of a scalar field were studied in
\cite{heu}. It was proved that an initial anisotropic universe isotropizes into a FLRW universe for specific initial conditions if the scalar field
potential has a large positive value. An exact
solution of  field equations for an exponential potential in some particular Bianchi
spacetimes has been found in \cite{b1,b2,b3}. These exact solutions lead to isotropic
homogeneous spacetimes as it was found in \cite{coley1,coley2}. An anisotropic solution of special interest is Kasner spacetime.  Kasner solution is
essential for the description of BKL singularity  when the
contribution of Ricci scalar of  three-dimensional spatial hypersurface
in the field equations is negligible \cite{bkl}. For other applications
of  Kasner universe and Bianchi I spacetimes in
gravitational physics see
\cite{kas1,kas2,kas3,kas4,barcl,barcl2,anan01,anan02} and references therein.
In \cite{Mitsopoulos:2019afs} the conformal algebra of Bianchi III and Bianchi V spacetimes which admit a proper conformal Killing vector were studied. In  \cite{Paliathanasis:2016rho}  method of Lie symmetries was applied for the Wheeler-De Witt equation in Bianchi Class A cosmologies for minimally coupled scalar field gravity and Hybrid Gravity in GR. Using these symmetries several invariant solutions were determined and classified according to the form of scalar field potential.

\subsection{LRS Bianchi III,  Bianchi I  and Kantowski-Sachs models}
Due to 
\begin{align}
  & \lim_{k\rightarrow -1}  k^{-1} \sin^2 (\sqrt{k} \vartheta)=\sinh^2 (\vartheta), \\
  & \lim_{k\rightarrow 0}  k^{-1} \sin^2 (\sqrt{k} \vartheta)= \vartheta^2,\\
   & \lim_{k\rightarrow 1}  k^{-1} \sin^2 (\sqrt{k} \vartheta)=\sin^2 (\vartheta), 
\end{align}
the  metric element for LRS Bianchi III,  Bianchi I  and KS models can be written as \cite{Nilsson:1995ah}
\begin{align}
\label{metric}
   &  ds^2= - dt^2 + \left[{e_1}^1(t)\right]^{-2} dr^2 \nonumber \\
   & + \left[{e_2}^2(t)\right]^{-2}  \left[ d \vartheta^2 + k^{-1} \sin^2 (\sqrt{k} \vartheta)d \zeta^2\right],
\end{align}
where ${e_1}^1$, ${e_2}^2$ and ${e_3}^3= \sqrt{k} {e_2}^2/\sin(\sqrt{k} \vartheta)$ are functions of $t$ which are components of the frame vectors \cite{Coley:2008qd}:
$\mathbf{e}_0= \partial_t, \quad  \mathbf{e}_1 = {e_1}^1\partial_r, \quad \mathbf{e}_2={e_2}^2 \partial_\vartheta, \quad \mathbf{e}_3={e_3}^3 \partial_\zeta$. That is, we obtain  
LRS Bianchi III for $k=-1$,  Bianchi I  for $k=0$ and  KS for $k=+1$ \cite{Fadragas:2013ina}.
Comparing with reference \cite{Nilsson:1995ah}
we have settled parameters $a=f=0$ and ${e_1}^1(t)=D_2(t)^{-1}, \; {e_2}^2(t)=D_2(t)^{-1}$ in their metric and we have used the identifications $(\vartheta, \zeta)=(y, z)$.  The line elements for spatially selfsimilar LRS models have been given by Wu in \cite{Chao:1980ky}. We only focus on SH but anisotropic class with the exception of SH LRS Bianchi V, that is:  LRS Bianchi III,  Bianchi I  and  KS. 
\newline 
It is useful to define a representative length  along  worldlines of the 4-velocity vector  $\mathbf{u}=\partial_t$ describing the volume expansion
(contraction), the behavior of congruence completely denoted  $\ell(t)$ and defined by 
\begin{equation}
\frac{\dot \ell(t)}{\ell(t)}= H(t):=     -\frac{1}{3}\frac{d}{d t} \ln\left[ {e_1}^1(t) ( {e_2}^2(t))^2\right],
\label{HubbleGen}
\end{equation}
where the Hubble parameter $H(t)$ 
and the anisotropic parameter $\sigma_{+}(t)$ are given by 
\begin{equation}
\sigma_+ = \frac{1}{3}\frac{d}{d t} \ln\left[ {e_1}^1(t) ( {e_2}^2(t))^{-1}\right].
\end{equation} 
Taking variation of \eqref{eq1}  for the 1-parameter family of metrics \eqref{metric} leads to \cite{Fadragas:2013ina}: 
\begin{align}
    & 3 H^2 + k K = 3 {\sigma_+}^2+ \rho_m + \frac{1}{2}\dot\phi^2 +V(\phi), \label{EQQ32}
\\
    & -3 ({\sigma_+}+H)^2 -2 \dot{\sigma_+}- 2 \dot{H} - k K \nonumber \\
    & = (\gamma-1)\rho_m + \frac{1}{2}\dot\phi^2-V(\phi), \label{EQQ33}
\\
      & -3 {\sigma_+}^2 +3  {\sigma_+} H-3 H^2 + \dot{\sigma_+}- 2 \dot{H} \nonumber \\
      & = (\gamma-1)\rho_m + \frac{1}{2}\dot\phi^2-V(\phi). \label{EQQ34}
\end{align}
For modeling matter in our model we use a perfect fluid with  a barotropic EoS $p_m=(\gamma-1)\rho_m$ with pressure $p_{m}$, energy density  $\rho_{m}$ and barotropic index $\gamma\in[0,2]$.
\newline
The  Gauss curvature of  spatial 2-space and 3-curvature scalar are
\begin{equation}
 K= ( {e_2}^2(t))^{2}, \quad   \R= 2 k K.
\end{equation} 
Furthermore, evolution of $K$ is
\begin{equation}
\label{Gauss}
    \dot{K}= -2 ({\sigma_+}+ H)K, 
\end{equation}
while evolution for ${e_1}^1$ is given by \cite{Coley:2008qd}: \begin{equation}
\label{eqe1evol}
    \dot{{e_1}^1}= - (H-2 \sigma_+){e_1}^1. 
\end{equation}
From Eqs. \eqref{EQQ33}, \eqref{EQQ34}  the shear equation is obtained: 
\begin{equation}
 \dot{\sigma_+}= -3 H {\sigma_+} -\frac{k K}{3}. \label{EQQ39}
\end{equation}
Eqs. \eqref{EQQ32}, \eqref{EQQ33}, \eqref{EQQ34},  \eqref{EQQ39} give the Raychaudhuri equation: 
\begin{equation}
    \dot{H}=- H^2 -2 {\sigma_+}^2 -\frac{1}{6}(3 \gamma -2) \rho_m -\frac{1}{3}\dot{\phi}^2+\frac{1}{3} V(\phi).
\end{equation}
Finally, the matter and KG equations are:    
\begin{align}
& \dot{\rho_m}=-3 \gamma H \rho_m,\\
& \ddot{\phi}= -3 H \dot{\phi}- \frac{d V(\phi)}{d \phi}. \end{align}
\newline 
In this paper we will study LRS Bianchi III model. By convenience we write 
\begin{align}
\label{metricLRSBIII}
   &  ds^2= - dt^2 + A(t)^2 dr^2 
 + B(t)^2  \mathbf{g}_{H^2}.
\end{align}
where $\mathbf{g}_{H^2}=   d \vartheta^2 +  \sinh^2 (\vartheta)d \zeta^2$ denotes the 2-metric of negative constant curvature on hyperbolic 2-space.  The functions $A(t)$ and $B(t)$, interpreted as scale factors, are defined as $A(t)= {e_1}^1(t)^{-1}$ and $B(t)^2= K(t)^{-1}$.

\subsection{FLRW models}
The general line element for spherically symmetric (SS) models can be written as \cite{Coley:2008qd}
\begin{align}
\label{sphsymm}
   &  ds^2= - dt^2 + \left[{e_1}^1(t,r)\right]^{-2} dr^2 \nonumber \\
   & + \left[{e_2}^2(t,r)\right]^{-2} (d\y^2 + \sin^2 \y\, d\z^2).
\end{align}
SH-SS models that are not Kantowski-Sachs  are FLRW models  where the metric can be written as 
\begin{align}
\label{metricFLRW}
& ds^2 = - dt^2 + a^2(t) \Big[ dr^2+  f^2(r) (d\y^2 + \sin^2 \y\, d\z^2)\Big],
\\
\label{fx_FLRW}
& \text{with}\; f(r) = \sin r,\ r,\ \sinh r 
\end{align}
for closed, flat and open FLRW models, respectively. 
In comparison with metric \eqref{sphsymm},  frame coefficients here are given by $\ex = a^{-1}(t)$ and $\ey = a^{-1}(t) f^{-1}(r)$ where $a(t)$ is the scale factor. Anisotropic parameter $\sigma_{+}= \frac13\frac{\partial}{\partial t} \ln(\ex/\ey)$ vanishes and Hubble parameter \eqref{HubbleGen} can be written as $H = \frac{d}{d t} \ln\left[a(t)\right]$.
\newline Furthermore, $\R$ is
\be
    \R = \frac{6k}{a^2},\quad
    k = 1,0,-1,
\ee
for closed, flat and open FLRW models, respectively. Therefore, evolution/constraint equations reduce to
\begin{subequations}
	\label{Non_minProb1FLRW0-1}
	\begin{align}
	&\ddot{\phi}= -3 H \dot{\phi}-V'(\phi),
\\
	&\dot{\rho_m}=- 3\gamma H\rho_m,
	\\
	&\dot{a} = a H, 
	\\
	& \dot{H}=-\frac{1}{2}\left(\gamma \rho_m+{\dot \phi}^2\right)+\frac{k}{a^2},
	\\
	& 3H^2=\rho_m+\frac{1}{2}{\dot{\phi}}^2+V(\phi)-\frac{3 k}{a^2}.
	\end{align}
\end{subequations}
By setting $\dot\phi=\rho_m=0$ and $V(\phi)=\Lambda$ we obtain vacuum cases (with or without cosmological constant $\Lambda$), which are  de Sitter model ($\Lambda>0$, $k=0$),
the model with $\Lambda>0$, $k=1$,
the model with $\Lambda>0$, $k=-1$,
Milne model ($\Lambda=0$, $k=-1$)
and  Minkowski spacetime ($\Lambda=0$, $k=0$), which is also static.
The model with $\Lambda>0$, $k=1$ is past asymptotic to  de Sitter
model with negative $H$ and is future asympotic to  de Sitter model
with positive $H$.
The model with $\Lambda>0$, $k=-1$ (and positive $H$) is past asymptotic
to  Milne model and it is future asympotic to  de Sitter model with
positive $H$.

In this paper we will study open FLRW model. By convenience we write the metric \eqref{mOpenFLRW} for $k=-1$ as 
\begin{align}
\label{mOpenFLRW}
   &  ds^2= - dt^2 + a(t)^2 dr^2 
 + a(t)^2  S_{-1}(r)^2 d\Omega^2.
\end{align}
where $ S_{-1}(r)=\sinh(r), d\Omega^2= d\y^2 + \sin^2 \y\, d\z^2$. 

\section{Averaging scalar field cosmologies}
\label{SECT:II}
 Given the differential equation 
 $\dot{\mathbf{x}}= \mathbf{f}(t, \mathbf{x},\varepsilon)$ with $\mathbf{f}$ periodic in $t$. One approximation scheme which can be used to solve the full problem is solving the unperturbed problem $\dot{\mathbf{x}}= \mathbf{f}(t, \mathbf{x},0)$ by setting $\varepsilon=0$ and then  use the approximated unperturbed solution to formulate variational equations in standard form which can be averaged.  
The term averaging is related to approximation  of initial value problems in ordinary differential equations which involves perturbations (chapter 11,  \cite{Verhulst}). 

\subsection{Simple example}

For example, consider this simple equation 
 \begin{equation}
 \label{harm-osc}
 \ddot \phi + \omega^2 \phi = \varepsilon (-2 \dot \phi)
 \end{equation}
 with $\phi(0)$  and $\dot\phi(0)$  given. The unperturbed problem 
 $\ddot \phi +\omega^2 \phi = 0$  
admits solution 
 $\dot\phi(t)= r_0 \omega \cos (\omega t-\Phi_0), \;  \phi(t)= r_0 \sin (\omega t-\Phi_0)$,
 where $r_0$ and $\Phi_0$ are constants depending on the initial conditions. 
 Let be defined the amplitude-phase transformation (\cite{Verhulst}, chapter 11):
 \begin{small}
 \begin{equation}
 \dot{\phi}(t)= r(t) \omega \cos (\omega t-\Phi(t)), \;  \phi(t)  = r(t) \sin (\omega t-\Phi(t)),
 \end{equation}
 \end{small}
 such that
 \begin{small}
 \begin{equation}
 \label{eqAA25}
 r=\frac{\sqrt{\dot{\phi}^2(t)+\omega ^2 \phi^2(t)}}{\omega }, \;  \Phi =\omega t-\tan ^{-1}\left(\frac{\omega \phi
   (t)}{\dot \phi(t)}\right). 
 \end{equation}
 \end{small}
 Then,  equation \eqref{harm-osc} 
 becomes 
 \begin{equation}
 \label{eq4}
 \dot r= -2 r \varepsilon  \cos ^2(t-\Phi), \;  \dot\Phi = - \varepsilon \sin (2 (t-\Phi )).
 \end{equation}
From \eqref{eq4} it follows that $r$ and $\Phi$ are  slowly varying with time,
 and the system takes the form $\dot y= \varepsilon f(y)$.  The idea is consider only nonzero average of  right-hand-sides keeping $r$ and $\Phi$
 fixed  and leaving out  terms with average zero and ignoring slow-varying dependence of $r$ and $\Phi$ on $t$ through averaging process: 
 \begin{equation}
\label{timeavrg}
      \bar{\mathbf{f}}(\cdot):=\frac{1}{L} \int_{0}^L \mathbf{f}(\cdot, t) dt, \quad L=\frac{2 \pi}{\omega}.  
\end{equation}
Replacing $r$ and $\Phi$ by their averaged approximations $\bar{r}$ and $\bar{\Phi}$ we obtain the system
 \begin{align}
 \label{eq6}
 & \dot {\bar{r}} = - \varepsilon \omega \bar{r}, 
\quad  \dot{\bar{\Phi}} = 0. 
 \end{align}
   Solving \eqref{eq6} with initial conditions $\bar{r}(0)=r_0$ and $\bar{\Phi}(0)= \Phi_0$, we obtain  $\bar{\phi}= r_0 e^{-\varepsilon \omega t} \sin (\omega t-\Phi_0)$, which is an accurate approximation of the exact solution
   \begin{align*}
     & \phi(t)=  -r_0  e^{-t \varepsilon }  \sin (\Phi_0) \cos
   \left(t \sqrt{\omega ^2-\varepsilon ^2}\right) \nonumber \\
   & -\frac{r_0 e^{-t
   \varepsilon } \sin \left(t \sqrt{\omega ^2-\varepsilon ^2}\right) (\varepsilon 
   \sin (\Phi_0)-\omega  \cos (\Phi_0))}{\sqrt{\omega
   ^2-\varepsilon ^2}},
   \end{align*} due to 
\begin{align*}
\bar{\phi}(t) - \phi(t)  =  \frac{r_0 \varepsilon  e^{-t \varepsilon } \sin (\Phi_0) \sin (t  \omega )}{\omega } + \mathcal{O}\left(\varepsilon  e^{-t \varepsilon }\right)
\end{align*}
as $\varepsilon \rightarrow 0^+$.

\subsection{General class of systems with a time-dependent perturbation parameter}
Generalizing \eqref{harm-osc}, now let us consider the KG system
\begin{align}
\label{KGharmonic}
   & \ddot \phi + \omega^2 \phi = -3 H \dot \phi, \\
   &\dot{H}= -\frac{1}{2}\dot\phi^2. \label{Friedmann}
\end{align} The similarity between \eqref{harm-osc} and \eqref{KGharmonic}  suggests treating the latter as a perturbed harmonic
oscillator as well, and applying averaging in an analogous way; taking into consideration that, in contrast to $\varepsilon$, $H$ is time-dependent and itself is governed by    evolution equation \eqref{Friedmann}. If it is valid, then a surprising feature of such approach is the possibility of exploiting the fact that $H$ is strictly decreasing and goes to zero, therefore, it can be promoted  to a time-dependent perturbation parameter in \eqref{KGharmonic}; controlling the magnitude of the error between solutions of full and time-averaged problems. 
Hence, with strictly decreasing $H$ the error should decrease as well. Therefore, it is possible to obtain information about  large-time behaviour of more complicated full system via an analysis of simpler averaged system equations. 
\newline 
With this in mind, in \cite{Fajman:2021cli} the long-term behavior of solutions of a general class of systems in standard form  
\begin{equation}
\label{standard51}
\left(\begin{array}{c}
       \dot{H} \\
        \dot{\mathbf{x}}
  \end{array}\right)= H \left(\begin{array}{c}
       0 \\
       \mathbf{f}^1 (\mathbf{x}, t)
  \end{array}\right) + H^2\left(\begin{array}{c}
       f^{[2]} (\mathbf{x}, t) \\
       \mathbf{0}
  \end{array}\right),
  \end{equation}
was studied; where $H>0$ is strictly decreasing in $t$ and $\lim_{t\rightarrow \infty}H(t)=0$. 

The following Theorem by \cite{Fajman:2021cli} gives local-in-time asymptotics for system \eqref{standard51}. Let the norm $\|\cdot\|$ denotes the standard discrete $\ell^1$- norm $\|\mathbf{u}\| := \sum_i^n |u_i|$ for $\mathbf{u}\in \mathbb{R}^n$. Let also $L_{\mathbf{x}, t}^\infty$ denotes the standard $L^{\infty}$ space in both $t$ and $\mathbf{x}$ variables with  norm defined as  $\|\mathbf{f}\|_{L_{\mathbf{x}, t}^\infty}:= \sup_{\mathbf{x}, t}|\mathbf{f}(\mathbf{x}, t)|.$ 

\begin{thm}[Theorem 3.1 of \cite{Fajman:2021cli}]
\label{localintime}
Suppose  $H(t)>0$ is strictly decreasing in $t$ and $\lim_{t\rightarrow \infty} H(t)=0.$ Fix any $\epsilon>0$ with $\epsilon<H(0)$ and define $t_*>0$ such that $\epsilon=H(t_*).$ Suppose that $\|\mathbf{f}^1\|_{L_{\mathbf{x}, t}^\infty},\quad \|f^{[2]}\|_{L_{\mathbf{x}, t}^\infty}<\infty$ and that $\mathbf{f}^1(\mathbf{x}, t)$ is Lipschitz continuous and $f^{[2]}$ is continuous with respect to $x$ for all $t\geq t_*.$ Also, assume that $\mathbf{f}^1$ and $f^{[2]}$ are $T$-periodic for some $T>0.$ Then for all $t>t_*$ with $t=t_*+\mathcal{O}\Big(H(t_*)^{-\delta}\Big)$ for any given $\delta \in (0,1)$ we have $$\mathbf{x}(t)-\mathbf{z}(t)=\mathcal{O}\Big(H(t_*)^{\min\{1,2-2\delta\}}\Big),$$  where $\mathbf{x}$ is the solution of system \eqref{standard51} with initial condition $\mathbf{x}(0)=\mathbf{x}_0$ and $\mathbf{z}(t)$ is the solution of the time-averaged system 
\begin{equation*}
    \dot{\mathbf{z}}=H(t_*)\bar{\mathbf{f}}^1(\mathbf{z}),\quad \text{for} \quad t>t_*,
\end{equation*} 
with initial condition $\mathbf{z}(t_*)=\mathbf{x}(t_*)$ where the time-averaged vector $\bar{\mathbf{f}}^1$ is defined as 
\begin{equation*}
   \bar{\mathbf{f}}^1(\mathbf{z})=\frac{1}{T}\int_{t_*}^{t_*+T}\mathbf{f}^1(\mathbf{z},s)ds.
\end{equation*}
\end{thm}
Therefore, Hubble parameter $H$ can be used as a time-dependent perturbation parameter. 

In this paper we study systems which are not in the standard form \eqref{standard51} but can be expressed as a series with center in $H=0$ according to the equation
\begin{align}
\label{nonstandtard}
 \left(\begin{array}{c}
       \dot{H} \\
        \dot{\mathbf{x}}
  \end{array}\right)= &\left(\begin{array}{c}
    0 \\
       \mathbf{f}^0 (\mathbf{x}, t)
  \end{array}\right)+ H \left(\begin{array}{c}
       0 \\
       \mathbf{f}^1 (\mathbf{x}, t)
  \end{array}\right) \nonumber \\
  & + H^2\left(\begin{array}{c}
       f^{[2]} (\mathbf{x}, t)  \\
       \mathbf{0}
  \end{array}\right)+ \mathcal{O}(H^3),
  \end{align}
depending on a parameter $\omega$ which is a free frequency that can be tuned to make $\mathbf{f}^0 (\mathbf{x}, t)= \mathbf{0}$. Therefore,  systems  can be expressed in the standard form \eqref{standard51}. 
\subsection{LRS Bianchi III}
\label{SECT3.4}
Firstly, we consider LRS Bianchi III metrics for the generalized harmonic potential \eqref{pot} minimally coupled to matter with field equations:
\begin{align}
& \ddot{\phi}= -3 H \dot{\phi} - V'(\phi), \\
& \dot{\rho_m}= -3 \gamma H \rho_m,  \\
& \dot{K} = -2 ({\sigma_+} +H) K,  \\
& \dot{H}= -H^2 -2 {\sigma_+}^2 -\frac{1}{6} (3 \gamma -2) \rho_m -\frac{1}{3} {\dot{\phi}^2} +\frac{1}{3} V(\phi), \\
& \dot{{\sigma_+}}= -3 H {\sigma_+} +\frac{K}{3},\\
    & 3 H^2 = 3 {\sigma_+}^2 +\rho_m+\frac{1}{2}{\dot{\phi}}^2 +V(\phi)+K.
\end{align}
Defining 
\begin{align}
& \Omega=\sqrt{\frac{\omega^2 \phi^2+{\dot\phi}^2}{6 H^2}}, \;   \Sigma=\frac{{\sigma_+}}{H},  \;  \Omega_k=\frac{K}{3 H^2}, \nonumber \\
&  \Phi= t \omega -\tan^{-1}\left(\frac{\omega \phi}{\dot \phi}\right),
\end{align}
we obtain the full system 
\begin{widetext}
\begin{subequations}
\label{BIIIunperturbed1}
\begin{small}
\begin{align}
  &\dot{\Omega}=  -\frac{b  \mu ^3}{\sqrt{6} H } \cos (t \omega -\Phi  ) \sin  \left(\frac{\sqrt{6} H  \sin (t \omega -\Phi  ) \Omega  }{f \omega }\right)  -\frac{b f \gamma  \Omega   \mu ^3}{H }  \sin ^2 \left(\frac{\sqrt{\frac{3}{2}} H  \sin (t \omega
   -\Phi  ) \Omega  }{f \omega }\right) \nonumber \\
   & +\frac{\left(\omega ^2-2 \mu ^2\right)\Omega  }{2 \omega }  \sin (2 t \omega -2 \Phi  )     + 3 H  \left(\left(\gamma  \left(\frac{
   \mu ^2}{\omega ^2}-\frac{1}{2}\right)+1\right) \Omega  ^3- \Omega  \right) \cos ^2(t \omega -\Phi  ) \nonumber \\
   & +\frac{1}{2} H  \Omega   \Big(-3 (\gamma -2) \Sigma  ^2-\frac{6 \gamma  \mu ^2 \Omega  ^2}{\omega ^2}   +3 \gamma +(2-3 \gamma )
    {\Omega_k} \Big),
\end{align}
\begin{align}
   & \dot{\Sigma}= -\frac{b f \gamma   \Sigma   \mu ^3}{H } \sin ^2\left(\frac{\sqrt{\frac{3}{2}} H  \sin (t \omega -\Phi  ) \Omega  }{f \omega }\right)    -\frac{3}{2} (\gamma -2) 
   H  \Sigma   \Omega  ^2 \cos ^2(t \omega -\Phi  ) \nonumber \\
   & +\frac{1}{4} H  \Bigg(4  {\Omega_k}-6 (\gamma -2) \Sigma  ^3     +2 \Big(-\frac{6 \gamma  \mu ^2  \Omega  ^2}{\omega ^2} \sin ^2(t \omega -\Phi  )  +3 \gamma -3 \gamma   {\Omega_k} +2  {\Omega_k} -6\Big) \Sigma \Bigg),
\end{align}
\begin{align}
   & \dot{\Omega_k}=
   \frac{b f \gamma  {\Omega_k}  \mu ^3}{H }  \left(-1+\cos \left(\frac{\sqrt{6} H  \sin (t
   \omega -\Phi  ) \Omega  }{f \omega }\right)\right)  -3 (\gamma -2) H  \Omega  ^2  {\Omega_k}  \cos ^2(t \omega -\Phi  ) \nonumber \\
   & +\frac{1}{2} H  \Big(-6 (\gamma -2) \Sigma  ^2-4
   \Sigma  -2 (3 \gamma -2) ( {\Omega_k} -1)      -\frac{12 \gamma  \mu ^2 \Omega  ^2}{\omega ^2} \sin ^2(t \omega -\Phi  ) \Big)  {\Omega_k} ,
\end{align}
\begin{align}& \dot{\Phi}= -\frac{b  \mu ^3}{\sqrt{6} H  \Omega  } \sin (t \omega -\Phi  ) \sin \left(\frac{\sqrt{6} H 
   \sin (t \omega -\Phi  ) \Omega  }{f \omega }\right)     +\frac{\left(\omega ^2-2 \mu ^2\right) }{\omega } \sin ^2(t \omega -\Phi  )   -3 \cos (t \omega -\Phi  ) H  \sin (t \omega -\Phi  ),   
   \end{align}
   \end{small}
and the Raychauhuri equation is
\begin{align}
   & \dot{H}= -(1+q) H^2, \label{RaychBIII}\end{align}
   \end{subequations}  
    \end{widetext} 
where the deceleration parameter is given by 
\begin{align}
& q=    -\frac{b f \gamma   \mu ^3}{H ^2} \sin ^2 \scriptscriptstyle \left(\frac{\sqrt{\frac{3}{2}} H  \sin (t \omega -\Phi
    ) \Omega  }{f \omega }\right)   \nonumber \\
    & -\frac{3}{2} (\gamma -2) \cos ^2(t \omega -\Phi  ) \Omega  ^2 -\frac{3}{2} (\gamma -2) \Sigma  ^2  \nonumber \\
    &   -\frac{3 \gamma  \mu ^2  \Omega
    ^2}{\omega ^2} \sin ^2(t \omega -\Phi  ) - \frac{1}{2} (3 \gamma -2) ( {\Omega_k} -1).
\end{align}
\newline 
Defining $\mathbf{x}= \left(\Omega, \Sigma, \Omega_k, \Phi \right)^T$, the system \eqref{BIIIunperturbed1} can be symbolically written  as a Taylor series of the form \eqref{nonstandtard}. Notice that the term 
\begin{small}
\begin{equation}
\mathbf{f}^0 (t, \mathbf{x}) 
= \left(
\begin{array}{c}
 \frac{\Omega (t) \left(f \omega ^2-\mu ^2 (b \mu +2 f)\right) \sin (2 t \omega -2 \Phi (t))}{2 f \omega } \\
 0 \\
 \frac{\left(-b \mu ^3-2 f \mu ^2+f \omega ^2\right) \sin ^2(t \omega -\Phi (t))}{f \omega } \\
 0 \\
\end{array}
\right)
\end{equation} 
\end{small}
\newline in  expression \eqref{nonstandtard} is eliminated imposing the condition $b \mu ^3+2 f \mu ^2-f \omega ^2=0$,  which defines an angular frequency $\omega \in\mathbb{R}$. Then, order zero terms 
in the series expansion around $H=0$ are eliminated  assuming $\omega ^2>2 \mu ^2$ and setting $f=\frac{b \mu ^3}{\omega ^2-2 \mu ^2}$, which is equivalent to tune $\omega$.
Hence, we obtain: 
\begin{widetext}
\begin{align}
& \dot{\mathbf{x}}= H \mathbf{f}(t, \mathbf{x})+ \mathcal{O}(H^2), \label{equx}\\
 & \dot{H}= -\frac{3}{2} H^2 \Big(\gamma(1- \Sigma ^2- {\Omega_k}-\Omega^2)   +2 \Sigma^2 +\frac{2}{3}{\Omega_k}  + 2 \Omega
   ^2 \cos^2(t \omega -\Phi) \Big) +\mathcal{O}(H^3), \label{EQ:81b}
\end{align}
where 
\begin{align}
\label{EQ:108}
 & \mathbf{f}(t, \mathbf{x}) =
   \left(\begin{array}{c}
\frac{1}{2}  \Omega  \Big(-3 (\gamma -2) \Sigma ^2+(2-3 \gamma )  {\Omega_k}   +3 \left(\Omega ^2-1\right) (-\gamma +2 \cos^2(t \omega -\Phi))\Big) \\\\
\frac{1}{2} \Bigg( {\Omega_k} ((2-3 \gamma ) \Sigma
   +2)   +3 \Sigma  \Big(-(\gamma -2) \Sigma ^2+\gamma   +\Omega ^2 (-\gamma +2 \cos^2(t \omega -\Phi))-2\Big)\Bigg)\\\\
   {\Omega_k} \Big(-3 \gamma  \left(\Sigma ^2+\Omega ^2+ {\Omega_k}-1\right)  
    +6 \Sigma ^2-2
   \Sigma + 6 \Omega ^2 \cos^2(t \omega -\Phi)  +2  {\Omega_k}-2\Big) \\\\
-\frac{3}{2} \sin (2 t \omega -2\Phi)
      \end{array}
   \right).
\end{align}
Replacing $\dot{\mathbf{x}}= H \mathbf{f}(t, \mathbf{x})$ where $\mathbf{f}(t, \mathbf{x})$ is defined by \eqref{EQ:108} with $\dot{\mathbf{y}}= H  \bar{\mathbf{f}}(\mathbf{y})$, $\mathbf{y}= \left(\bar{\Omega}, \bar{\Sigma}, \bar{\Omega}_k, \bar{\Phi} \right)^T$ and $\bar{\mathbf{f}}(\mathbf{y})$ given by time-averaging \eqref{timeavrg}
we obtain:
 \begin{align}
&\dot{\bar{\Omega}}=\frac{1}{2} H \bar{\Omega}  \Big(-3 \gamma  \left(\bar{\Sigma} ^2+\bar{\Omega} ^2+\bar{\Omega}_{k}-1\right)    +6 \bar{\Sigma} ^2+3 \bar{\Omega} ^2+2 \bar{\Omega}_{k}-3\Big), \label{IIIeq24}
\\
    &\dot{\bar{\Sigma}}=\frac{1}{2} H \Bigg(\bar{\Sigma}  \Big(-3 \gamma  \left(\bar{\Sigma} ^2+\bar{\Omega}
   ^2+\bar{\Omega}_{k}-1\right)   +6 \bar{\Sigma} ^2+3 \bar{\Omega} ^2+2 \bar{\Omega}_{k}-6\Big)+2 \bar{\Omega}_{k}\Bigg),  \label{IIIeq25}
\\
&\dot{{\bar{\Omega}_{k}}}= -H \bar{\Omega}_{k} \Big(3 \gamma  \left(\bar{\Sigma} ^2+\bar{\Omega} ^2+\bar{\Omega}_{k}-1\right)     -6 \bar{\Sigma} ^2+2 \bar{\Sigma} -3
   \bar{\Omega} ^2-2 \bar{\Omega}_{k}+2\Big),  \label{IIIeq26}
\\
&\dot{\bar{\Phi}}=0,  \label{IIIeq27}
\\
& \dot{H}=  -\frac{1}{2} H^2 \Big(3 \gamma  \left(1-{\bar{\Sigma}}^2-\bar{\Omega} ^2-\bar{\Omega}_k\right)     + 6 {\bar{\Sigma}}^2+ 3 \bar{\Omega}^2+ 2 \bar{\Omega}_k\Big)  \label{IIIeq28}.
\end{align}
\end{widetext}
Proceeding in analogous way as in references \cite{Alho:2015cza,Alho:2019pku} we implement a local nonlinear transformation:   
\begin{small}
\begin{align}
&\mathbf{x}_0:=\left(\Omega_{0}, \Sigma_{0}, \Omega_{k0}, \Phi_{0}\right)^T  \mapsto \mathbf{x}:=\left(\Omega, \Sigma, \Omega, \Phi\right)^T \nonumber \\
& \mathbf{x}=\psi(\mathbf{x}_0):=\mathbf{x}_0 + H \mathbf{g}(H, \mathbf{x}_0,t), \label{AppBIIIquasilinear211}
\\
& \mathbf{g}(H, \mathbf{x}_0,t)= \left(\begin{array}{c}
    g_1(H , \Omega_{0}, \Sigma_{0}, \Omega_{k0}, \Phi_{0}, t)\\
    g_2(H , \Omega_{0}, \Sigma_{0}, \Omega_{k0}, \Phi_{0}, t)\\
    g_3(H , \Omega_{0}, \Sigma_{0}, \Omega_{k0}, \Phi_{0}, t)\\
    g_4(H , \Omega_{0}, \Sigma_{0}, \Omega_{k0}, \Phi_{0}, t)\\
 \end{array}\right).   \label{eqT55}
\end{align}
\end{small}
Taking time derivative in both sides of \eqref{AppBIIIquasilinear211} with respect to $t$ we obtain 
\begin{small}
\begin{align}
    & \dot{\mathbf{x}_0}+ \dot{H} \mathbf{g}(H, \mathbf{x}_0,t)  \nonumber \\
    & + H \Bigg(\frac{\partial }{\partial t} \mathbf{g}(H, \mathbf{x}_0,t) + \dot{H} \frac{\partial }{\partial H} \mathbf{g}(H, \mathbf{x}_0,t)    + D_{\mathbf{x}_0} \mathbf{g}(H, \mathbf{x}_0,t) \cdot \dot{\mathbf{x}_0}\Bigg) \nonumber \\
    & = \dot{\mathbf{x}}, \label{EQT56}
    \end{align}
    \end{small}
    where 
    \begin{equation}
        D_{\mathbf{x}_0} \mathbf{g}(H, \mathbf{x}_0,t)= \left(\begin{array}{cccc}
             \frac{\partial g_1}{\partial \Omega_0}&  \frac{\partial g_1}{\partial \Sigma_0} & \frac{\partial g_1}{\partial \Omega_k}&  \frac{\partial g_1}{\partial \Phi_0}\\
             \frac{\partial g_2}{\partial \Omega_0}&  \frac{\partial g_2}{\partial \Sigma_0} & \frac{\partial g_2}{\partial \Omega_k} &  \frac{\partial g_2}{\partial \Phi_0}\\
                \frac{\partial g_3}{\partial \Omega_0}&  \frac{\partial g_3}{\partial \Sigma_0} & \frac{\partial g_3}{\partial \Omega_k} &  \frac{\partial g_3}{\partial \Phi_0}\\
                   \frac{\partial g_4}{\partial \Omega_0}&  \frac{\partial g_4}{\partial \Sigma_0} & \frac{\partial g_4}{\partial \Omega_k} &  \frac{\partial g_4}{\partial \Phi_0}\\
        \end{array}
                \right)
    \end{equation}    
is the  Jacobian matrix of $\mathbf{g}(H, \mathbf{x}_0,t)$ for the vector  $\mathbf{x}_0$.  The function $\mathbf{g}(H, \mathbf{x}_0,t)$ is conveniently chosen. 
\newline By substituting \eqref{equx} and \eqref{AppBIIIquasilinear211} in \eqref{EQT56} we obtain 
\begin{small}
\begin{align}
       & \Bigg(\mathbf{I}_4 + H D_{\mathbf{x}_0} \mathbf{g}(H, \mathbf{x}_0,t)\Bigg) \cdot \dot{\mathbf{x}_0}= H \mathbf{f}(\mathbf{x}_0 + H \mathbf{g}(H, \mathbf{x}_0,t),t) \nonumber \\
       & -H \frac{\partial }{\partial t} \mathbf{g}(H, \mathbf{x}_0,t) -\dot{H} \mathbf{g}(H, \mathbf{x}_0,t) -H \dot{H} \frac{\partial }{\partial H} \mathbf{g}(H, \mathbf{x}_0,t), 
\end{align}
\end{small}
where 
$\mathbf{I}_4= \left(\begin{array}{cccc}
             1 & 0 & 0 & 0\\
             0 & 1 & 0 & 0\\
             0 & 0 & 1 & 0\\
              0 & 0 & 0 & 1\\
        \end{array}
                \right)$ is the $4\times 4$ identity matrix.
                
Then we obtain 
\begin{widetext}
  \begin{small}  
  \begin{align}
 & \dot{\mathbf{x}_0} = \Bigg(\mathbf{I}_4 + H D_{\mathbf{x}_0} \mathbf{g}(H, \mathbf{x}_0,t)\Bigg)^{-1} \cdot \Bigg(H \mathbf{f}(\mathbf{x}_0 + H \mathbf{g}(H, \mathbf{x}_0,t),t)-H \frac{\partial }{\partial t} \mathbf{g}(H, \mathbf{x}_0,t) -\dot{H} \mathbf{g}(H, \mathbf{x}_0,t) -H \dot{H} \frac{\partial }{\partial H} \mathbf{g}(H, \mathbf{x}_0,t)\Bigg).  
\end{align}
\end{small}
Using eq. \eqref{EQ:81b}, we have $ \dot{H}= \mathcal{O}(H^2)$. Hence,
\begin{small}
\begin{align}
    & \dot{\mathbf{x}_0} = \underbrace{\Bigg(\mathbf{I}_4 - H D_{\mathbf{x}_0} \mathbf{g}(0, \mathbf{x}_0,t) +  \mathcal{O}(H^2)\Bigg)}_{4\times 4 \: \text{matrix}} \cdot \underbrace{\Bigg(H \mathbf{f}(\mathbf{x}_0, t)-H \frac{\partial }{\partial t} \mathbf{g}(0, \mathbf{x}_0,t) +   \mathcal{O}(H^2)\Bigg)}_{4\times 1 \; \text{vector}}= \underbrace{H \mathbf{f}(\mathbf{x}_0, t)-H \frac{\partial }{\partial t} \mathbf{g}(0, \mathbf{x}_0,t) +   \mathcal{O}(H^2)}_{4\times 1 \; \text{vector}}.\label{eqT59}
    \end{align} 
    \end{small}
\end{widetext}
\noindent The strategy is to use eq. \eqref{eqT59} for choosing conveniently $\frac{\partial }{\partial t} \mathbf{g}(0, \mathbf{x}_0,t)$ to prove that 
\begin{align}
 & \dot{\Delta\mathbf{x}_0}= -H G(\mathbf{x}_0, \bar{\mathbf{x}}) +   \mathcal{O}(H^2), \label{EqY60}
  \end{align}
where $\bar{\mathbf{x}}=(\bar{\Omega},  \bar{\Sigma}, \bar{\Phi})^T$ and  $\Delta\mathbf{x}_0=\mathbf{x}_0 - \bar{\mathbf{x}}$. The function $G(\mathbf{x}_0, \bar{\mathbf{x}})$ is unknown at this stage. 
\newline 
By construction we neglect dependence of $\partial g_i/ \partial t$ and $g_i$ on $H$, i.e., assume $\mathbf{g}=\mathbf{g}(\mathbf{x}_0,t)$ because dependence of $H$ is dropped out along with higher order terms eq. \eqref{eqT59}. Next, we solve a partial differential equation  for $\mathbf{g}(\mathbf{x}_0,t)$ given by:  
\begin{align}
     & \frac{\partial }{\partial t} \mathbf{g}(\mathbf{x}_0,t) = \mathbf{f}(\mathbf{x}_0, t) - \bar{\mathbf{f}}(\bar{\mathbf{x}}) + G(\mathbf{x}_0, \bar{\mathbf{x}}). \label{eqT60}
\end{align}
\noindent  where we have considered $\mathbf{x}_0$, and $t$ as independent variables. 
\newline
The right hand side of \eqref{eqT60} is almost periodic of period $L=\frac{2\pi}{\omega}$ for large times. Then, implementing the average process \eqref{timeavrg} on right hand side of \eqref{eqT60}, where slow-varying dependence of quantities $\Omega_{0}, \Sigma_{0}, \Omega_{k0},  \Phi_0$ and  $\bar{\Omega},  \bar{\Sigma}, \bar{\Omega_k}, \bar{\Phi}$  on $t$ are ignored through  averaging process, we obtain \begin{align}
    & \frac{1}{L}\int_0^{L} \Bigg[\mathbf{f}(\mathbf{x}_0, s) - \bar{\mathbf{f}}(\bar{\mathbf{x}}) +G(\mathbf{x}_0, \bar{\mathbf{x}}) \Bigg] ds  \nonumber \\
    & = \bar{\mathbf{f}}( {\mathbf{x}}_0)-\bar{\mathbf{f}}(\bar{\mathbf{x}} )+G(\mathbf{x}_0, \bar{\mathbf{x}}). \label{newaverage}
\end{align}
Defining 
\begin{equation}
  G(\mathbf{x}_0, \bar{\mathbf{x}}):=  -\left(\bar{\mathbf{f}}( {\mathbf{x}}_0)-\bar{\mathbf{f}}(\bar{\mathbf{x}})\right)
\end{equation} the average \eqref{newaverage} is zero so that $\mathbf{g}(\mathbf{x}_0,t)$ is bounded.
\newline 
Finally, eq. \eqref{EqY60} transforms to 
\begin{align}
 & \dot{\Delta\mathbf{x}_0}= H \left(\bar{\mathbf{f}}( {\mathbf{x}}_0)-\bar{\mathbf{f}}(\bar{\mathbf{x}})\right) +   \mathcal{O}(H^2)  \label{EqY602}
  \end{align}
and eq. \eqref{eqT60} 
is simplified to 
\begin{align}
     & \frac{\partial }{\partial t} \mathbf{g}(\mathbf{x}_0,t) = \mathbf{f}(\mathbf{x}_0, t) - \bar{\mathbf{f}}( \mathbf{x}_0). \label{eqT602}
\end{align}
 Theorem \ref{BIIILFZ11} establish the existence of the vector \eqref{eqT55}. 

\begin{thm}
\label{BIIILFZ11} Let $\bar{\Omega}, \bar{\Sigma}, \bar{\Omega}_k,  \bar{\Phi}$ and $H$ be  defined  functions that  satisfy  averaged  equations
 \eqref{IIIeq24}, \eqref{IIIeq25}, \eqref{IIIeq26}, \eqref{IIIeq27}, \eqref{IIIeq28}. Then, there exist continuously differentiable functions $g_1, g_2, g_3$ and $g_4$,  such that   $\Omega, \Sigma, \Omega_k$ and $\Phi$  are locally given by \eqref{AppBIIIquasilinear211}, where $\Omega_{0}, \Sigma_{0}, \Omega_{k0}, \Phi_0$ are order zero approximations of them as $H\rightarrow 0$. Then,  functions $\Omega_{0}, \Sigma_{0}, \Omega_{k0}, \Phi_0$ and averaged solution $\bar{\Omega},  \bar{\Sigma}, \bar{\Omega}_k, \bar{\Phi}$  have the same limit as $t\rightarrow \infty$. 
Setting $\Sigma=\Sigma_0=0$ are derived the analogous results for negatively curved FLRW model. 
\end{thm}

\subsection{FLRW metric with $k=-1$}
\label{SECT:IIIA}

Secondly, we consider FLRW metric \eqref{metricFLRW} with $k=-1$ for generalized harmonic potential \eqref{pot} minimally coupled to matter with field equations \eqref{Non_minProb1FLRW0-1} with the substitution $k=-1$. Defining
\begin{align}
& \Omega=\sqrt{\frac{\omega^2 \phi^2+{\dot{\phi}}^2}{6 H^2}},  \quad \Omega_k=-\frac{k}{a^2 H^2}, \nonumber\\
& \Phi= t \omega -\tan^{-1}\left(\frac{\omega \phi}{\dot{\phi}}\right),
\end{align}
the full system is deduced from \eqref{BIIIunperturbed1} by setting $\Sigma=0$. Then, 
\begin{widetext}
\begin{subequations}
\label{unperturbed1FLRW}
\begin{small}
\begin{align}
   & \dot{\Omega}= -\frac{b \gamma  f \mu ^3 \Omega}{H} \sin ^2 \left(\frac{\sqrt{\frac{3}{2}} H \Omega  \sin (t \omega -\Phi )}{f \omega }\right)   -\frac{b \mu ^3 }{\sqrt{6} H} \cos (t \omega -\Phi ) \sin \left(\frac{\sqrt{6} H \Omega
    \sin (t \omega -\Phi )}{f \omega }\right) \nonumber\\
    & +H \cos ^2(t \omega -\Phi ) \left(\Omega ^3 \left(\gamma  \left(\frac{3 \mu ^2}{\omega ^2}-\frac{3}{2}\right)+3\right)-3 \Omega \right)  +\frac{1}{2}
   H \Omega  \left(3 \gamma -\frac{6 \gamma  \mu ^2 \Omega ^2}{\omega ^2}+(2-3 \gamma ) \Omega_{k}\right) \nonumber\\
   & +\frac{\left(\omega ^2-2 \mu ^2\right) \Omega  \sin (2 t \omega -2 \Phi )}{2 \omega },
\\
   & \dot{\Omega_k}=\frac{b
   \gamma  f \mu ^3 \Omega_{k}}{H}  \left(-1+\cos  \left(\frac{\sqrt{6} H \Omega  \sin (t \omega -\Phi )}{f \omega }\right)\right)    -H \Omega_{k}  \left(\frac{6 \gamma  \mu ^2 \Omega ^2 \sin ^2(t
   \omega -\Phi )}{\omega ^2}+(3 \gamma -2) (\Omega_{k}-1)\right) \nonumber \\
   & -3 (\gamma -2) H \Omega ^2 \Omega_{k} \cos ^2(t \omega -\Phi ),
\\
   & \dot \Phi= -3 H \sin (t \omega -\Phi ) \cos (t \omega -\Phi )    +\frac{\left(\omega ^2-2 \mu ^2\right) }{\omega } \sin ^2(t \omega -\Phi )  -\frac{b \mu ^3}{\sqrt{6} H \Omega } \sin (t \omega -\Phi ) \sin  \left(\frac{\sqrt{6}
   H \Omega  \sin (t \omega -\Phi )}{f \omega }\right) ,
\\
   & \dot{H}= -(1+q)H^2, 
   \end{align}
   \end{small}
   with deceleration parameter 
   \begin{small}
   \begin{align}
   \label{eq.49e}
   & q =-1+\frac{3 \gamma }{2}  -\frac{3 \gamma  \mu ^2 \Omega ^2 \sin ^2(t \omega -\Phi )}{\omega ^2}    -\frac{3}{2} (\gamma -2) \Omega ^2 \cos ^2(t \omega -\Phi )   -\frac{3}{2} \gamma  \Omega_{k}+\Omega_{k}   -\frac{b \gamma  f \mu ^3}{H^2} \sin ^2   \left(\frac{\sqrt{\frac{3}{2}} H \Omega  \sin (t \omega -\Phi )}{f
   \omega }\right).
\end{align}
\end{small}
\end{subequations}
Setting $f=\frac{b \mu ^3}{\omega ^2-2 \mu ^2}>0$, we obtain:
\begin{align}
& \dot{\mathbf{x}}= H \mathbf{f}(t, \mathbf{x}) + \mathcal{O}(H^2), \;   \mathbf{x}= \left(\Omega, \Omega_k, \Phi \right)^T,\\
   &  \dot{H}= -\frac{1}{2} \left(3
   \gamma\left(1-\Omega^2- \Omega_{k}\right)+2 \Omega_{k}\right)  -3 \Omega^2 \cos ^2(t \omega -\Phi)+  \mathcal{O}(H^3),
\end{align} 
where
\begin{align}
    \label{EQ:50}
 &  f(t, \mathbf{x}) = 
   \left(\begin{array}{c}
   \frac{1}{2}   \Omega  \left(3 \gamma -3 \gamma  \left(\Omega^2+\Omega_{k}\right)+2 \Omega_{k}\right)  +3  \Omega \left(\Omega^2-1\right) \cos ^2(t \omega -\Phi) \\\\
- \Omega_{k} \left(3 \gamma  \Omega^2+(3 \gamma -2) (\Omega_{k}-1)\right)   +6  \Omega^2 \Omega_{k} \cos ^2(t \omega -\Phi)\\\\
    -\frac{3}{2} \sin (2 t \omega -2 \Phi)
      \end{array}
   \right).
\end{align}
\end{widetext}
Replacing $\dot{\mathbf{x}}= H \mathbf{f}(t, \mathbf{x})$ with  $\mathbf{f}(t, \mathbf{x})$ defined by \eqref{EQ:50} with $\dot{\mathbf{y}}= H  \bar{f}(\mathbf{y})$, $\mathbf{y}= \left(\bar{\Omega}, {\bar{\Omega}_{k}}, \bar{\Phi} \right)^T$  and using the time averaging \eqref{timeavrg} we obtain the time-averaged system: 
\begin{align}
    &\dot{\bar{\Omega}}=-\frac{1}{2} H \; \bar{\Omega}  \left(3 (\gamma -1) \left(\bar{\Omega} ^2-1\right)+(3 \gamma -2)
   {{\bar{\Omega}_{k}}}\right), \label{eq46}\\
   &\dot{\bar{\Omega}} _k=-H \; {{\bar{\Omega}_{k}}} \left(3 (\gamma -1) \bar{\Omega} ^2-3 \gamma +(3 \gamma -2) {{\bar{\Omega}_{k}}}+2\right) \label{eq47},
\\
   &\dot{\bar{\Phi}}=0. \label{eq48}
\end{align}
Theorem \ref{BIIILFZ11} applies to Bianchi III, and the invariant set $\Sigma=0$ corresponds to  negatively curved FLRW models.

\section{Qualitative analysis of averaged systems}
\label{SECT:III}

Theorem \ref{BIIILFZ11} proved in   \ref{appendixeq28} implies that $\Omega , \Sigma, \Omega_k$ and $\Phi$  evolve according to time-averaged system \eqref{IIIeq24}, \eqref{IIIeq25}, \eqref{IIIeq26}, \eqref{IIIeq27} as $H\rightarrow 0$. Hence, the full equations of time-dependent system \eqref{BIIIunperturbed1} and their corresponding time-averaged versions have the same late-time dynamics as $H\rightarrow 0$. Therefore, the simplest time-averaged system determines the future asymptotic of full system. In  particular, depending on  values of  barotropic index $\gamma$, the generic late-time attractors  of physical interests are found. With this approach, the oscillations entering full system through  KG equation can be controlled and smoothed out as Hubble factor $H$ - acting as a time-dependent perturbation parameter- tends monotonically to zero. These results are supported by numerical simulations  given in \ref{numerics}. 

\subsection{LRS Bianchi III}
\label{LRSBIII}
From the averaged system \eqref{IIIeq24},
\eqref{IIIeq25}, \eqref{IIIeq26} and  \eqref{IIIeq27} we obtain Hubble normalized averaged system 
\begin{widetext}
\begin{subequations}
\label{BIIIavrgsyst}
\begin{align}
  &    \frac{d\bar{\Omega}}{d\tau}= \frac{1}{2} \bar{\Omega}  \Big(3 \gamma  \left(1-\bar{\Sigma} ^2-\bar{\Omega} ^2-\bar{\Omega}_{k}\right) +6 \bar{\Sigma} ^2+3 \bar{\Omega} ^2+2 \bar{\Omega}_{k}-3\Big), \label{BIIIguidingC1}\\
& \frac{d\bar{\Sigma}}{d\tau}= \frac{1}{2}  \Bigg(\bar{\Sigma}  \Big(3 \gamma  \left(1-\bar{\Sigma} ^2-\bar{\Omega}
  ^2-\bar{\Omega}_{k}\right)   +6 \bar{\Sigma} ^2+3 \bar{\Omega} ^2+2 \bar{\Omega}_{k}-6\Big)+2 \bar{\Omega}_{k}\Bigg), \label{BIIIguidingC2}\\
& \frac{d{\bar{\Omega}_k}}{d\tau}= \bar{\Omega}_{k} \Big(3 \gamma  \left(1-\bar{\Sigma} ^2-\bar{\Omega} ^2-\bar{\Omega}_{k}\right)   +6 \bar{\Sigma} ^2-2 \bar{\Sigma} +3
  \bar{\Omega} ^2+2 \bar{\Omega}_{k}-2\Big),\label{BIIIguidingC3}\\
& \frac{d{{\bar{\Phi}}}}{d\tau}=0,   \\   
& \frac{d{H}}{d\tau} =  -\frac{1}{2} H \Big(3 \gamma  \left(1-{\bar{\Sigma}}^2-\bar{\Omega} ^2-\bar{\Omega}_k\right)   + 6 {\bar{\Sigma}}^2+ 3 \bar{\Omega}^2+ 2 \bar{\Omega}_k\Big), \quad  \frac{d{ {t}}}{d\tau} = 1/{ {H}}.
\end{align} 
\end{subequations}
\end{widetext}
We have $\bar{\Sigma} ^2+\bar{\Omega} ^2+\bar{\Omega}_{k}+ \bar{\Omega}_{m}=1$. Therefore, from energy condition $\bar{\Omega}_{m}\geq 0$ the phase space is: 
\begin{equation}
    \left\{(\bar{\Omega}, \bar{\Sigma}, \bar{\Omega}_{k})\in \mathbb{R}^3: \bar{\Sigma} ^2+\bar{\Omega} ^2+\bar{\Omega}_{k}\leq  1, \bar{\Omega}_{k}\geq 0\right\}.
\end{equation}

 The equilibrium points of the guiding system \newline \eqref{BIIIguidingC1}, \eqref{BIIIguidingC2}, \eqref{BIIIguidingC3}  are:
 \begin{enumerate}
 \item $T: (\bar{\Omega},\bar{\Sigma},\bar{\Omega}_k)=(0,-1,0)$ with eigenvalues \newline $\left\{6,\frac{3}{2},6-3 \gamma \right\}$.  
     \begin{enumerate}
         \item It is a source for $0\leq \gamma <2.$
         \item It is nonhyperbolic for $\gamma=2.$
     \end{enumerate}
Defining  a representative length along worldlines of the 4-velocity field as 
\begin{equation}
\frac{\ell(t)}{\ell_0}= \left[ {e_1}^1(t) ( {e_2}^2(t))^2\right]^{-\frac{1}{3}}, \; \tau =  \ln\left(\frac{\ell(t)}{\ell_0}\right),
\end{equation} 
such that 
\begin{equation}
    H=\frac{\dot \ell}{\ell}. 
\end{equation}
We denote by convention $t=0$ the current time where $ \left(\frac{\ell(0)}{\ell_0}\right)^3= \frac{1}{{e_1}^1(0) ( {e_2}^2(0))^2} =1$ and $\tau(0)=0$ and evaluating Raychaudhuri equation \eqref{RaychBIII} at $T$ we obtain: 
\begin{equation}
\left\{\begin{array}{c}
     \dot{H}= -3 H^2  \\\\
     \dot{\ell}=\ell H
\end{array}\right.
\implies \left\{\begin{array}{c} H(t)= \frac{H_{0}}{3 H_{0} t+1} \\\\
\ell(t)= {\ell_0}
   \sqrt[3]{3 H_{0} t+1}
\end{array}\right..
\end{equation}
 $\Sigma=-1$ implies $\sigma_{+}=-H=- \frac{H_{0}}{3 H_{0} t+1}$. This implies that $K$ is constant. 
 Indeed, from 
 \begin{equation}
   \dot K = -2 ({\sigma_+} +H) K, 
 \end{equation}
it follows $K=( {e_2}^2(t))^{2}=c_2^{-1}$. 
Substituting back in equation  \eqref{eqe1evol} is obtained:
\begin{align}
   & \dot{{e_1}^1}=-\frac{3H_{0}}{3 H_{0} t+1} {e_1}^1, \; {e_1}^1(0)=c_2. 
\end{align}
Hence, 
\begin{align}
& {{e_1}^1}(t)=\frac{c_2}{3 H_{0} t+1}.
\end{align}
That is, line element \eqref{metricLRSBIII} becomes
\begin{align}
   &  ds^2= - dt^2 + \frac{\left(3 H_{0} t+1\right)^2}{c_2^2} dr^2 
 + c_2  \mathbf{g}_{H^2}.
\end{align}
Therefore, the corresponding solution can be expressed in form of Taub-Kasner solution ($p_1=1, p_2= 0, p_3= 0$) where the scale factors of Kasner solution are $t^{p_i}, i=1,2,3$ with $p_1+p_2+p_3=1, p_1^2+p_2^2+p_3^2=1$
\cite{WE} (Sect 6.2.2 and p 193, Eq.  (9.6)).

     \item $Q: (\bar{\Omega},\bar{\Sigma},\bar{\Omega}_k)=(0,1,0)$ with eigenvalues \newline $\left\{2,\frac{3}{2},6-3 \gamma \right\}.$ 
     \begin{enumerate}
         \item It is a source for $0\leq \gamma <2.$
         \item It is nonhyperbolic for $\gamma=2.$
     \end{enumerate}
Evaluating Raychaudhuri equation \eqref{RaychBIII} at $Q$ and solving it we obtain 
\begin{equation}
 H(t)= \frac{H_{0}}{3 H_{0} t+1}.
\end{equation}
 $\Sigma=1$ implies $\sigma_{+}=H=\frac{H_{0}}{3 H_{0} t+1}$. 
 Hence, Gauss equation  \eqref{Gauss} and evolution equation \eqref{eqe1evol} become 
 \begin{equation}
    \dot{K}= -\frac{4 H_{0} K}{3 H_{0}
   t+1}, \; K(0)=c_1^{-1}
\end{equation}
and 
\begin{equation}
    \dot{{e_1}^1}=\frac{H_{0}}{3 H_{0} t+1}{e_1}^1, \; {e_1}^1(0)=c_1. 
\end{equation}
Then, by integration, 
\begin{align}
    & {e_1}^1(t)= c_1 \sqrt[3]{3 H_{0} t+1},\\
    & K(t)= \frac{1}{c_1\left(3 H_{0} t+1\right)^{4/3}}.
\end{align}
That is, line element \eqref{metricLRSBIII} becomes
\begin{align}
   &  ds^2= - dt^2 + c_1^{-2}\left( {3 H_{0} t+1}\right)^{-\frac{2}{3}} dr^2 \nonumber \\
   & +  {c_1^{-1}}{\left(3 H_{0} t+1\right)^{4/3}} \mathbf{g}_{H^2}.
\end{align}
Therefore, the corresponding solution can be expressed in  form of non-flat LRS Kasner ($p_1=-\frac{1}{3}, p_2= \frac{2}{3}, p_3= \frac{2}{3}$) Bianchi I solution (\cite{WE} Sect. 6.2.2 and Sect. 9.1.1 (2)). 

 \item $D: (\bar{\Omega},\bar{\Sigma},\bar{\Omega}_k)=(0,\frac{1}{2},\frac{3}{4})$ with eigenvalues \newline $\left\{-\frac{3}{2},0,3-3 \gamma \right\}.$ It is a nonhyperbolic saddle for $0\leq \gamma <1.$
     \begin{enumerate}
         \item For $\gamma=1$ the eigenvalues are $\{-\frac{3}{2},0,0\}$
         \item For $\gamma>1$ two eigenvalues are negative.
     \end{enumerate}
 Evaluating Raychaudhuri equation \eqref{RaychBIII} at $D$ we obtain 
\begin{equation}
\left\{\begin{array}{c}
     \dot{H}= -\frac{3}{2} H^2  \\\\
     \dot{\ell}=\ell H
\end{array}\right.
\implies \left\{\begin{array}{c} H(t)= \frac{2 H_{0}}{3
   H_{0} t+2}\\\\
\ell(t)= {\ell_0} \left(\frac{3 H_{0}
   t}{2}+1\right)^{2/3} 
\end{array}\right..
\end{equation}
$\Sigma=1/2$ implies $\sigma_{+}=\frac{1}{2}H=\frac{H_{0}}{3 H_{0} t+2}$. 
 Hence, Gauss equation  \eqref{Gauss} and evolution equation \eqref{eqe1evol} become
\begin{align}
   & \dot{{e_1}^1}=0, \; {e_1}^1(0)=c_1, \\
   & \dot{K} =-\frac{6 H_{0}
   K }{3 H_{0} t+2}, \; K(0)=\frac{1}{c_1}.
\end{align}
Hence, 
\begin{equation}
 {e_1}^1=c_1, \; K=   \frac{4}{c_1 (3
   H_{0} t+2)^2}.
\end{equation}
That is, line element \eqref{metricLRSBIII} becomes
\begin{align}
   &  ds^2= - dt^2 + c_1^{-2} dr^2 +  \frac{(3
   H_{0} t+2)^2}{4 c_1} \mathbf{g}_{H^2}. \label{metricD}
\end{align}
Therefore, the corresponding solution can be expressed in  form of Bianchi III form of flat spacetime (\cite{WE} p 193, Eq. (9.7)). 
     
     \item $F: (\bar{\Omega},\bar{\Sigma},\bar{\Omega}_k)=(1,0,0)$ with eigenvalues \newline $\left\{-\frac{3}{2},1,3-3 \gamma \right\}.$ 
 This point is always a saddle because it has a negative and a positive eigenvalue. For $\gamma =1$ it is a nonhyperbolic saddle.  
 
 Setting  $\psi(t)= t \omega -\Phi(t)$ and  evaluating Raychaudhuri equation \eqref{RaychBIII} at $F$ we obtain  
     \begin{align}
   &   \dot{H}=   \frac{b^2 \gamma  \mu ^6 \left(\cos
   \left(\frac{\sqrt{6} H \left(2 \mu ^2-\omega ^2\right) \sin (\psi)}{b \mu ^3 \omega }\right)-1\right)}{2(2 \mu ^2-\omega
   ^2)} \nonumber\\
   & +\frac{3 H^2 \left(2 \gamma  \mu ^2 \sin ^2(\psi)+\omega ^2
   \left((\gamma -2) \cos ^2(\psi)-\gamma \right)\right)}{2 \omega
   ^2}.
     \end{align}
     Therefore, 
     \begin{equation}
     \dot{H}   \sim -3 H^2 \cos ^2(t \omega -\Phi),
     \end{equation} for large $t$. 
     In average, $\Phi$ is constant, setting for simplicity $\Phi=0$ and integrating we obtain 
     \begin{equation}
         H(t)=\frac{4 H_{0} \omega }{6 H_{0} t \omega +3 H_{0} \sin (2 t
   \omega )+4 \omega }, 
     \end{equation}
     where $H_{0}$ is the current value of $H(t)$. Finally, $H(t)\sim \frac{2}{3 t}$ for large $t$.  
     
     Gauss equation  \eqref{Gauss} and evolution equation \eqref{eqe1evol} become
     \begin{equation}
         \dot{e_1^1}=-\frac{2 {e_1^1}}{3 t}, \; \dot{K}=-\frac{4 K}{3    t},
     \end{equation}
     with general solution
     \begin{equation}
     \dot{e_1^1}(t)= \frac{c_1}{t^{2/3}}, \; K(t)= 
   \frac{c_2}{t^{4/3}}. 
     \end{equation}
         That is, line element \eqref{metricLRSBIII} becomes
\begin{align}
   &  ds^2= - dt^2 + c_1^{-2} {t^{4/3}} dr^2    +  {c_2^{-1}}{t^{4/3}} \mathbf{g}_{H^2}. \label{metricF}
\end{align}  
 Hence for large $t$  the equilibrium point  can be associated with Einstein-de-Sitter solution (\cite{WE}, Sec 9.1.1 (1)) with $\gamma= 1$.

     \item $F_0: (\bar{\Omega},\bar{\Sigma},\bar{\Omega}_k)=(0,0,0)$ with eigenvalues \newline $\left\{\frac{3 (\gamma -2)}{2},\frac{3 (\gamma
   -1)}{2},3 \gamma -2\right\}.$ 
   \begin{enumerate}
       \item It is a sink for $0\leq \gamma <\frac{2}{3}.$
       \item It is a saddle for $\frac{2}{3}<\gamma <1$ or $1<\gamma <2.$
       \item It is nonhyperbolic for $\gamma =\frac{2}{3}$ or $\gamma =1$ or $\gamma =2.$
   \end{enumerate}
Evaluating Raychaudhuri equation \eqref{RaychBIII} at $F_0$ we obtain 
\begin{equation}
\label{scaling1}
\left\{\begin{array}{c}
      \dot{H}= -\frac{3}{2} \gamma H^2\\\\ 
      \dot{\ell}=\ell H  
       \end{array} \right.\implies 
       \left\{\begin{array}{c}
       H(t)= \frac{2 H_{0}}{3 \gamma  H_{0} t+2} \\\\
       \ell(t)=
   \ell_{0} \left(\frac{3 \gamma  H_{0} t}{2}+1\right)^{\frac{2}{3
   \gamma }}
\end{array}\right..
\end{equation}
          That is, line element \eqref{metricLRSBIII} becomes
\begin{align}
\label{scaling2}
   &  ds^2= - dt^2 + \ell_{0}^2 \left(\frac{3 \gamma  H_{0} t}{2}+1\right)^{\frac{4}{3
   \gamma }} \left (dr^2 + \mathbf{g}_{H^2}\right).
\end{align}   
The corresponding solution is a  matter dominated FLRW universe
   with $\bar{\Omega}_m=1$. 
   
     \item $MC: (\bar{\Omega},\bar{\Sigma},\bar{\Omega}_k)=(0,\frac{3 \gamma }{2}-1,-\frac{9 \gamma
   ^2}{4}+6 \gamma -3)$ with eigenvalues \newline $\left\{\frac{3 (\gamma -1)}{2},\frac{3}{4}
   \left(\gamma +\sqrt{2-\gamma } \sqrt{\gamma 
   (24 \gamma -41)+18}-2\right), \right. $\newline
   $\left. \frac{3}{4}
   \left(\gamma -\sqrt{2-\gamma } \sqrt{\gamma 
   (24 \gamma -41)+18}-2\right)\right\}.$ 
   By definition $\Omega_k\geq 0$, therefore, we impose   restriction $\frac{2}{3}\leq \gamma \leq 2$.
   \begin{enumerate}
       \item It is a sink for $\frac{2}{3}<\gamma <1.$
       \item It is a saddle for  $1<\gamma <2.$
       \item It is nonhyperbolic for $\gamma =\frac{2}{3}$ or $\gamma =1$ or $\gamma =2.$
   \end{enumerate}

   The corresponding solution is a matter-curvature scaling solution with $\bar{\Omega}_m=3(1-\gamma)$. We obtain same expressions in \eqref{scaling1}   for $\ell, H$ and for $ds^2$ we obtain the expression: 
   \begin{align}
\label{scaling2b}
   &  ds^2= - dt^2 + \ell_{0}^2 \left(\frac{3 \gamma  H_{0} t}{2}+1\right)^{\frac{4}{3
   \gamma }} \left( dr^2 +  \mathbf{g}_{H^2}\right).
\end{align}
 \end{enumerate}

\begin{figure*}[p]
    \centering
    \subfigure[\label{BIIIphaseplot3DCC} $\gamma=0$.]{\includegraphics[width=0.75\textwidth]{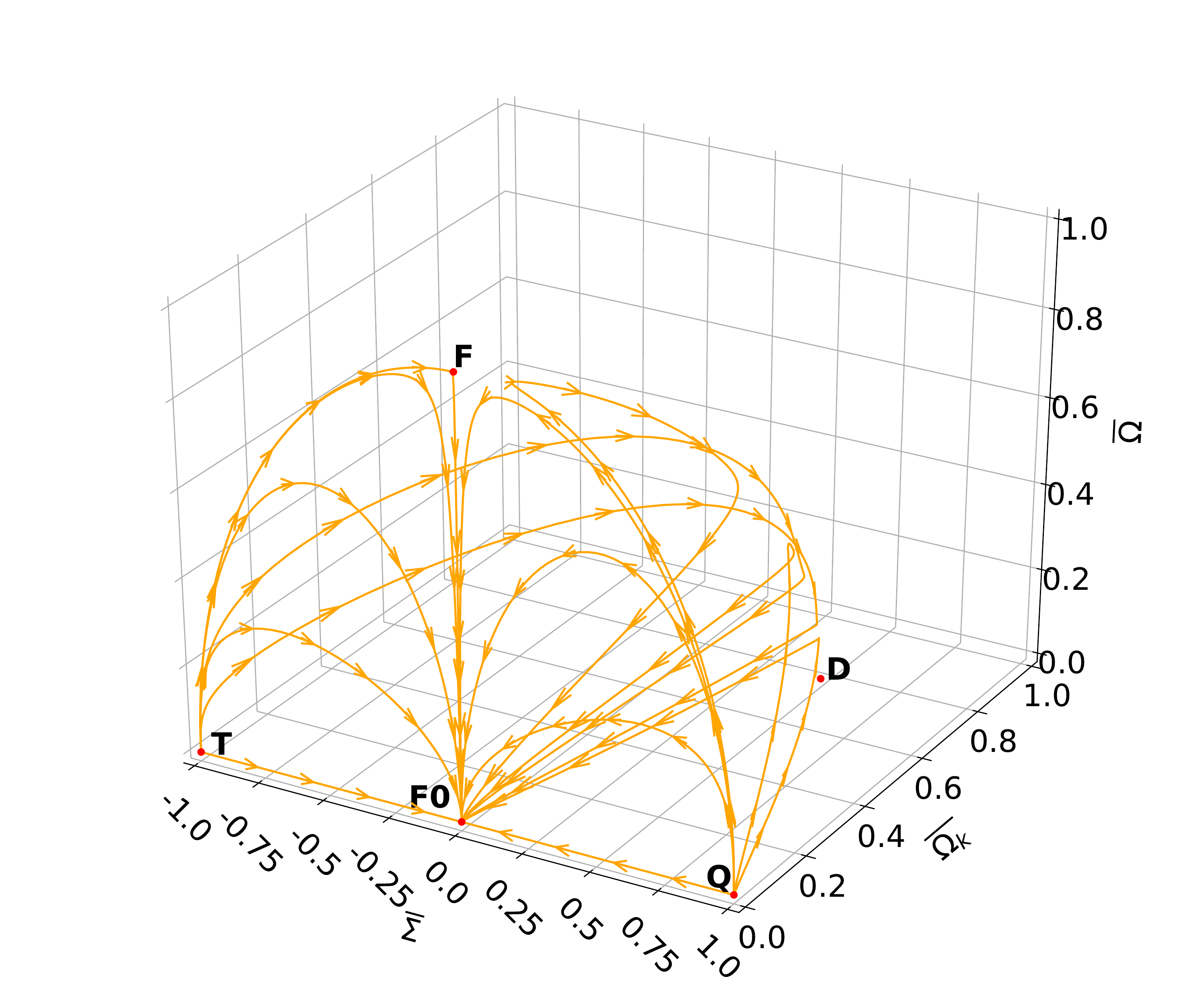}}
      \subfigure[\label{BIIIphaseplot3DBif} $\gamma=\frac{2}{3}$.]{\includegraphics[width=0.75\textwidth]{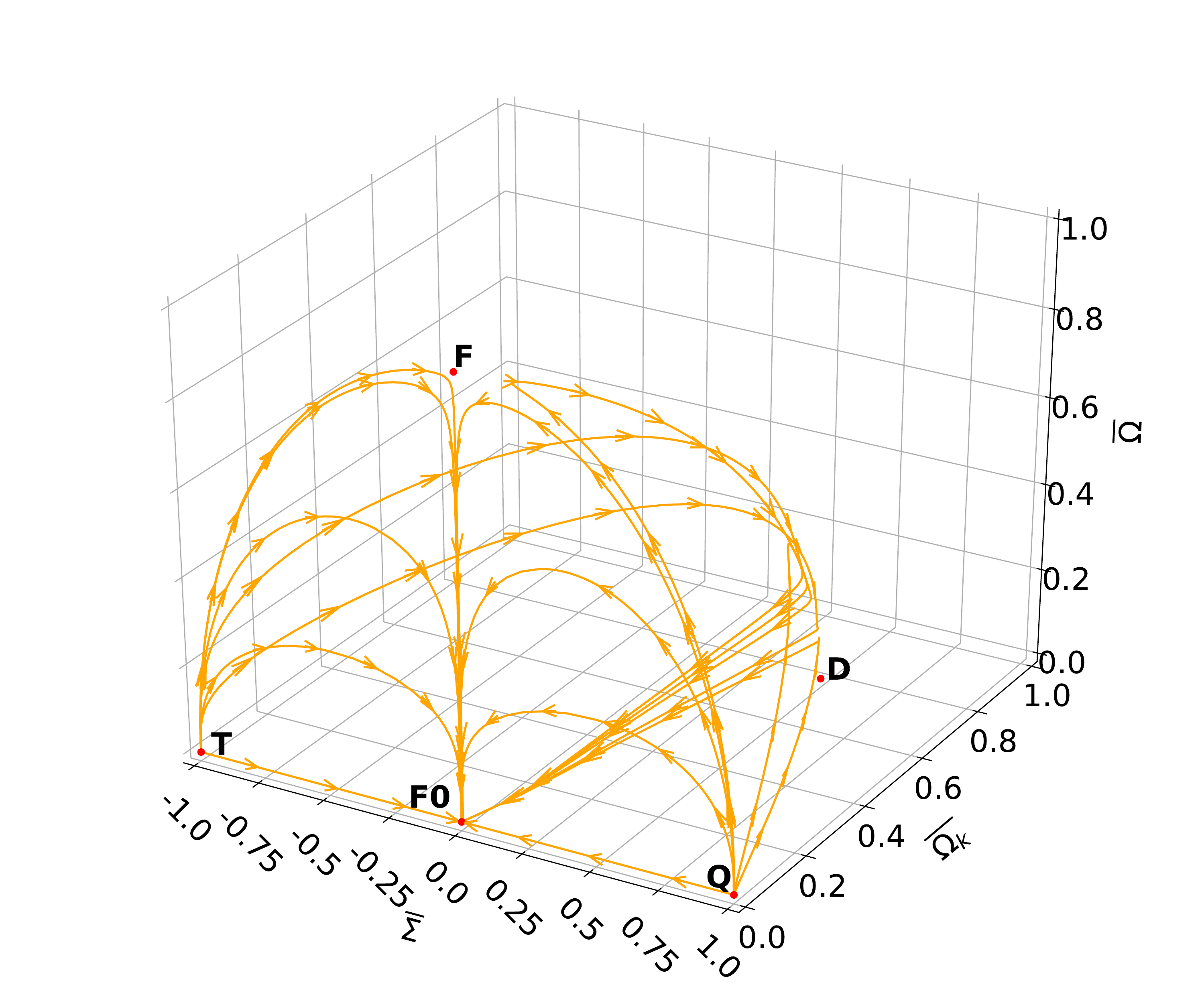}}
       \caption{Phase space of the guiding system  \eqref{BIIIguidingC1}, \eqref{BIIIguidingC2}, \eqref{BIIIguidingC3} for  $\gamma=0, \frac{2}{3}$.}
    \label{BIIIphaseplot3D}
\end{figure*}
\noindent
In figure \ref{BIIIphaseplot3DCC}  some orbits in the phase space of the guiding system \eqref{BIIIguidingC1}, \eqref{BIIIguidingC2}, \eqref{BIIIguidingC3} for $\gamma=0$ corresponding to cosmological constant are presented. The attractor is $F_0$ where scalar field mimics a cosmological constant. The equilibrium point $D$ is a saddle.

\noindent
In figure \ref{BIIIphaseplot3DBif}  some orbits of the phase space of the guiding system  \eqref{BIIIguidingC1}, \eqref{BIIIguidingC2}, \eqref{BIIIguidingC3} for  $\gamma=\frac{2}{3}$ are presented. The point $F_0$ coincides with $MC$;  it is asymptotically stable as proved in  \ref{AppCenter2} by means of Center Manifold theory.  The equilibrium point $D$ is a saddle.

   \begin{figure*}[p]
    \centering
    \subfigure[\label{BIIIphaseplot3DMC1} $\gamma=0.8$.]{\includegraphics[width=0.75\textwidth]{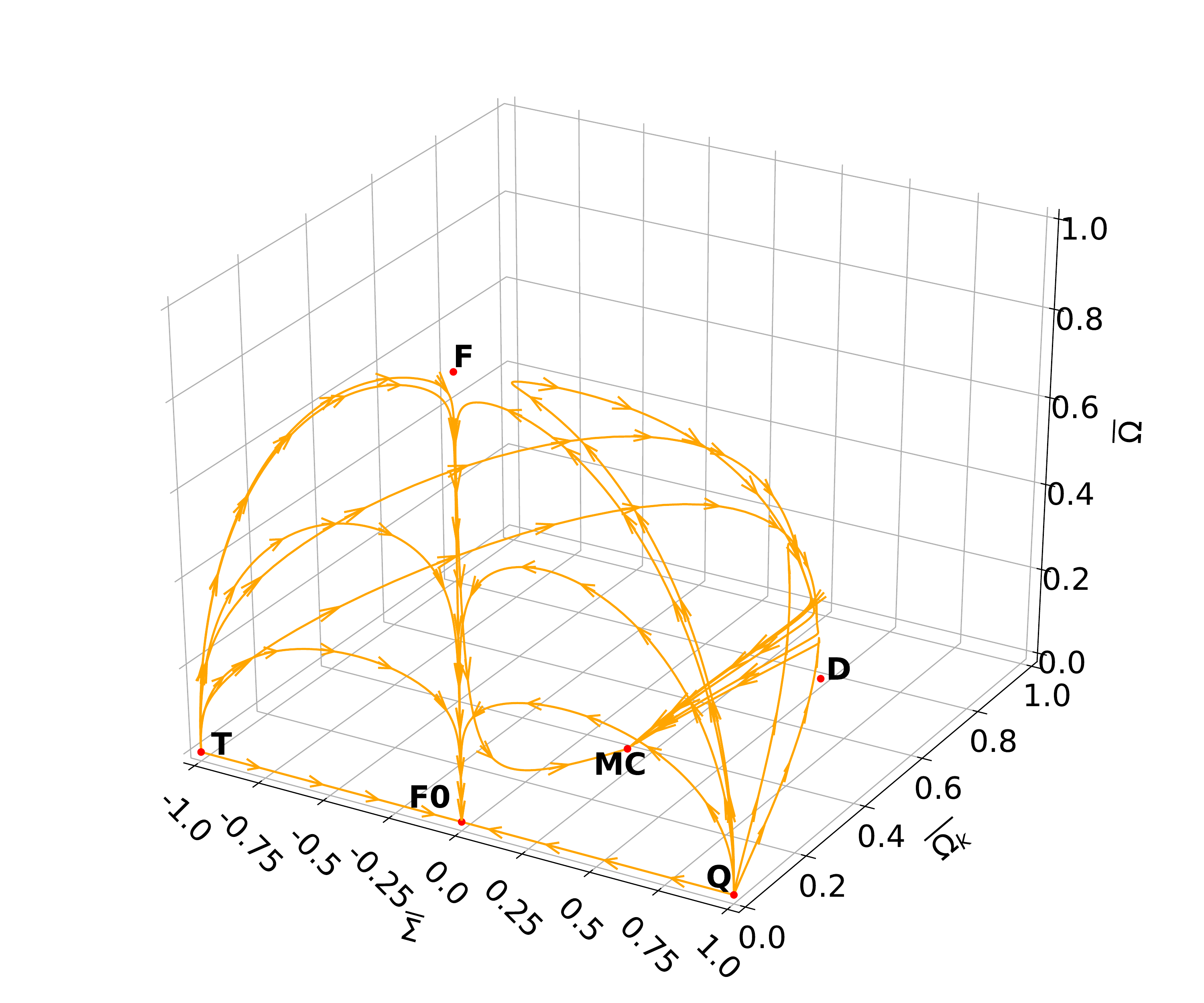}}
    \subfigure[\label{BIIIphaseplot3DMC2} $\gamma=0.9$.]{\includegraphics[width=0.75\textwidth]{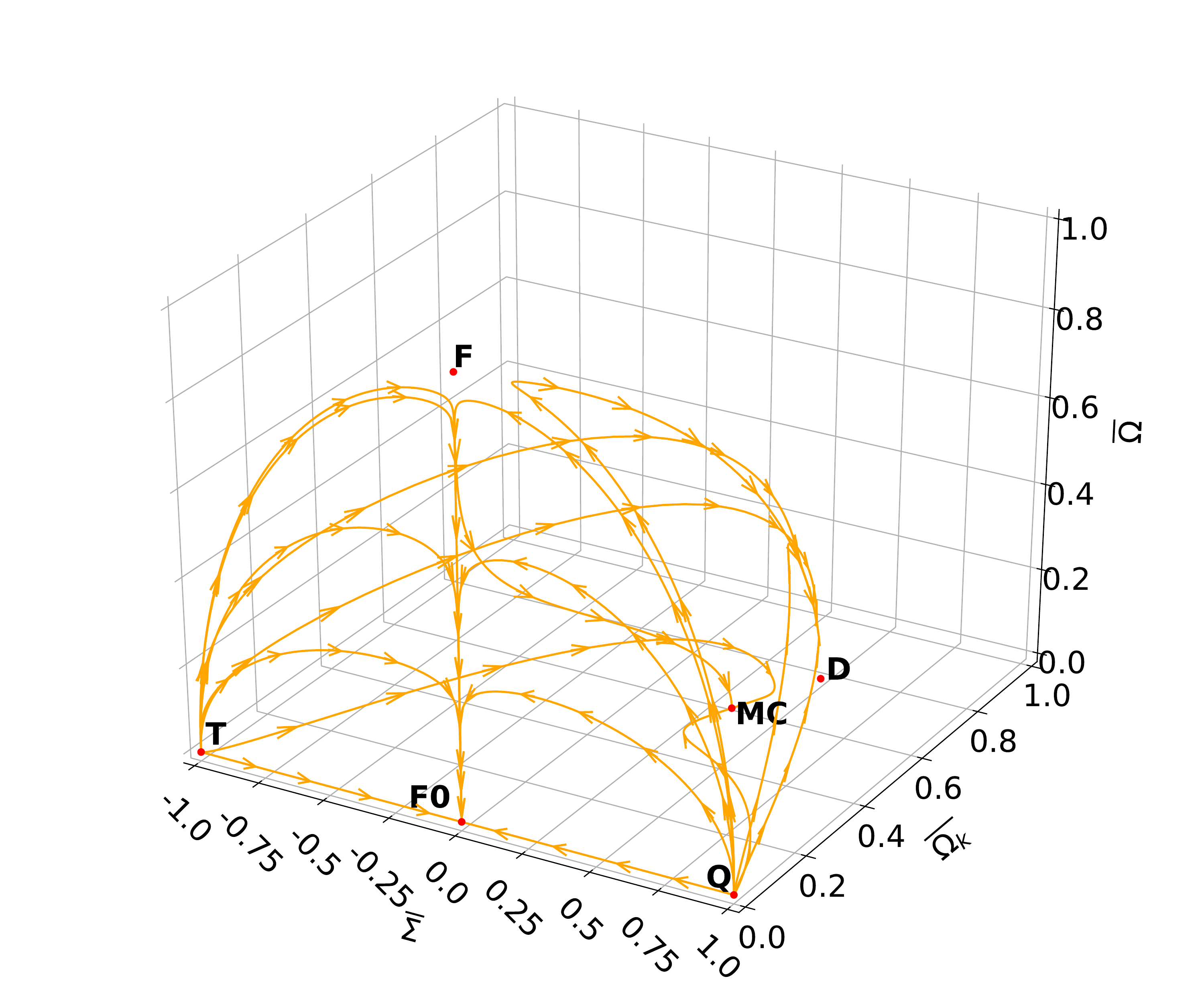}}
    \caption{Phase space of the guiding system  \eqref{BIIIguidingC1}, \eqref{BIIIguidingC2}, \eqref{BIIIguidingC3} for some values of $\gamma=0.8, 0.9$.}
    \label{BIIIphaseplot3DMC}
\end{figure*}

In figures \ref{BIIIphaseplot3DMC1} and \ref{BIIIphaseplot3DMC2}  some orbits in the phase space of the guiding system  \eqref{BIIIguidingC1}, \eqref{BIIIguidingC2}, \eqref{BIIIguidingC3} for $\gamma=0.8$ and $\gamma=0.9$ are presented. 
In both cases $MC$ is a stable node and $D$ is a saddle.

\begin{figure*}
    \centering
    \includegraphics[width=0.75\textwidth]{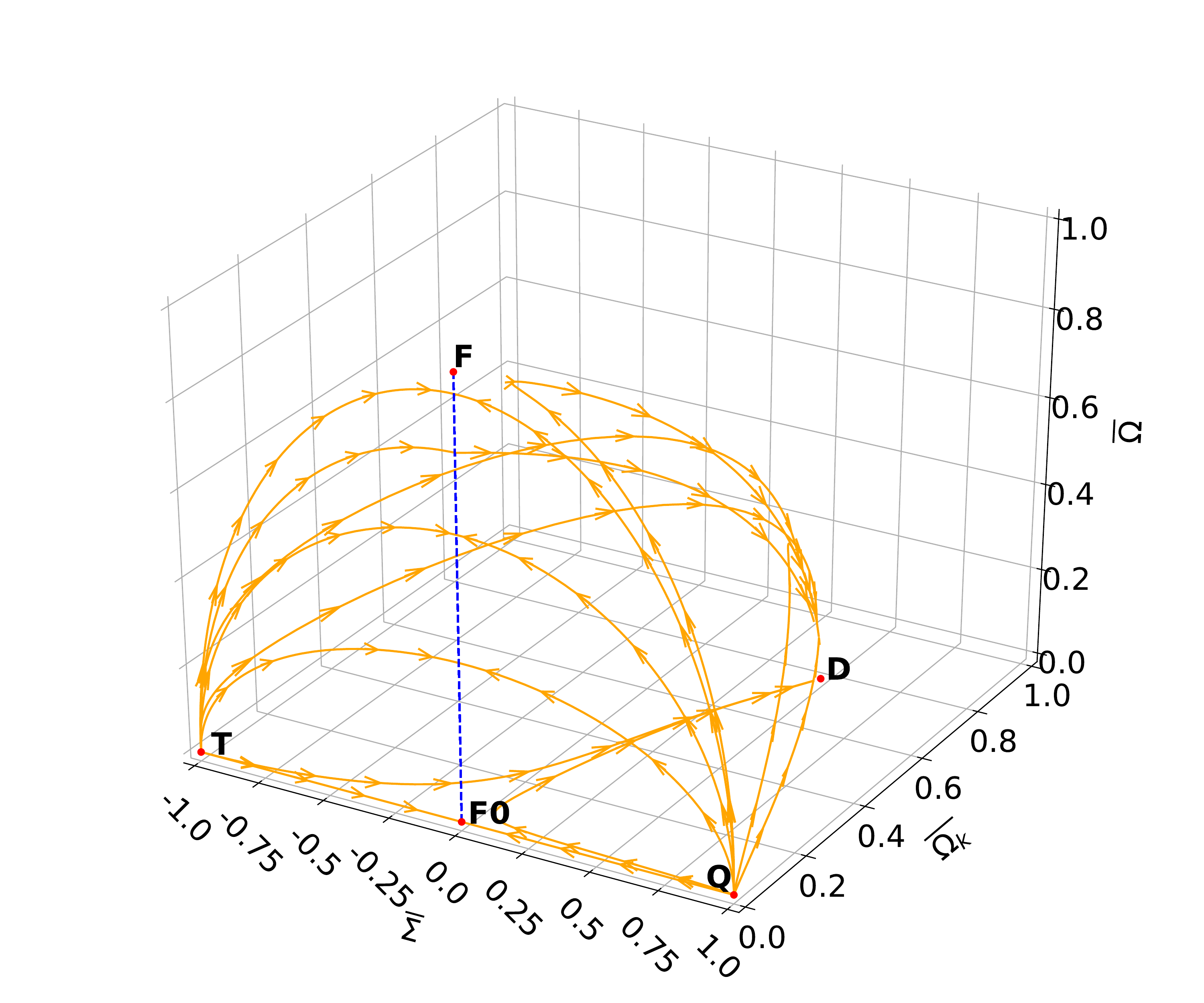}
    \caption{ \label{BIIIphaseplot3DDust} Phase space of the guiding system  \eqref{BIIIguidingC1}, \eqref{BIIIguidingC2}, \eqref{BIIIguidingC3} for $\gamma=1$.}
   \end{figure*}
   
   It is worth to notice that for  $\gamma=1$ the system admits the lines of equilibrium points  $(\bar{\Omega}, \bar{\Sigma}, \bar{\Omega}_k)= (\bar{\Omega}^*, 0,  0)$ and $D^*:= (\bar{\Omega}, \bar{\Sigma}, \bar{\Omega}_k)= (\bar{\Omega}^*, \frac{1}{2}, \frac{3}{4})$, where $\bar{\Omega}^*$  is an arbitrary number which satisfies $\bar{\Omega}^*\in[0,1]$. Therefore, the Bianchi III flat spacetime $D$, as well as $F_0$ are not isolated fixed points anymore. Additionally, $MC$ coincides with $D$.
In figure \ref{BIIIphaseplot3DDust}  some orbits in the phase space of the guiding system  \eqref{BIIIguidingC1}, \eqref{BIIIguidingC2}, \eqref{BIIIguidingC3} for $\gamma=1$ which corresponds to dust are presented. The attractor on the invariant set $\bar{\Omega}_k=0$ is the line that contains $F_0$ and $F$.  

According to the center manifold analysis in  \ref{AppCenter1} and supported by Figure  \ref{fig:my_labelA225-226} for $\gamma=1$  it is shown that $D$ is unstable (saddle type) for $\frac{1}{8} (4 \bar{\Sigma} +4 \bar{\Omega}_{k}-5)\neq 0$.  However, if we restrict the analysis to $\frac{1}{8} (4 \bar{\Sigma} +4 \bar{\Omega}_{k}-5)<0$  which is the physical region of the phase space, $D$ is asymptotically stable and behaves as a local attractor. 

   \begin{figure*}[p]
    \centering
    \subfigure[\label{BIIIphaseplot3DRad} $\gamma=\frac{4}{3}$.]{\includegraphics[width=0.75\textwidth]{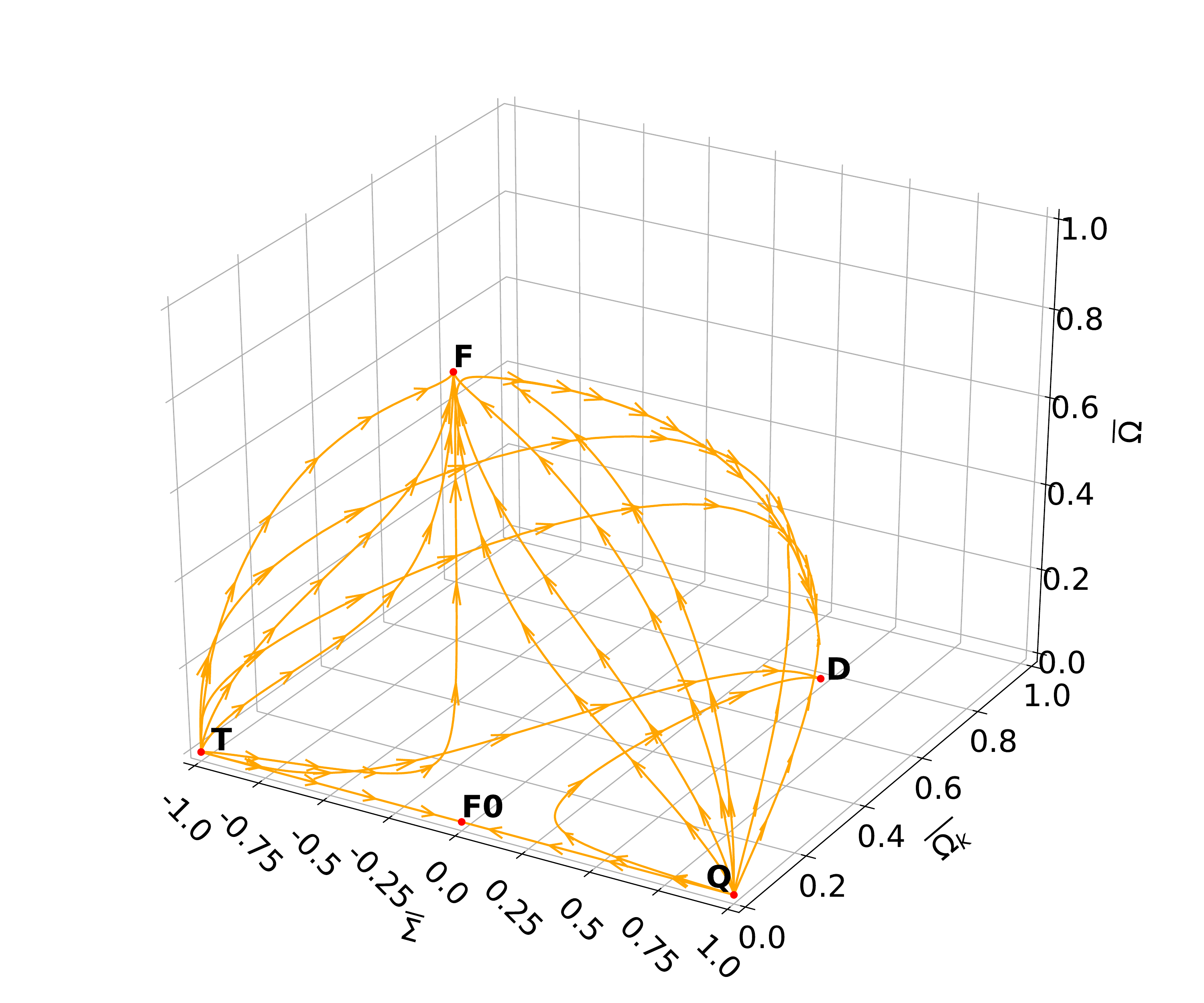}}
    \subfigure[\label{BIIIphaseplot3DStiff} $\gamma=2$.]{\includegraphics[width=0.75\textwidth]{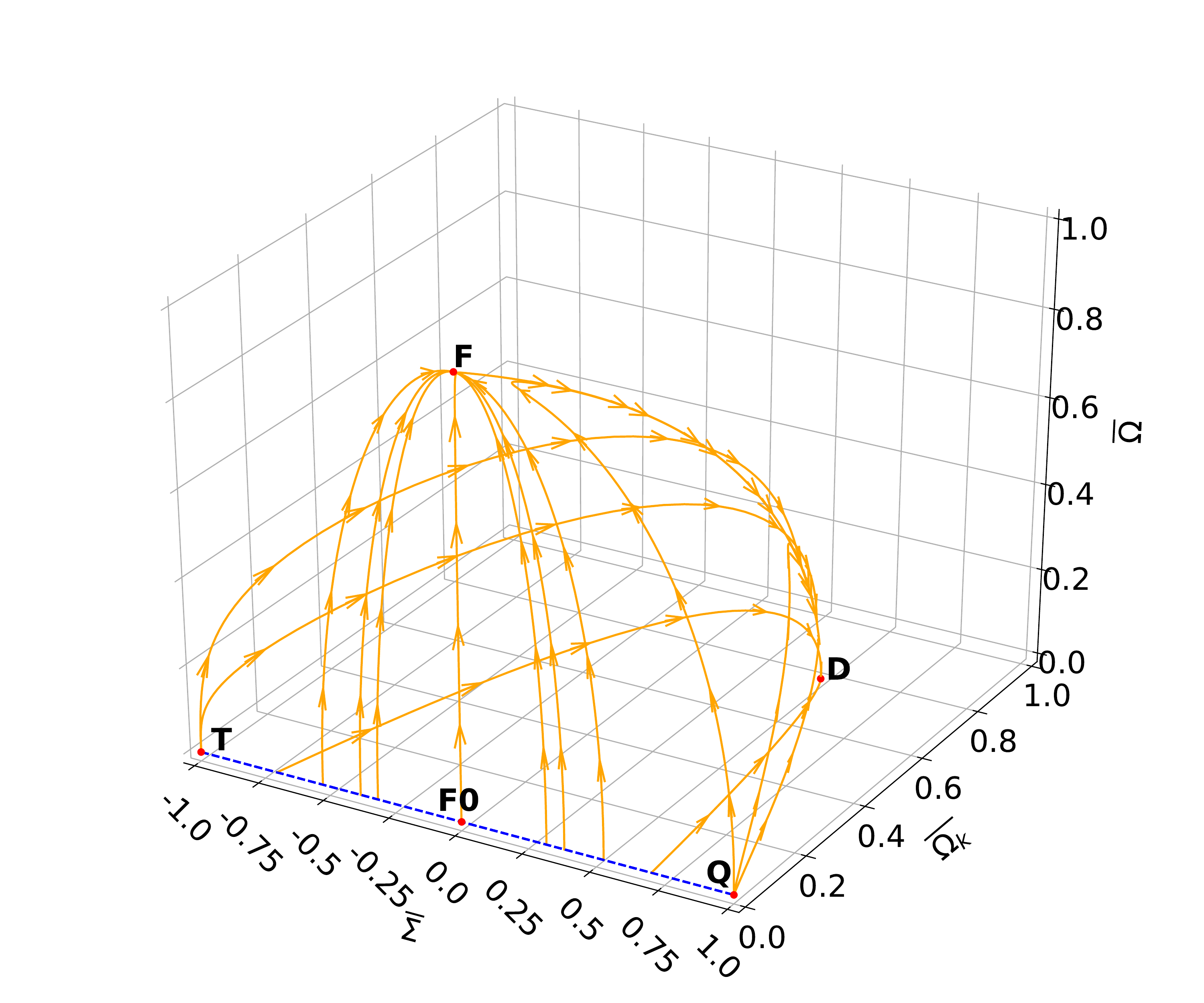}}
    \caption{Phase space of the guiding system  \eqref{BIIIguidingC1}, \eqref{BIIIguidingC2}, \eqref{BIIIguidingC3} for  $\gamma=\frac{4}{3}, 2$.}
    \label{BIIIphaseplot3Db}
\end{figure*}

In figure \ref{BIIIphaseplot3DRad}  some orbits in the phase space of  guiding system  \eqref{BIIIguidingC1}, \eqref{BIIIguidingC2}, \eqref{BIIIguidingC3} for $\gamma=\frac{4}{3}$ corresponding to radiation are presented. In  figure \ref{BIIIphaseplot3DRad} some orbits in the phase space of  guiding system  \eqref{BIIIguidingC1}, \eqref{BIIIguidingC2}, \eqref{BIIIguidingC3} for $\gamma=2$ which corresponds to stiff matter are presented. In  both figures, the attractor on the invariant set $\bar{\Omega}_k=0$ is  $F$. 
For $\gamma>1$, $D$ is locally asymptotically stable according to the center manifold analysis in \ref{AppCenter1}. 
For $\gamma=2$ the line connecting $T, F_0, Q$ is invariant and unstable.

In table \ref{Metrics}  exact solutions associated with the equilibrium points of reduced
averaged system  \eqref{BIIIguidingC1}, \eqref{BIIIguidingC2} and \eqref{BIIIguidingC3} are summarized. $A(t)$ and $B(t)$ denote scale factors of the metric \eqref{metricLRSBIII} where $c_1, c_2, a_0 \in \mathbb{R}^+$. 

\begin{table*}[t]
    \centering    
    \caption{Exact solutions associated with  equilibrium points of reduced
averaged system  \eqref{BIIIguidingC1}, \eqref{BIIIguidingC2} and \eqref{BIIIguidingC3}. $A(t)$ and $B(t)$ denote  scale factors of metric \eqref{metricLRSBIII}. $c_1, c_2, \ell_0 \in \mathbb{R}^+$. }
\resizebox{\textwidth}{!}{\begin{tabular}{clll}\hline
         Point & $A(t)$ & $B(t)$ & Solution  \\\hline 
        $T$ & $ \frac{\left(3 H_{0} t+1\right)}{c_2}$ &  $\sqrt{c_2}$ & Taub-Kasner solution ($p_1=1, p_2= 0, p_3= 0$) \\ \hline 
        $Q$ & $c_1^{-2}\left( {3 H_{0} t+1}\right)^{-1/3}$  & $ {c_1^{-1}}{\left(3 H_{0} t+1\right)^{2/3}}$ & non-flat LRS Kasner ($p_1=-\frac{1}{3}, p_2= \frac{2}{3}, p_3= \frac{2}{3}$) Bianchi I solution \\ \hline 
        $D$ & $c_1^{-1}$ & $\frac{(3 H_{0} t+2)}{2 \sqrt{c_1}}$ & Bianchi III form of flat spacetime\\\hline 
        $F$ & $c_1^{-1} {t^{2/3}}$   & ${c_2^{-1/2}}{t^{2/3}}$ & Einstein-de-Sitter solution \\\hline
        $F_0$ &  $\ell_{0} \left(\frac{3 \gamma  H_{0} t}{2}+1\right)^{\frac{2}{3
   \gamma }}$ & $ \ell_{0} \left(\frac{3 \gamma  H_{0} t}{2}+1\right)^{\frac{2}{3
   \gamma }}$ & Matter dominated FLRW universe \\\hline 
       $MC$ &  $\ell_{0} \left(\frac{3 \gamma  H_{0} t}{2}+1\right)^{\frac{2}{3
   \gamma }}$ & $ \ell_{0} \left(\frac{3 \gamma  H_{0} t}{2}+1\right)^{\frac{2}{3
   \gamma }}$ & Matter-curvature scaling solution \\\hline 
    \end{tabular}}
    \label{Metrics}
\end{table*}
    \begin{table}[t]
    \centering    
    \caption{Exact solutions associated with equilibrium points of reduced
averaged system  \eqref{guidingB1}-\eqref{guidingB2}. $a(t)$ denotes a scale factor of metric \eqref{mOpenFLRW} and $ a_0 \in \mathbb{R}^+$. }
\footnotesize\setlength{\tabcolsep}{5pt}
\begin{tabular}{cll}
\hline\\
         Point & $a(t)$ & Solution  \\\hline 
          $F$ & $ a_{0}  \left(\frac{3    H_{0} t}{2}+1\right)^{\frac{2}{3
     }}$ & Einstein-de-Sitter solution \\\hline
        $F_0$ &  $ a_{0} \left(\frac{3 \gamma  H_{0} t}{2}+1\right)^{\frac{2}{3
   \gamma }}$ & Matter dominated FLRW universe \\\hline 
       $C$ &  $a_{0}  \left(H_0 t+1\right)  $  & Milne solution \\\hline 
    \end{tabular}
    \label{MetricsOpenFLRW}
\end{table}
 \subsubsection{Late-time behavior}
The results from the linear stability analysis, the Center Manifold calculations in  \ref{AppCenter} and combined with Theorem \ref{BIIILFZ11} lead to: 
\begin{thm}
\label{thm8}
The late time attractors of full system \eqref{BIIIunperturbed1} and averaged system  \eqref{BIIIavrgsyst} for Bianchi III line element are:
\begin{enumerate}
    \item[(i)]  The matter dominated FLRW universe $F_0$    with  line element \eqref{scaling2}   if  $0< \gamma \leq  \frac{2}{3}$. $F_0$ represents a  quintessence fluid for $0<\gamma<\frac{2}{3}$ or a zero-acceleration model for $\gamma=\frac{2}{3}$. 
In the limit $\gamma=0$ we have  de Sitter solution. 
    \item[(ii)] The matter-curvature scaling solution $MC$ with $\bar{\Omega}_m=3(1-\gamma)$ and line element \eqref{scaling2b}
   if  $\frac{2}{3}<\gamma <1.$
    
    \item[(iii)] The Bianchi III flat spacetime $D$  with metric \eqref{metricD}
 if $1\leq \gamma\leq 2$.
    
\end{enumerate}
\end{thm}

\subsection{FLRW metric with $k=-1$.}
\label{FLRWflatopen}

In this case the  time-averaged system is \eqref{eq46}, \eqref{eq47},  \eqref{eq48}.
\begin{subequations}
\label{avrgsystFLRW}
\begin{align}
    &\frac{d \bar{\Omega}}{d \tau}=-\frac{1}{2} \bar{\Omega}  \left(3 (\gamma -1) \left(\bar{\Omega} ^2-1\right)+(3 \gamma -2)
   {{\bar{\Omega}_{k}}}\right),\label{guidingB1}
\\
  &\frac{d {{\bar{\Omega}_{k}}}}{d \tau}=- {{\bar{\Omega}_{k}}} \left(3 (\gamma -1) \bar{\Omega} ^2-3 \gamma +(3 \gamma -2) {{\bar{\Omega}_{k}}}+2\right). \label{guidingB2}
\\
  &    \frac{d{{\bar{\Phi}}}}{d\tau}=0,  
\\
  &   \frac{d{H}}{d\tau} =  -\frac{1}{2} H \left[ 3 \gamma (1- \Omega _k- \Omega^2) + 3 \Omega^2 + 2 \Omega_k \right]     \label{eqH}.
\end{align}
\end{subequations}
and $\frac{d{ {t}}}{d\tau} = 1/{ {H}}.$ We have $\bar{\Omega} ^2+\bar{\Omega}_{k}+ \bar{\Omega}_{m}=1$. 
Therefore, from condition $\bar{\Omega}_{m}\geq 0$ the phase space is 
\begin{equation}
    \left\{(\bar{\Omega},  \bar{\Omega}_{k})\in \mathbb{R}^2: \bar{\Omega} ^2+\bar{\Omega}_{k}\leq  1, \bar{\Omega}_{k}\geq 0\right\}.
\end{equation}
For $\gamma=1$,  guiding system \eqref{guidingB1}-\eqref{guidingB2} is reduced to 
\begin{align}\label{guidingB1gamma1}
&\frac{d \bar{\Omega}}{d \tau}=-\frac{1}{2} \bar{\Omega} \;  {{\bar{\Omega}_{k}}}, \; \frac{d {{\bar{\Omega}_{k}}}}{d \tau}=  {{\bar{\Omega}_{k}}} \left(1- {{\bar{\Omega}_{k}}}\right).
\end{align}
The solution is 
\begin{align}
    &{\bar{\Omega}}(\tau)=\frac{\Omega _0 }{\sqrt{{e^\tau \Omega _{k0}-\Omega
   _{k0}+1}}},  \nonumber \\
   & {{\bar{\Omega}_{k}}}(\tau)=\frac{e^\tau \Omega _{k0}}{e^\tau \Omega _{k0}-\Omega _{k0}+1}.
\end{align}
\begin{figure}[t!]
    \centering
    \includegraphics[width=0.45\textwidth]{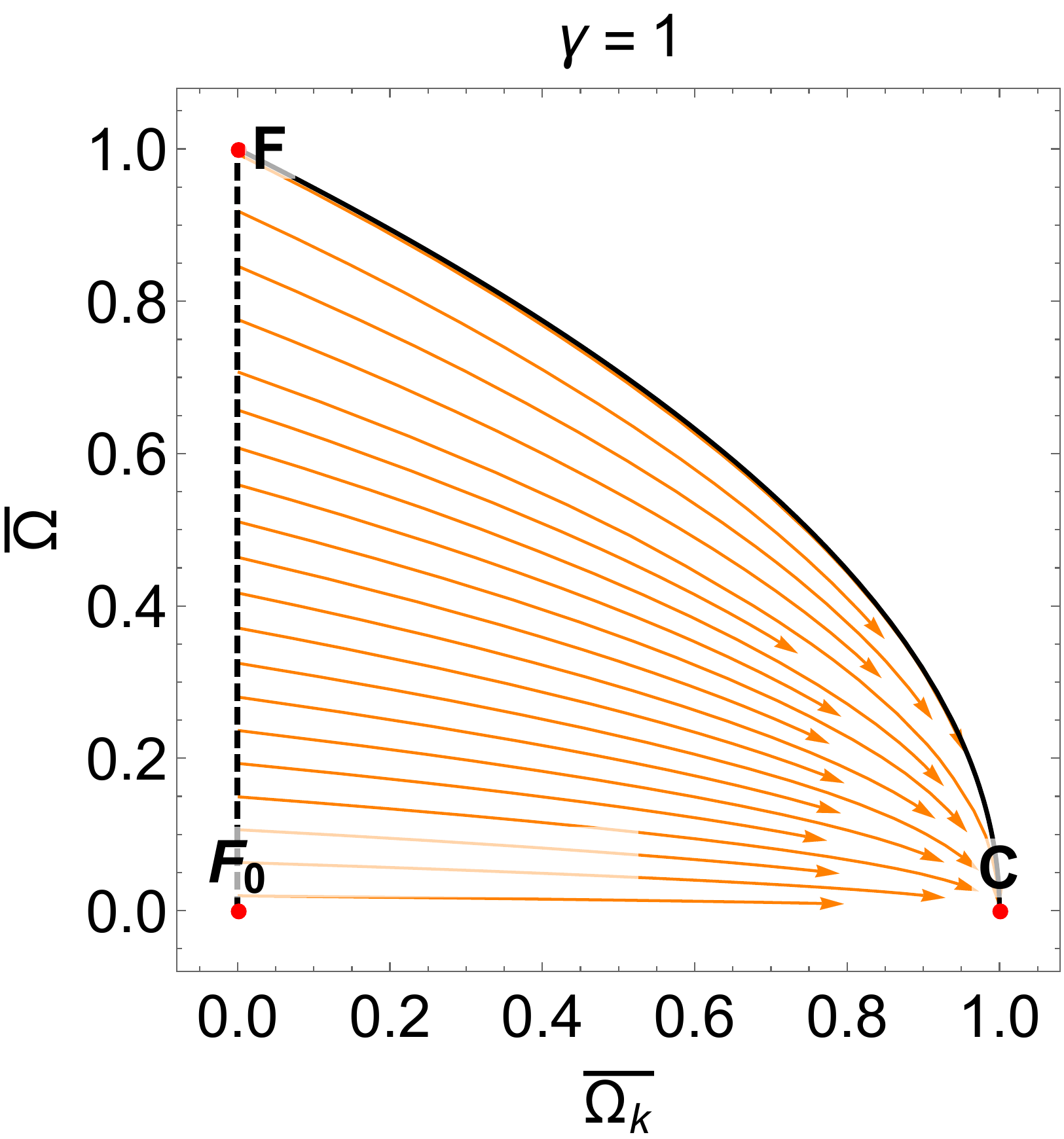}
    \caption{Phase plane of system \eqref{guidingB1gamma1} for $\gamma=1$}
    \label{fig:figu2}
\end{figure}

\begin{figure*}[t!]
    \centering
    \subfigure[]{\includegraphics[width=0.45\textwidth]{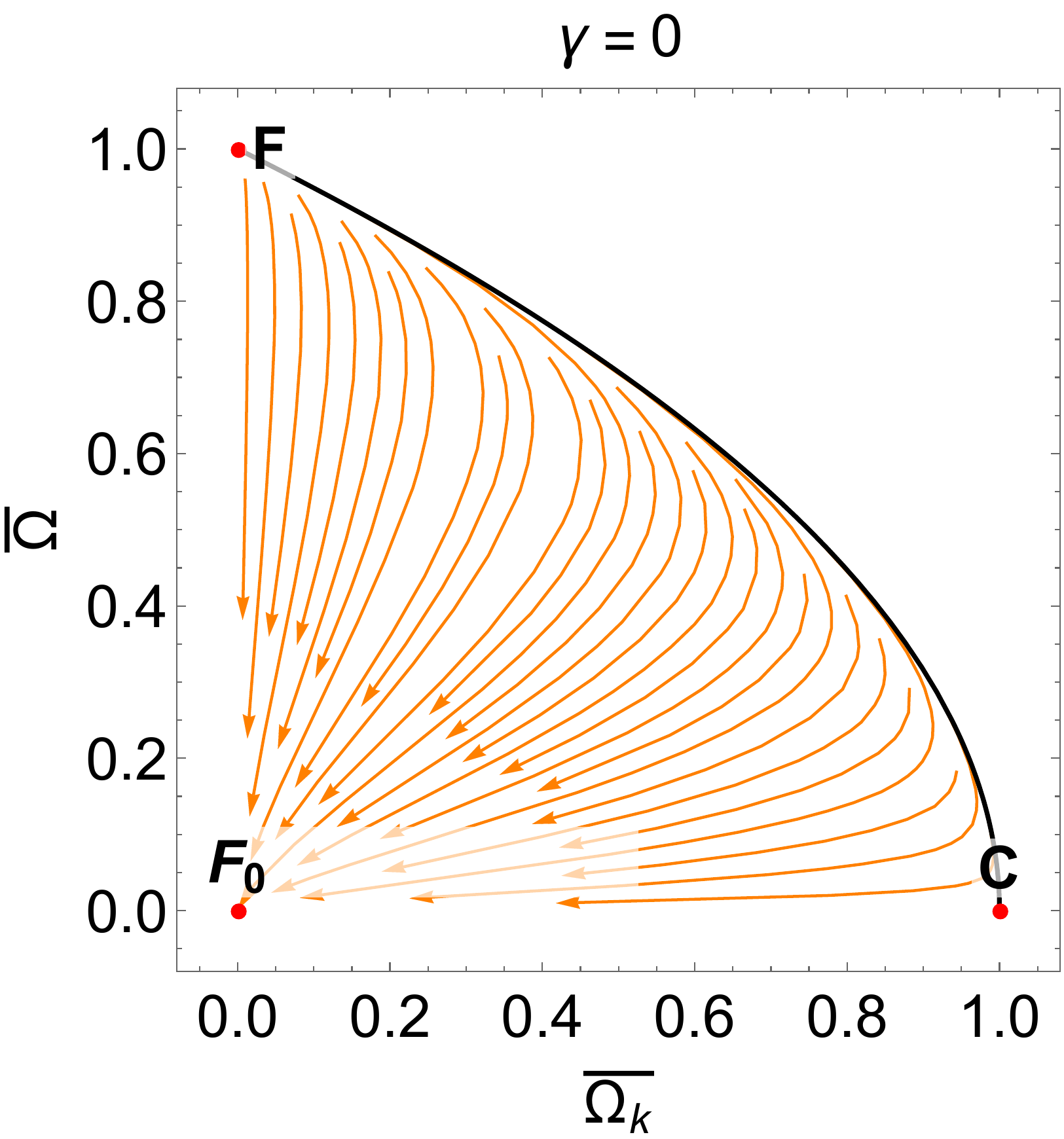}}
    \subfigure[]{\includegraphics[width=0.45\textwidth]{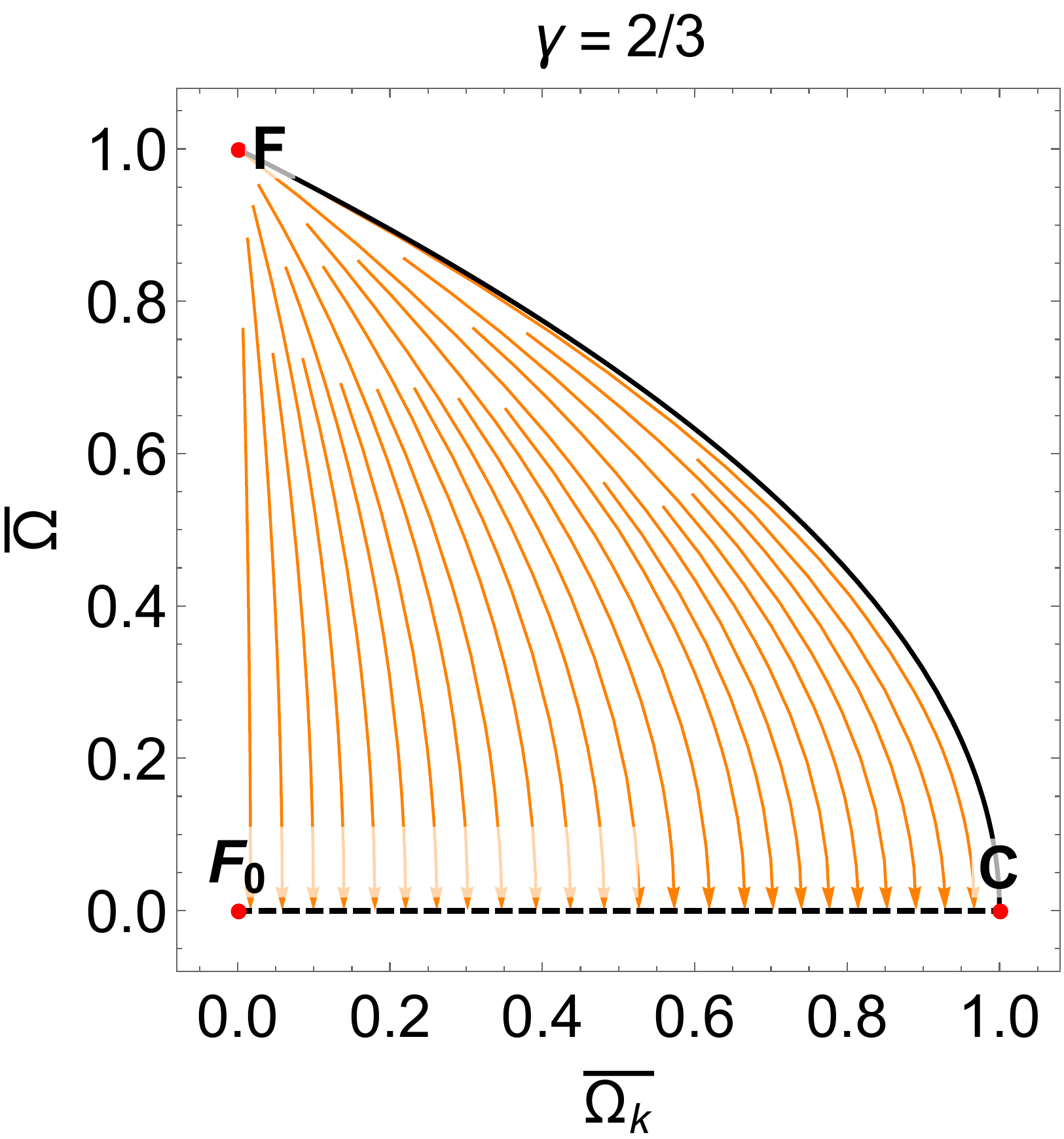}}
    \subfigure[]{ \includegraphics[width=0.45\textwidth]{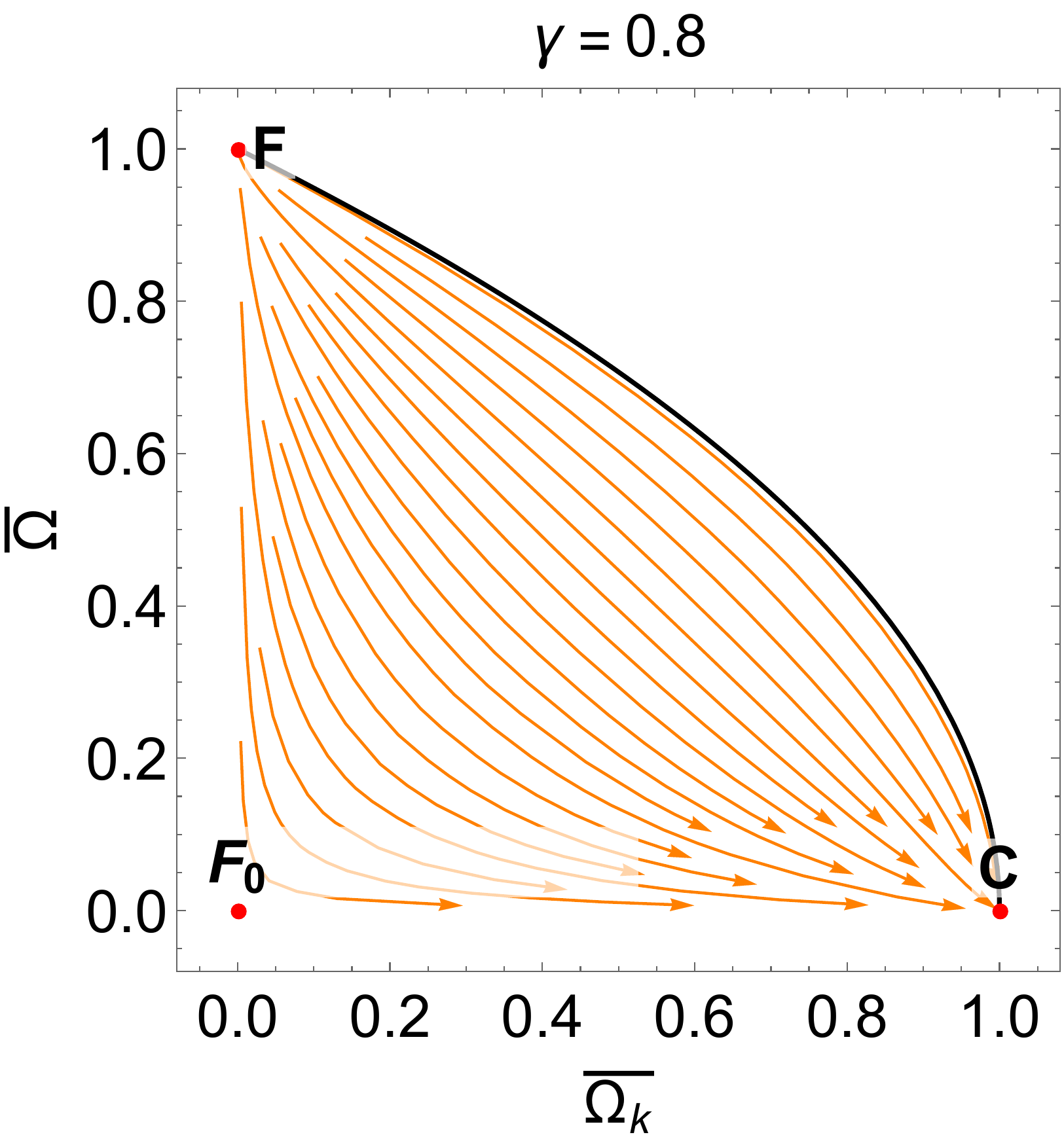}}
    \subfigure[]{\includegraphics[width=0.45\textwidth]{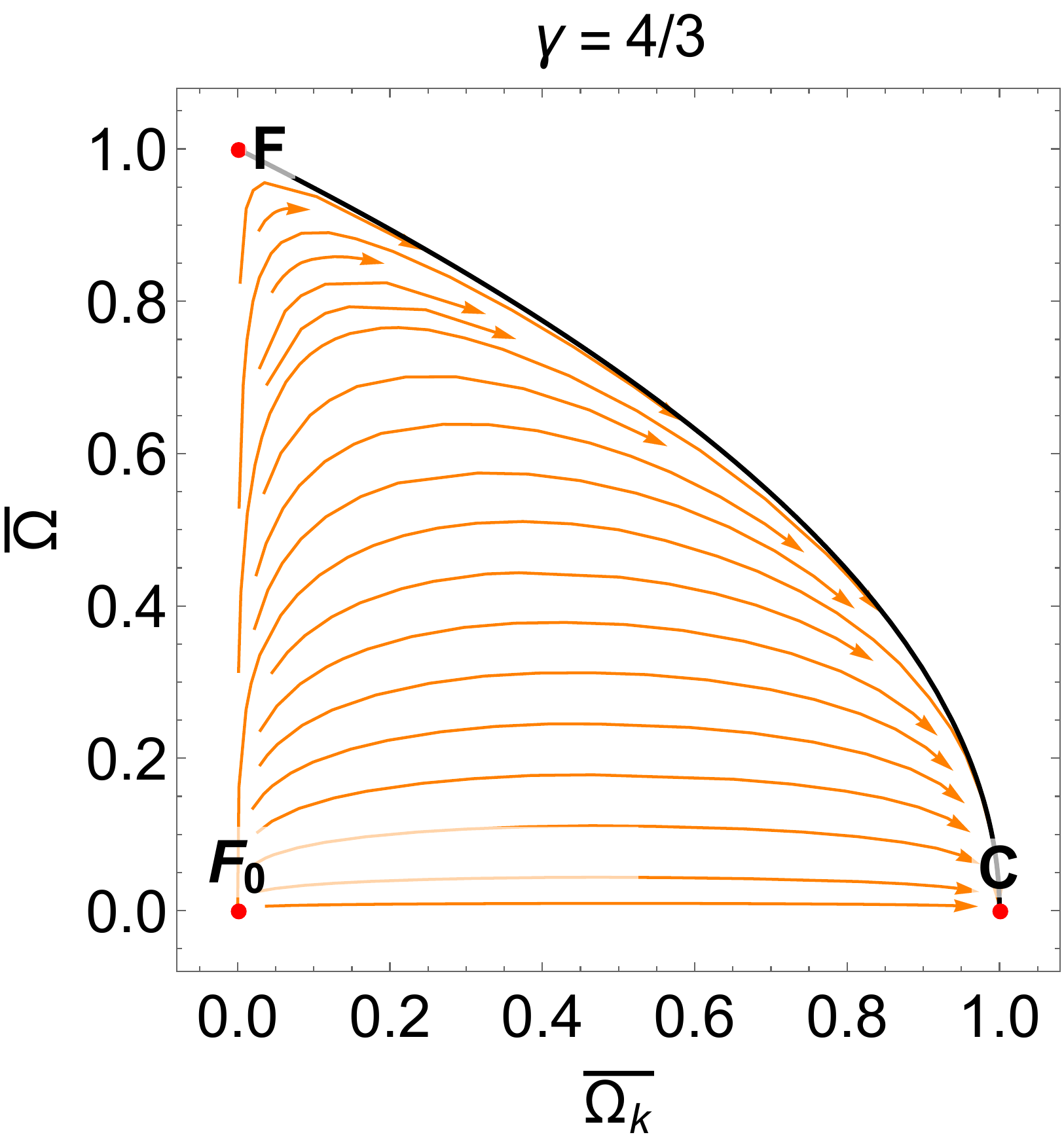}}
    \caption{Phase plane for system \eqref{guidingB1}, \eqref{guidingB2} for different choices of $\gamma$.}
    \label{fig:figu1}
\end{figure*}

\noindent where $\bar{\Omega}(0)=\Omega_0$, ${\bar{\Omega}_{k}}(0)=\Omega_{k0}$.
 
The equilibrium point $C: (\bar{\Omega}, {\bar{\Omega}_{k}})= (0,1)$ with eigenvalues $\{-1,-\frac{1}{2}\}$ is the late-time attractor. 
\newline
The line of equilibrium points   $\bar{ \Omega}(\Omega_0)=\Omega_0, \;  {\bar{\Omega}_{k}}(\Omega_0)=0$ where $\Omega_0 \in \mathbb{R}$ with eigenvalues $\{1,0\}$ is normally hyperbolic. Therefore, by considering eigenvalues with non zero real parts, it is found  that the line is unstable. This line contains the points $F_0, F$. In  Figure \ref{fig:figu2}  the phase plane of system \eqref{guidingB1gamma1} for $\gamma=1$  where the unstable line is ${\bar{\Omega}_{k}}= 0$ and the attractor is $C$ is presented.
\newline
For $\gamma\neq 1$, the 2D guiding system \eqref{guidingB1}, \eqref{guidingB2} has the following equilibrium points:
\begin{enumerate}    
    \item $F_0: (\bar{\Omega}, {\bar{\Omega}_{k}})=(0,0)$ with eigenvalues \newline $\left\{\frac{3 (\gamma -1)}{2},3
   \gamma -2\right\}$.
   \begin{enumerate}
       \item It is a sink for $0<\gamma <\frac{2}{3}$.
       \item It is nonhyperbolic for $\gamma =\frac{2}{3}$.
       \item It is a saddle for $\frac{2}{3}<\gamma <1$. 
       \item It is source for $1<\gamma <2$.
    \end{enumerate}
Evaluating Eq.  \eqref{eqH}  at $F_0$ we obtain 
\begin{equation}
\left\{\begin{array}{c}
      \dot{H}= -\frac{3}{2} \gamma H^2\\\\ 
      \dot{a}=a H  
       \end{array} \right.\implies 
       \left\{\begin{array}{c}
       H(t)= \frac{2 H_{0}}{3 \gamma  H_{0} t+2} \\\\
       a(t)=
   a_{0} \left(\frac{3 \gamma  H_{0} t}{2}+1\right)^{\frac{2}{3
   \gamma }}
\end{array}\right..
\end{equation}
          That is,  line element \eqref{mOpenFLRW} becomes 
    \begin{small}
    \begin{align}
\label{scaling}
   &  ds^2= - dt^2 + a_{0}^2 \left(\frac{3 \gamma  H_{0} t}{2}+1\right)^{\frac{4}{3
   \gamma }} \left (dr^2 +\sinh^2 r d\Omega^2 \right),
\end{align}
\end{small}
\newline where
$d\Omega^2 = d\y^2 + \sin^2 \y\, d\z^2$
is the metric for a two-sphere. 
The corresponding solution is a matter dominated FLRW universe, i.e., $\bar{\Omega}_m=1$. 
   
    \item $F: (\bar{\Omega}, {\bar{\Omega}_{k}})=(1,0)$ with eigenvalues \newline $\{1,-3 (\gamma -1)\}$.
    \begin{enumerate}
        \item It is a source for $0<\gamma <1$, 
        \item It is a saddle for $1<\gamma <2$. 
    \end{enumerate}
Evaluating Eq. \eqref{eqH}  at $F_1$ we obtain 
\begin{equation}
\left\{\begin{array}{c}
      \dot{H}= -\frac{3}{2}  H^2\\\\ 
      \dot{a}=a H  
       \end{array} \right.\implies 
       \left\{\begin{array}{c}
       H(t)= \frac{2 H_{0}}{3    H_{0} t+2} \\\\
       a(t)=
   a_{0} \left(\frac{3    H_{0} t}{2}+1\right)^{\frac{2}{3
     }}
\end{array}\right..
\end{equation}
Then, line element \eqref{mOpenFLRW} becomes 
\begin{align}
   &  ds^2= - dt^2 + a_{0}^2 \left(\frac{3    H_{0} t}{2}+1\right)^{\frac{4}{3
     }} dr^2  \nonumber \\
   & +  a_{0}^2 \left(\frac{3    H_{0} t}{2}+1\right)^{\frac{4}{3
     }} \left( dr^2+
\sinh^2 r d\Omega^2 \right). \label{metricFopenFLRW}
\end{align}  
 Hence for large $t$ the equilibrium point  can be associated with  Einstein-de-Sitter solution. 
    \item $C: (\bar{\Omega}, {\bar{\Omega}_{k}})=(0,1)$ with eigenvalues \newline $\left\{-\frac{1}{2},2-3 \gamma
   \right\}$.
   \begin{enumerate}
         \item It is a saddle for $0<\gamma <\frac{2}{3}$.    
         \item It is nonhyperbolic for $\gamma =\frac{2}{3}$.
         \item It is a sink for $\frac{2}{3}<\gamma <2$.
   \end{enumerate}
   Evaluating the deceleration parameter \eqref{eq.49e} at  $C$ we have $q=0$. Then, 
  \begin{equation}
\left\{\begin{array}{c}
     \dot{H}= - H^2  \\\\
     \dot{a}=a H
\end{array}\right.
\implies \left\{\begin{array}{c} H(t)= \frac{H_{0}}{
   H_{0} t+1}\\\\
a(t)= a_0 (H_0 t+1)
\end{array}\right..
\end{equation}
The line element \eqref{mOpenFLRW} becomes
\begin{align}
   &  ds^2= - dt^2 + a_{0}^2 \left(H_0 t+1\right)^{2} \left( dr^2+
\sinh^2 r d\Omega^2 \right) \label{Milne}
\end{align}
This is a curvature dominated Milne  solution ($\Omega=\Omega_m=0, \Omega_k=1, k=-1$) (\cite{Milne}; \cite{WE} Sect. 9.1.6, Eq. (9.8), \cite{Misner:1974qy,Carroll:2004st,Mukhanov:2005sc}). 
    \end{enumerate}
In table \ref{MetricsOpenFLRW} exact solutions associated with equilibrium points of the reduced averaged system  \eqref{guidingB1}-\eqref{guidingB2} where $a(t)$ denote scale factors of metric \eqref{mOpenFLRW} and $a_0 \in \mathbb{R}^+$  are presented.        
    
\noindent
In Figure \ref{fig:figu1} the phase plane of the system \eqref{guidingB1}, \eqref{guidingB2} for $\gamma=0, \frac{2}{3}, 0.8$ and $\gamma=\frac{4}{3}$ is portrayed.

\subsubsection{Late-time behaviour}
The results from the linear stability analysis combined with Theorem \ref{BIIILFZ11} (for $\Sigma=0$) lead to: 
\begin{thm}
\label{thm10}
The late time attractors of  the full system  \eqref{unperturbed1FLRW} and the averaged system \eqref{avrgsystFLRW} are: 
\begin{enumerate}
    \item[(i)] The matter dominated FLRW universe $F_0$
   with line element  \eqref{scaling} 
for $0< \gamma \leq \frac{2}{3}$. $F_0$ represents a  quintessence fluid or a zero-acceleration model for $\gamma=\frac{2}{3}$. 
In the limit $\gamma=0$ we have  de Sitter solution. 

\item[(ii)]   The Milne solution $C$ with $\bar{\Omega}_k=1, k=-1$ with  line element \eqref{Milne} 
for $\frac{2}{3}<\gamma <2$. 
    \end{enumerate}
\end{thm}
\section{Conclusions}
\label{Conclusions}
In this paper we  have used asymptotic methods and averaging theory to explore the solution's space of  scalar field cosmologies with generalized harmonic potential \eqref{pot} in vacuum or minimally coupled to matter. We have studied systems that can be expressed in the standard form \eqref{standard51},  
where $H$ is the Hubble parameter and is positive  strictly decreasing in $t$ and $\lim_{t\rightarrow \infty}H(t)=0$. 
Defining  $Z= 1-\Sigma ^2-\Omega ^2- \Omega_{k}$ which is monotonic as $H\rightarrow 0$, due to there is a continuous function $\alpha$ such that $\dot{Z}= -H Z \alpha +\mathcal{O}\left(H^2\right)$ it was proved that sign of $1-\Sigma ^2-\Omega ^2- \Omega_{k}$ is invariant as $H\rightarrow 0$.  Then, from  equation  \eqref{EQ:81b}
it was proved that $H$ is a  monotonic decreasing function of  $t$  if $0<\Omega^2+\Sigma^2+{\Omega_k}<1$. 
This  implies $\lim_{t\rightarrow \infty} H(t)=0$ based on the invariance of  initial surface for large $t$. Hence, from Theorem \ref{BIIILFZ11} was deduced that $\Omega , \Sigma, \Omega_k$, and $\Phi $  evolve according to the time-averaged system \eqref{IIIeq24}, \eqref{IIIeq25}, \eqref{IIIeq26}, \eqref{IIIeq27} as $H\rightarrow 0$.  
Then, the stability of a periodic solution can be established as it matches  exactly the stability of  stationary solution of  time-averaged system. We have given a rigorous demonstration of Theorem   \ref{BIIILFZ11} in \ref{appendixeq28} based on the construction of a smooth local near-identity nonlinear transformation, well-defined as $H$ tends to zero. We have used properties of the sup norm, and the theorem of the mean values  for a vector function $\bar{\mathbf{f}}: \mathbb{R}^3\longrightarrow \mathbb{R}^3$. We have explained preliminaries of  the method of proof in Section \ref{SECT3.4}.

In particular,  in LRS Bianchi III  late time attractors of full system \eqref{BIIIunperturbed1} and averaged system  \eqref{BIIIavrgsyst} for Bianchi III line element are:
\begin{enumerate}
    \item[(i)]  The matter dominated FLRW universe $F_0$    with line element \eqref{scaling2}  if  $0<\gamma \leq \frac{2}{3}$. $F_0$ represents a  quintessence fluid or a zero-acceleration model for $\gamma=\frac{2}{3}$. 
In the limit $\gamma=0$ we have  de Sitter solution. 
    
    \item[(ii)] The matter-curvature scaling solution $CS$ with $\bar{\Omega}_m=3(1-\gamma)$ and line element \eqref{scaling2b}  if  $\frac{2}{3}<\gamma <1.$
    
    \item[(iii)] The Bianchi III flat spacetime $D$  with line element \eqref{metricD}   if $1<\gamma\leq 2$.
 \end{enumerate}

For FLRW metric with $k=-1$, late time attractors of  full system  \eqref{unperturbed1FLRW} and  averaged system  are: 
\begin{enumerate}
    \item[(i)] The matter dominated FLRW universe $F_0$
   with line element \eqref{scaling}
for $0<\gamma \leq \frac{2}{3}$. $F_0$ represents a  quintessence fluid or a zero-acceleration model for $\gamma=\frac{2}{3}$. 
In the limit $\gamma=0$ we have de Sitter solution. 

\item[(ii)]   The Milne solution $C$ with  line element \eqref{Milne}  
for $\frac{2}{3}<\gamma <2$. 
    \end{enumerate}
Summarizing, in LRS Bianchi III late-time attractors are: a  matter dominated flat FLRW universe  if  $0\leq \gamma \leq \frac{2}{3}$ (mimicking de Sitter, quintessence or zero acceleration solutions), a matter-curvature scaling solution  if  $\frac{2}{3}<\gamma <1$ and  Bianchi III flat spacetime  for $1\leq \gamma\leq 2$.  For FLRW metric with $k=-1$ late time attractors are: the matter dominated FLRW universe  if  $0\leq \gamma \leq \frac{2}{3}$ (mimicking de Sitter, quintessence or zero acceleration solutions) and  Milne solution if $\frac{2}{3}<\gamma <2$.  In all metrics,  matter dominated flat  FLRW universe represents quintessence fluid if $0< \gamma < \frac{2}{3}$. 
\newline 
Continuing the Program ``Averaging Generalized Scalar Field Cosmologies'', the cases:  (II)  Bianchi I and flat  FLRW model and (III) KS and closed FLRW are  studied in two companion papers \cite{Leon:2021rcx,Leon:2021hxc}, respectively. In  \cite{Leon:2021rcx} using the same approach used here we found the oscillations entering the full system through KG equation can be controlled and smoothed out when the Hubble factor $H$ is used as a time-dependent parameter since it tends monotonically to zero and preserves its sign during the evolution. However, in  case (III) this approach is not valid given that $H$ is not necessarily monotonically decreasing to zero, and it can change its sign. Therefore, we have developed an alternative procedure in \cite{Leon:2021hxc}.
For LRS Bianchi I and  flat FLRW metrics as well as for LRS Bianchi III and open FLRW, we use Taylor expansion with respect to $H$ near $H=0$ such that the  resulting system can be expressed in standard form \eqref{standard51} after selecting a convenient angular frequency $\omega$ in the transformation \eqref{eqAA25}.  
\noindent Next, we have taken the time-averaged of  previous system, obtaining a system that can be easily studied using dynamical systems tools. Using the last approach, we have formulated Theorems \ref{thm8} and  \ref{thm10} about  late-time behavior of our model, whose proofs are based on  Theorem \ref{BIIILFZ11} center manifold calculations and linear stability analysis. 

\noindent 
As in paper \cite{Fajman:2020yjb}, our analytical results were strongly supported by numerics in  \ref{numerics} as well. We showed that asymptotic methods and averaging theory are powerful tools to investigate scalar field cosmologies with generalized harmonic potential; which have evident advantages, e.g., it is not needed to analyze the full dynamics to determine the stability of the full oscillation, but only the late-time behavior of the time-averaged (simpler) system has to be analyzed.  Interestingly, for LRS Bianchi III and open  FLRW model, when matter fluid  corresponds to a cosmological constant, $H$ tends asymptotically to constant values depending on the initial conditions which is consistent to de Sitter expansion (see Figures \ref{fig:BIIICC3DS} and \ref{fig:OpenFLRWCC3D}). In  addition, for open  FLRW and any $\gamma<\frac{2}{3}$ and $\Omega_k>0$, $\Omega_k \rightarrow 0$. On the other hand, when $\gamma>\frac{2}{3}$ and $\Omega_k>0$ the universe becomes curvature dominated asymptotically ($\Omega_k \rightarrow 1$). In   \ref{numerics} are presented evidences that the main theorem of section \ref{SECT:II} is valid for LRS Bianchi III and for  FLRW metrics with negative curvature.

\section*{Acknowledgements}

This research was funded by  Agencia Nacional de Investigaci\'on y Desarrollo- ANID  through the program FONDECYT Iniciaci\'on grant no.
11180126 and by Vicerrector\'{\i}a de Investigaci\'on y Desarrollo Tecnol\'ogico at
Universidad Cat\'olica del Norte. Ellen de los Milagros Fern\'andez Flores  is acknowledged for proofreading this manuscript and  improving the English. We thank anonymous referee for his/her comments which have helped us  improve our work.

\appendix

\section{Proof of Theorem \ref{BIIILFZ11}}
\label{appendixeq28}

\begin{lem}[\textbf{Gronwall's Lemma (Integral form)}]
\label{Gronwall}
Let be $\xi(t)$ a nonnegative function, summable over  $[0,T]$ which satisfies almost everywhere the integral inequality $$\xi(t)\leq C_1 \int_0^t \xi(s)ds +C_2, \;  C_1, C_2\geq 0.$$
       Then, 
      $$\xi(t)\leq C_2  e^{C_1 t},$$
        almost everywhere for $t$ in $0\leq t\leq T$.
 In particular, if    
     $$\xi(t)\leq C_1 \int_0^t \xi(s)ds, \;  C_1\geq 0$$
        almost everywhere for $t$ in $0\leq t\leq T$. Then,  $
           \xi \equiv 0$  
        almost everywhere for $t$ in $0\leq t\leq T$.
 \end{lem}
 
\begin{lem}[Mean value theorem]
\label{lemma6}
 Let $U \subset \mathbb{R}^n$ be open, $\mathbf{f}: U \rightarrow \mathbb{R}^m$ continuously differentiable, and $\mathbf{x}\in U$, $\mathbf{h}\in \mathbb{R}^m$ vectors such that the line segment $\mathbf{x}+z \; \mathbf{h}$,  $0 \leq z \leq 1$ remains in $U$. Then we have:
\begin{equation}
    \mathbf{f}(\mathbf{x}+\mathbf{h})-\mathbf{f}(\mathbf{x}) = \left (\int_0^1 D\mathbf{f}(\mathbf{x}+z \; \mathbf{h})\,dz\right)\cdot \mathbf{h},
\end{equation} where  $D \mathbf{f}$ denotes the Jacobian matrix  of $\mathbf{f}$ and the integral of a matrix is to be understood componentwise.
\end{lem}

\textbf{Proof of Theorem \ref{BIIILFZ11} }.
Defining $Z= 1-\Sigma ^2-\Omega ^2- \Omega_{k}$ it follows 
from 
\begin{align*}
   & \dot{Z}= -H Z \Big(3 (\gamma -2) \Sigma ^2+3 (\gamma -1) \Omega ^2   +(3 \gamma -2) \Omega_{k} \nonumber \\
   & -3 \Omega ^2 \cos (2 (\Phi -t \omega ))\Big)+\mathcal{O}\left(H^2\right)\end{align*}  that the sign of $1-\Sigma ^2-\Omega ^2- \Omega_{k}$
 is invariant as $H\rightarrow 0$. 
 From  equation  \eqref{EQ:81b}
it follows that $H$ is a  monotonic decreasing function of  $t$  if $0<\Omega^2+\Sigma^2+{\Omega_k}<1$. 
 These allow to define recursively  bootstrapping sequences 
\begin{align}
    & \left\{\begin{array}{c}
       t_0=t_{*}   \\ \\
        H_0=H(t_{*}) 
    \end{array}\right., \nonumber \\
    & 
    \left\{\begin{array}{c}
       {t_{n+1}}= {t_{n}} +\frac{1}{H_n}   \\ \\
       H_{n+1}= H(t_{n+1})  
    \end{array}\right.,
\end{align}
such that  $\lim_{n\rightarrow \infty}H_n=0$ y $\lim_{n\rightarrow \infty} t_n=\infty$.

\noindent 
Given  expansions \eqref{AppBIIIquasilinear211}, eqs. \eqref{eqT59} become
\begin{widetext}
\begin{small}
\begin{align*}
 & \dot{\Omega_0}=  \frac{1}{2} \Bigg({\Omega_0} \Big(6 {\Sigma_{0}}^2+3 {\Omega_0}^2+2 {\Omega_{k0}}   -3 \gamma 
   \left({\Sigma_{0}}^2+{\Omega_0}^2+{\Omega_{k0}}-1\right)   +3 \left({\Omega_0}^2-1\right) \cos (2 (t \omega -\Phi_{0}))-3\Big)\Bigg)
   H  - \frac{\partial g_1}{\partial t}H +\mathcal{O}\left(H^2\right),
\end{align*}
\begin{align*}
 & \dot{\Sigma_0}=   \frac{1}{2} \Bigg(3 {\Sigma_{0}} \cos (2 (t \omega -\Phi_{0})) {\Omega_0}^2 +2 {\Omega_{k0}}    +{\Sigma_{0}} \Big(6 {\Sigma_{0}}^2+3 {\Omega_0}^2+2 {\Omega_{k0}} -6    -3 \gamma  \left( {\Sigma_{0}}^2+{\Omega_0}^2+{\Omega_{k0}}-1\right)\Big)\Bigg) H    - \frac{\partial g_2}{\partial t}H +\mathcal{O}\left(H^2\right),
\end{align*}
\begin{align*}
 & \dot{\Omega_{k0}}=\Bigg({\Omega_{k0}} \Big(6 {\Sigma_{0}}^2-2
   {\Sigma_{0}}+3 {\Omega_0}^2+2 {\Omega_{k0}}   
    -3 \gamma  \left({\Sigma_{0}}^2+{\Omega_0}^2+{\Omega_{k0}}-1\right)     +3 {\Omega_0}^2  \cos (2 (t \omega -\Phi_{0}))-2\Big) \Bigg) H    - \frac{\partial g_3}{\partial t}H +\mathcal{O}\left(H^2\right),
\\
    & \dot{\Phi_0}=-\left(\frac{3}{2} \sin (2 (t
   \omega - \Phi_{0}))+\frac{\partial g_4}{\partial t}\right)
   H+\mathcal{O}\left(H^2\right).
\end{align*}
\end{small}
Furthermore, eqs. \eqref{eqT602} become
\begin{subequations}
\label{eq28}
\begin{align}
& \frac{\partial g_1}{\partial t}=  \frac{3}{2} {\Omega_0} \left({\Omega_0}^2-1\right) \cos (2 ({\Phi_0}-t \omega )),
\\
& \frac{\partial g_2}{\partial t}=\frac{3}{2} {\Sigma_0} {\Omega_0}^2 \cos (2 (\Phi_{0}-t \omega )),
\\
& \frac{\partial g_3}{\partial t}= 3 {\Omega_0}^2 {\Omega_{k0}} \cos (2 ({\Phi_0}-t \omega
   )),
\\
&  \frac{\partial g_4}{\partial t}=-\frac{3}{2} \sin (2 ({\Phi_0}-t \omega )).
\end{align}
\end{subequations}
Then, explicit expressions of the $g_i, i=1, \ldots, 4$ are straightforwardly found by  integration of \eqref{eq28}:
\begin{align}
& g_1(H , \Omega_{0}, \Sigma_{0}, \Omega_{k0}, \Phi_{0}, t) =  \frac{3 {\Omega_0} \left(1-{\Omega_0}^2\right) \sin (2 (\Phi_0-t \omega ))}{4 \omega } +C_1(\Omega_{0}, \Sigma_{0}, \Omega_{k0}, \Phi_{0}), 
\\
& g_2(H , \Omega_{0}, \Sigma_{0}, \Omega_{k0}, \Phi_{0}, t)   =  -\frac{3 {\Sigma_0} {\Omega_0}^2 \sin (2 ({\Phi_{0}}-t \omega ))}{4 \omega } +C_2(\Omega_{0}, \Sigma_{0}, \Omega_{k0}, \Phi_{0}),
\\
& g_3(H , \Omega_{0}, \Sigma_{0}, \Omega_{k0}, \Phi_{0}, t)  =  -\frac{3 {\Omega_0}^2 {\Omega_{k0}} \sin (2 ({\Phi_{0}}-t \omega ))}{2 \omega }   +C_3(\Omega_{0}, \Sigma_{0}, \Omega_{k0}, \Phi_{0}),
\\
&g_4(H , \Omega_{0}, \Sigma_{0}, \Omega_{k0}, \Phi_{0}, t)   = \frac{3 \cos (2 (\Phi_0-t \omega ))}{4 \omega } +C_4(\Omega_{0}, \Sigma_{0}, \Omega_{k0}, \Phi_{0}), 
\end{align}
where we can set four integration functions $C_i(\Omega_{0}, \Sigma_{0}, \Omega_{k0}, \Phi_{0}), i = 1,2,3,4$ to zero. Functions   $g_i, i=1, \ldots, 4$ are continuously differentiable, such that their partial derivatives are bounded on $t\in [t_n, t_{n+1}]$.

\noindent
Let    $\Delta \Omega_0= \Omega_0 - \bar{\Omega}, \;  \Delta \Sigma_0= \Sigma_0 - \bar{\Sigma}, \;  \Delta \Omega_{k0}= \Omega_{k0}- {\bar{\Omega}_k},\;  \Delta \Phi_0= \Phi_0 - \bar{\Phi}$ be defined such that 
$\Omega_0(t_n)=\bar{\Omega}(t_n)= {\Omega_{n}}, \;   \Sigma_0(t_n)=\bar{\Sigma}(t_n)= {\Sigma}_{n}, \;  \Omega_{k0}(t_n)={\bar{\Omega}_k}(t_n)= {\Omega_k}_{n}, \;  \Phi_0(t_n)=\bar{\Phi}(t_n)= {\Phi}_{n},$
with 
$0<\Omega(t_n)+\Sigma(t_n)^2+{\Omega_k}(t_n)^2 <1.$

\noindent
Keeping the terms of second order in $H$,  system \eqref{EqY602} becomes
\begin{small}
\begin{subequations}
\label{eqA8}
\begin{align}
    & \dot{\Delta \Omega_0}= \frac{1}{2} H \left(\bar{\Omega } \left(3 (\gamma -2) \bar{\Sigma }^2+3 (\gamma -1) \left(\bar{\Omega }^2-1\right)+(3 \gamma -2) \bar{\Omega}_k\right)+\Omega_0 \left(3 \gamma  \left(1-\Sigma_{0}^2-\Omega_0^2-\Omega_{k0}\right)+6 \Sigma_0^2+3 \Omega_0^2+2 \Omega_{k0}-3\right)\right) \nonumber \\
    & +H^2 \sin (2 (\Phi_0-t \omega )) \Bigg(\frac{\Omega_0^3 \left(2 \mu
   ^2-\omega ^2\right)^3}{4 b^2 \mu ^6 \omega ^3}+\frac{\Omega_0 \left(9 b^2 \mu ^6 \omega ^2 \left(-4 \Omega_0^4+3 \Omega_0^2+1\right)-2 \Omega_0^2 \left(2 \mu ^2-\omega ^2\right)^3\right)
   \cos (2 (\Phi_0-t \omega ))}{8 b^2 \mu ^6 \omega ^3} \nonumber\\
   &+\frac{3 \left(\Omega_0^2-1\right) \Omega_0 \left(3 \gamma  \left(\Sigma_0^2+\Omega_0^2+\Omega_{k0}-1\right)-6
   \Sigma_0^2-3 \Omega_0^2-2 \Omega_{k0}\right)}{8 \omega }\Bigg),\\
   & \dot{\Delta \Sigma_0}= \frac{1}{2} H \left(3 \bar{\Sigma } \left((\gamma -2) \left(\bar{\Sigma }^2-1\right)+(\gamma -1) \bar{\Omega
   }^2\right)+\bar{\Omega}_k \left((3 \gamma -2) \bar{\Sigma }-2\right)+\Sigma_0 \left(-3 \gamma  \left(\Sigma_0^2+\Omega_0^2+\Omega_{k0}-1\right)+6 \Sigma_0^2+3
   \Omega_0^2+2 \Omega_{k0}-6\right)+2 \Omega_{k0}\right) \nonumber \\
   & +H^2 \sin (2 (\Phi_0-t \omega )) \left(\frac{3 \Omega_0^2 \left(3 \gamma  \Sigma_0 \left(\Sigma_0^2+ \Omega_{0}^2+\Omega_{k0}-1\right)-3 \Sigma_0 \left(2 \Sigma_0^2+\Omega_0^2\right)-2 (\Sigma_0+1) \Omega_{k0}\right)}{8 \omega }-\frac{9 \Sigma_0 \Omega_0^4 \cos (2
   (\Phi_0-t \omega ))}{2 \omega }\right),\\
   & \dot{\Delta \Omega_{k0}}= H \left(\bar{\Omega}_k \left(3 (\gamma -2) \bar{\Sigma }^2+3 (\gamma -1) \bar{\Omega }^2+(3 \gamma -2) \left(\bar{\Omega}_k-1\right)+2 \bar{\Sigma
   }\right)+\Omega_{k0} \left(-3 \gamma  \left(\Sigma_0^2+\Omega_0^2+\Omega_{k0}-1\right)+6 \Sigma_0^2-2 \Sigma_0+3 \Omega_0^2+2 \Omega_{k0}-2\right)\right) \nonumber \\
   & +H^2
   \sin (2 (\Phi_0-t \omega )) \left(\frac{3 \Omega_0^2 \Omega_{k0} \left(3 \gamma  \left(\Sigma_0^2+\Omega_0^2+\Omega_{k0}-1\right)-6 \Sigma_0^2+2 \Sigma_0-3
   \Omega_0^2-2 \Omega_{k0}\right)}{4 \omega }-\frac{45 \Omega_0^4 \Omega_{k0} \cos (2 (\Phi_0-t \omega ))}{4 \omega }\right),\\
   & \dot{\Delta \Phi_0}= H^2 \Bigg(\frac{3 \bar{\Omega }^2 \left(2 \mu ^2-\omega
   ^2\right)^3}{8 b^2 \mu ^6 \omega ^3}-\frac{\Omega_0^2 \left(2 \mu ^2-\omega ^2\right)^3 \sin ^4(\Phi_0-t \omega )}{b^2 \mu ^6 \omega ^3} \nonumber \\
   & +\frac{3  \cos (2 (\Phi_0-t \omega ))}{8 \omega } \left(-3 \gamma  \left(\Sigma_0^2+\Omega_{0}^2+\Omega_{k0}-1\right)+6 \Sigma_0^2+3 \Omega_0^2+2 \Omega_{k0} + 3\left(\Omega_0^2+2\right) \cos (2 (\Phi_0-t
   \omega ))\right) \Bigg). \label{A11}
\end{align}
\end{subequations}
\end{small}
\noindent
Denoting $\mathbf{x}_0=(\Omega_0, \Sigma_0, \Omega_{k0})^T$, $\bar{\mathbf{x}}=(\bar{\Omega}, \bar{\Sigma}, \bar{\Omega}_{k})^T$ the system \eqref{eqA8} can be written as a 3-dimensional system: 
\begin{align*}
 & \dot{\Delta\mathbf{x}_0}= H \left(\bar{\mathbf{f}}( {\mathbf{x}}_0)-\bar{\mathbf{f}}(\bar{\mathbf{x}})\right) +   \mathcal{O}(H^2), 
  \end{align*}
plus eq. \eqref{A11}, where the vector function $\bar{\mathbf{f}}$ 
is given explicitly (last row corresponding to $\Delta{\Phi}_0$ was omitted) by: 
\begin{align*}
    \bar{\mathbf{f}}(y_1, y_2, y_3)= \left(
\begin{array}{c}
 -\frac{1}{2} y_1 \left(-3 \gamma +3 (\gamma -1) y_1^2+3 (\gamma -2) y_2^2+3 \gamma  y_3-2 y_3+3\right) \\
 \frac{1}{2} \left(y_2 \left(3 \gamma -3 (\gamma -1) y_1^2-3 \gamma  y_3+2 y_3-6\right)-3 (\gamma -2) y_2^3+2 y_3\right) \\
 y_3 \left(-3 (\gamma -1) y_1^2-3 (\gamma -2) y_2^2-2 y_2-(3 \gamma -2) (y_3-1)\right) \\
\end{array}
\right).
\end{align*}
It is a vector function with polynomial components in variables $(y_1, y_2, y_3)$. Therefore, it is continuously differentiable in all its components.  

Let be $\Delta\mathbf{x}_0(t)= (\Omega_0-\bar{\Omega},{\Sigma_0}-  \bar{\Sigma}, \bar{\Omega}_k - \Omega_{k0})^T$ with $0\leq |\Delta\mathbf{x}_0|:=\max \left\{|\Omega_0-\bar{\Omega}|, |{\Sigma_0}-  \bar{\Sigma}|, |\Omega_{k0} - \bar{\Omega}_k| \right\}$  finite in the closed interval $[t_n,t]$. Using same initial conditions for $\mathbf{x}_0$ and $\bar{\mathbf{x}}$ we obtain by integration: 
\begin{align*}
 \Delta\mathbf{x}_0(t) = \int_{t_n}^t \dot{\Delta\mathbf{x}_0} d s =  \int_{t_n}^t \left(H \left(\bar{\mathbf{f}}( {\mathbf{x}}_0)-\bar{\mathbf{f}}(\bar{\mathbf{x}})\right) +   \mathcal{O}(H^2)\right) ds. 
\end{align*}

Using Lemma \ref{lemma6} we have 
\begin{equation}
   \bar{\mathbf{f}}( {\mathbf{x}}_0(s))-\bar{\mathbf{f}}(\bar{\mathbf{x}}(s)) = \underbrace{\left (\int_0^1 D   \bar{\mathbf{f}}\left(\bar{\mathbf{x}}(s)+ z \; \left({\mathbf{x}}_0(s) - \bar{\mathbf{x}}(s)\right)\right)\,d z\right)}_{\mathbf{A}(s)}\cdot \left({\mathbf{x}}_0(s) - \bar{\mathbf{x}}(s)\right),
\end{equation} where  $D\bar{\mathbf{f}}$ denotes the Jacobian matrix  of $\bar{\mathbf{f}}$ and the integral of a matrix is to be understood componentwise.
Omitting the dependence on $s$ we calculate   the components of  
\begin{equation}
 \mathbf{A}=   \left(
\begin{array}{ccc}
 a_{11} & a_{12} & a_{13} \\
 a_{21} & a_{22} & a_{23} \\
 a_{31} & a_{32} & a_{33} \\
\end{array}
\right),
\end{equation}
which are
\begin{subequations}
\label{A-compts}
\begin{align}
  a_{11}=&\frac{1}{4} \Bigg( -2 (\gamma -2) \bar{\Sigma }^2-2 (\gamma -2) \Sigma _0 \bar{\Sigma }-6 (\gamma -1) \bar{\Omega }^2+6 \Omega _0^2-6 (\gamma -1) \bar{\Omega } \Omega _0 \nonumber \\
  & +2 \left(2 \Sigma _0^2+3 \gamma
   +\bar{\Omega}_k+\Omega _{k0}-3\right)-\gamma  \left(2 \Sigma _0^2+6 \Omega _0^2+3 \bar{\Omega}_k+3 \Omega _{k0}\right)\Bigg), \\  
    a_{12}=&-\frac{1}{2} (\gamma -2) \left(\bar{\Sigma }
   \left(2 \bar{\Omega }+\Omega _0\right)+\Sigma _0 \left(\bar{\Omega }+2 \Omega _0\right)\right),\\
   a_{13}=& -\frac{1}{4} (3 \gamma -2) \left(\bar{\Omega }+\Omega _0\right),\\
   a_{21}=& -\frac{1}{2} (\gamma -1) \left(\bar{\Sigma } \left(2 \bar{\Omega }+\Omega _0\right)+\Sigma _0 \left(\bar{\Omega }+2 \Omega _0\right)\right),\\
   a_{22}= & \frac{1}{2} \Bigg(-\gamma  \bar{\Omega }^2+\bar{\Omega }^2-\gamma 
   \Omega _0 \bar{\Omega }+\Omega _0 \bar{\Omega }-\gamma  \Omega _0^2+\Omega _0^2+3 \gamma  \nonumber \\
   &-\frac{3 \gamma  \bar{\Omega}_k}{2}+\bar{\Omega}_k -3 (\gamma -2) \left(\bar{\Sigma }^2+\Sigma _0 \bar{\Sigma
   }+\Sigma _0^2\right)-\frac{3 \gamma  \Omega _{k0}}{2}+\Omega _{k0}-6\Bigg),\\
   a_{23}= &\frac{1}{4} \left((2-3 \gamma ) \bar{\Sigma }+(2-3 \gamma ) \Sigma _0+4\right),\\
   a_{31}= & -(\gamma -1) \left(\bar{\Omega } \left(2 \bar{\Omega}_k+\Omega _{k0}\right)+\Omega _0 \left(\bar{\Omega}_k+2 \Omega _{k0}\right)\right),\\
   a_{32}= & -\bar{\Omega}_k
   \left((\gamma -2) \left(2 \bar{\Sigma }+\Sigma _0\right)+1\right)-\left((\gamma -2) \left(\bar{\Sigma }+2 \Sigma _0\right)+1\right) \Omega _{k0}, \\
  a_{33}= & -(\gamma -2) \bar{\Sigma }^2-\left((\gamma -2) \Sigma _0+1\right)
   \bar{\Sigma }-(\gamma -1) \bar{\Omega }^2+\Omega _0^2+3 \gamma +2 \bar{\Omega}_k+\Sigma _0 \left(2 \Sigma _0-1\right) \nonumber \\
   & -(\gamma -1) \bar{\Omega } \Omega _0+2 \Omega _{k0}-\gamma  \left(\Sigma _0^2+\Omega
   _0^2+3 \bar{\Omega}_k+3 \Omega _{k0}\right)-2.
\end{align}
\end{subequations}
Taking   sup norm
$ |\Delta\mathbf{x}_0|=\max \left\{|\Omega_0-\bar{\Omega}|, |{\Sigma_0}-  \bar{\Sigma}|, |\Omega_{k0}-\bar{\Omega}_k| \right\}$ and the sup norm of a matrix
${|} \mathbf{A} {|}$ defined by $\max\{|a_{ij}|:  i=1,2,3, j=1,2,3\}$, where $a_{ij}$ are given in \eqref{A-compts} 
we have 
\begin{equation*}
    \Big{|} \mathbf{A}(s) \cdot \Delta\mathbf{x}_0(s) \Big{|}\leq 3 \Big{|} \mathbf{A}(s) \Big{|} \Big{|}\Delta\mathbf{x}_0(s)\Big{|}, \quad \forall s\in [t_n, t_{n+1}].
\end{equation*}
By continuity of polynomials $a_{i j} \left(\Omega_{0}, \Sigma_{0}, \Omega_{k0}, \Phi_0, \bar{\Omega},  \bar{\Sigma}, \bar{\Omega}_k, \bar{\Phi}\right)$ given in \eqref{A-compts} and by continuity of  functions $\Omega_{0}, \Sigma_{0}, \Omega_{k0}, \Phi_0$ and  $\bar{\Omega},  \bar{\Sigma}, \bar{\Omega}_k, \bar{\Phi}$  in $[t_n, t_{n+1}]$ the following finite constants are found:
\begin{equation*}
    L_1= 3 \max_{t\in[t_n,t_{n+1}]} \Big{|} \mathbf{A} (t)\Big{|},
\end{equation*}
\begin{align*}
& M_1= \max_{t\in[t_{n},t_{n+1}]}  \Bigg\{  \Bigg{|} \frac{\Omega_0^3 \left(2 \mu
   ^2-\omega ^2\right)^3}{4 b^2 \mu ^6 \omega ^3}+\frac{\Omega_0 \left(9 b^2 \mu ^6 \omega ^2 \left(-4 \Omega_0^4+3 \Omega_0^2+1\right)-2 \Omega_0^2 \left(2 \mu ^2-\omega ^2\right)^3\right)
   \cos (2 (\Phi_0-t \omega ))}{8 b^2 \mu ^6 \omega ^3} \nonumber\\
   &+\frac{3 \left(\Omega_0^2-1\right) \Omega_0 \left(3 \gamma  \left(\Sigma_0^2+\Omega_0^2+\Omega_{k0}-1\right)-6
   \Sigma_0^2-3 \Omega_0^2-2 \Omega_{k0}\right)}{8 \omega }\Bigg{|},\nonumber \\
   & \Bigg{|} \frac{3 \Omega_0^2 \left(3 \gamma  \Sigma_0 \left(\Sigma_0^2+ \Omega_{0}^2+\Omega_{k0}-1\right)-3 \Sigma_0 \left(2 \Sigma_0^2+\Omega_0^2\right)-2 (\Sigma_0+1) \Omega_{k0}\right)}{8 \omega }-\frac{9 \Sigma_0 \Omega_0^4 \cos (2
   (\Phi_0-t \omega ))}{2 \omega }\Bigg{|},\nonumber \\
   & \Bigg{|} \frac{3 \Omega_0^2 \Omega_{k0} \left(3 \gamma  \left(\Sigma_0^2+\Omega_0^2+\Omega_{k0}-1\right)-6 \Sigma_0^2+2 \Sigma_0-3
   \Omega_0^2-2 \Omega_{k0}\right)}{4 \omega }-\frac{45 \Omega_0^4 \Omega_{k0} \cos (2 (\Phi_0-t \omega ))}{4 \omega }\Bigg{|} \Bigg\},
\end{align*}
and
\begin{align*}
M_2= & \max_{t\in[t_{n},t_{n+1}]}\Bigg{|} \frac{3 \bar{\Omega }^2 \left(2 \mu ^2-\omega
   ^2\right)^3}{8 b^2 \mu ^6 \omega ^3}-\frac{\Omega_0^2 \left(2 \mu ^2-\omega ^2\right)^3 \sin ^4(\Phi_0-t \omega )}{b^2 \mu ^6 \omega ^3} \nonumber \\
   & +\frac{3  \cos (2 (\Phi_0-t \omega ))}{8 \omega } \left(-3 \gamma  \left(\Sigma_0^2+\Omega_{0}^2+\Omega_{k0}-1\right)+6 \Sigma_0^2+3 \Omega_0^2+2 \Omega_{k0} + 3\left(\Omega_0^2+2\right) \cos (2 (\Phi_0-t
   \omega ))\right) \Bigg{|}
\end{align*}
such that for all $t\in[t_n, t_{n+1}]$: 
\begin{align*}
 & \Big{|}\Delta\mathbf{x}_0(t) \Big{|} = \Bigg{|} \int_{t_n}^t \dot{\Delta\mathbf{x}_0} d s \Bigg{|} = \Bigg{|} \int_{t_n}^t \Bigg(H \left(\bar{\mathbf{f}}( {\mathbf{x}}_0)-\bar{\mathbf{f}}(\bar{\mathbf{x}})\right) +   \mathcal{O}(H^2)\Bigg) ds  \Bigg{|}\nonumber \\
 & \leq H_n \int_{t_n}^t \Big{|}  \bar{\mathbf{f}}( {\mathbf{x}}_0)-\bar{\mathbf{f}}(\bar{\mathbf{x}}) \Big{|} ds +   M_1 H_n^2 (t-t_n)
  \leq H_n \int_{t_n}^t  \Big{|} \mathbf{A}(s) \cdot \Delta\mathbf{x}_0(s) \Big{|} ds  +   M_1 H_n^2 (t-t_n)\nonumber \\
& \leq L_1 H_n \int_{t_n}^t  \Big{|}
 \Delta\mathbf{x}_0(s) \Big{|} ds +   M_1 H_n^2 (t-t_n) \leq  L_1 H_n \int_{t_n}^t  \Big{|}  \Delta\mathbf{x}_0(s) \Big{|} ds +   M_1 H_n, \; \text{due to}\; t-t_n\leq {t_{n+1}}- {t_{n}} =\frac{1}{H_n}.
\end{align*}
Using   Gronwall's Lemma \ref{Gronwall}, we have for $t \in[t_n, t_{n+1}]$: 
\begin{align*}
 & \Big{|} \Delta \mathbf{x}_0(t)  \Big{|} \leq   M_1  H_n   e^{L_1  H_n(t-t_n)} \leq    M_1  {H_n}e^{L_1}, \; \text{due to}\; t-t_n\leq {t_{n+1}}- {t_{n}} =\frac{1}{H_n}.
 \end{align*}
 Then, 
 \begin{align*}
& \Big{|} \Delta \Omega_0(t) \Big{|} \leq    M_1 e^{L_1} {H_n}, \;  \Big{|} \Delta \Sigma_0(t) \Big{|} \leq     M_1 e^{L_1} {H_n}, \; \Big{|} \Delta \Omega_{k0}(t) \Big{|} \leq   M_1 e^{L_1} {H_n}.
\end{align*}
Furthermore, from eq. \eqref{A11} we have
\begin{align*}
& |\Delta \Phi_0(t)|= |\Phi_0(t)- \bar{\Phi}(t)| = \Bigg{|} \int_{t_n}^{t} \left( \dot{\Phi_0}(s)- \dot{\bar{\Phi}}(s)\right) d s\Bigg{|}  \nonumber \\
&= \Bigg{|} \bigints_{t_n}^{t}  H^2 \Bigg\{ \frac{3 \bar{\Omega }^2 \left(2 \mu ^2-\omega
   ^2\right)^3}{8 b^2 \mu ^6 \omega ^3}-\frac{\Omega_0^2 \left(2 \mu ^2-\omega ^2\right)^3 \sin ^4(\Phi_0-s \omega )}{b^2 \mu ^6 \omega ^3} \nonumber \\
   & +\frac{3  \cos (2 (\Phi_0-s \omega ))}{8 \omega } \left(3 \gamma  \left(1-\Sigma_0^2-\Omega_{0}^2-\Omega_{k0}\right)+6 \Sigma_0^2+3 \Omega_0^2+2 \Omega_{k0} + 3\left(\Omega_0^2+2\right) \cos (2 (\Phi_0-s
   \omega ))\right) \Bigg\} d s\Bigg{|} \nonumber \\
  &  \leq  M_2 H_n^2 (t-t_n) + \Big{|}\mathcal{O}({ {H_n}}^3)\Big{|}   \leq   M_2 H_n, \; \text{due to}\; t-t_n\leq {t_{n+1}}- {t_{n}} =\frac{1}{H_n}.
   \end{align*}
\end{widetext}
Finally, taking   limit as $n\rightarrow \infty$, we obtain $H_n\rightarrow  0$. Then,
as $H_n \rightarrow 0$,   functions $\Omega_{0}, \Sigma_{0}, \Omega_{k0}, \Phi_0$ and  $\bar{\Omega},  \bar{\Sigma}, \bar{\Omega}_k, \bar{\Phi}$  have the same limit as $\tau\rightarrow \infty$.
\newline 
Setting $\Sigma=\Sigma_0=0$ are derived  analogous results for negatively curved FLRW model. 
$\square$

\section{Center Manifold Calculations}
\label{AppCenter}

  \subsection{Center manifold of $F_0$ for $\gamma=\frac{2}{3}$}
 \label{AppCenter2}
 Letting $\gamma =\frac{2}{3}$   and defining  new variables
   \begin{align}
    &   x=\bar{\Sigma}
   -\frac{\bar{\Omega}_{k}}{2}, \;    y=\bar{\Omega}, \;    z=\bar{\Omega}_k,
   \end{align}
we obtain new equations \begin{align}
   \label{XNewsystemforDa}
      & x'=A x+f_1(x,y,z),\\
        \label{XNewsystemforDb}
      & y'=B y+f_2(x,y,z),\\
        \label{XNewsystemforDc}
      & z'=C z +f_3(x,y,z),
   \end{align}
      with 
   \begin{align}
   & A= - 2, \;  B= -\frac{1}{2}, \;  C= 0,
\\
       & f_1(x,y,z)=2 x^3+x^2 z+\frac{x
   y^2}{2}-\frac{x z^2}{2} \nonumber\\
   & +x
   z-\frac{y^2
   z}{4}-\frac{z^3}{4}+\frac{z
   ^2}{2}, 
\\
       & f_2(x,y,z)= 2 x^2 y+2 x y
   z+\frac{y^3}{2}+\frac{y
   z^2}{2}, 
\\
       & f_3(x,y,z)=4 x^2 z+4 x z^2-2 x z+y^2
   z+z^3-z^2.
   \end{align}
  Note that if $J_1$ is the linearization matrix of system \eqref{XNewsystemforDa},  \eqref{XNewsystemforDb}, \eqref{XNewsystemforDc} then the eigensystem of  $J_1(0,0,0)$ is $$\left(
\begin{array}{ccc}
 -2 & -\frac{1}{2} & 0 \\
 \{1,0,0\} & \{0,1,0\} &
   \{0,0,1\} \\
\end{array}
\right).$$
          \begin{figure}[t]
    \centering
    \includegraphics[width=0.4\textwidth]{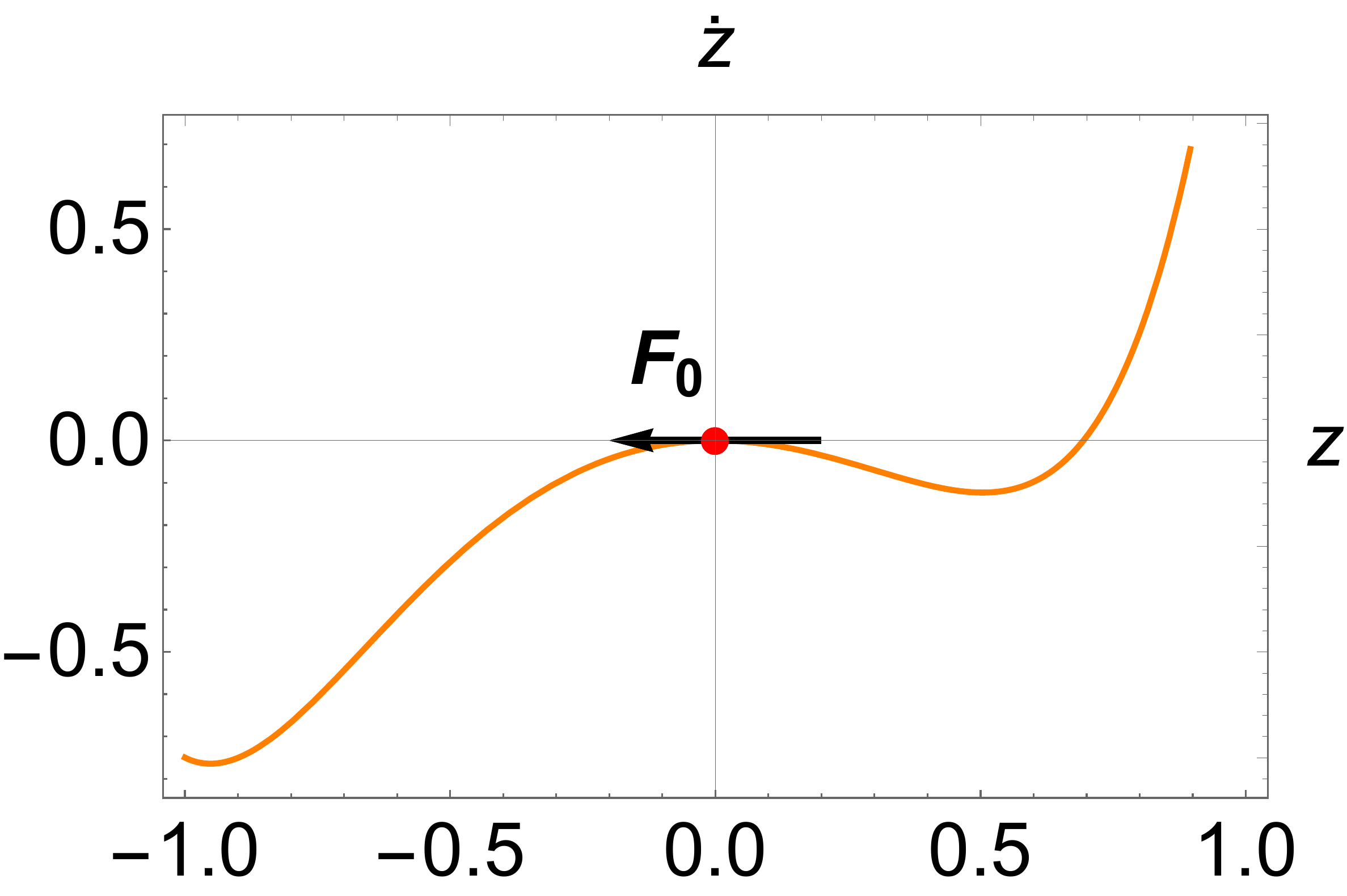}
    \caption{One dimensional flow for \eqref{XcentermanifoldF0} for $z\in [-1,1]$ the origin is stable if $z\geq 0$. Note that $z\geq 0$ corresponds to $\bar{\Omega}_k\geq 0$.}
    \label{Xfig:my_label0x0x0}
\end{figure} 
This implies that the local center manifold of the origin for \eqref{XNewsystemforDa},  \eqref{XNewsystemforDb}, \eqref{XNewsystemforDc} is given by the graph 
\begin{align}
\label{XgraphmanifoldD}
 W^{c}_{loc}(\mathbf{0})=  &  \Bigg\{(x,y,z)\in \mathbb{R}^3: x=h_1(z), y= h_2(z), \nonumber \\
 & h_1(0)=h_2(0)=0, \nonumber \\
 & h_1'(0)=h_2'(0)=0, |z|<\delta\Bigg\} 
\end{align}
for some $\delta>0$. 

\noindent
Therefore, we can use Taylor series to define 
\begin{align}
\label{Xexpansion1}
   & h_1(z)=a_1 z^2+a_2
   z^3+a_3 z^4+a_4
   z^5+a_5
   z^6+\mathcal{O}\left(z^7\right),\\
   \label{Xexpansion2}
  & h_2(z)=b_1 z^2+b_2
   z^3+b_3 z^4+b_4
   z^5+b_5
   z^6+\mathcal{O}\left(z^7\right).
\end{align} 
The following quasi-linear differential equations 
\begin{align}
    \mathcal{N}(h_1(z))\equiv h_1'(z)(C z+f_3(h_1(z),h_2(z),z))& \nonumber \\-A h_1(z)-f_1(h_1(z),h_2(z),z),& \\ \mathcal{N}(h_2(z))\equiv h_2'(z)(C z+f_3(h_1(z),h_2(z),z)) \nonumber &\\-B h_2(z)-f_2(h_1(z),h_2(z),z),
\end{align}
must be solved for  $a_i, b_i  , i=1, \ldots 5$ up to order six to approximate the center manifold. We notice that  $b_i, i=1, \ldots 5$ are zero and 
$ a_1= \frac{1}{4}, \; a_2= \frac{1}{4}, \; a_3=\frac{5}{16}, \;
    a_4=\frac{7}{16}$.
Hence,  we obtain
   \begin{align}
& h_1(z)=\frac{z^2}{4}+\frac{z^3}{4}+\frac{5
   z^4}{16}+\frac{7
   z^5}{16}+\frac{21 z^6}{32}+ \mathcal{O}(z^7), \\
&   h_2(z)= \mathcal{O}(z^7),
   \end{align}
and the dynamics on the center manifold is given by:
   \begin{equation}
   \label{XcentermanifoldF0}
      z'=-z^2+\frac{z^3}{2}+\frac{z^4}{
   2}+\frac{5 z^5}{8}+\frac{7
   z^6}{8}+\mathcal{O}\left(z^7\right).
   \end{equation}
   Then, $F_0$ is locally asymptotically stable for  $z\geq 0$. Note that $z\geq 0$ corresponds to $\bar{\Omega}_k\geq 0$ given that $z=\bar{\Omega}_k.$

\noindent   
In Figure \ref{Xfig:my_label0x0x0} a one dimensional flow for \eqref{XcentermanifoldF0} for $z\in [-1,1]$ is  represented. The origin is locally asymptotically stable if $z\geq 0$.

 \subsection{Center manifold of $D$ for $\gamma\geq 1$}
 \label{AppCenter1}
 Letting $\gamma\notin \left\{\frac{4}{3}, \frac{3}{2}\right\}$  and defining new variables 
   \begin{align}
    &   x=\frac{\gamma 
   (12 \bar{\Sigma} -4 \bar{\Omega}_k
   -3)-16 \bar{\Sigma} +8
   \bar{\Omega}_k+2}{24-16
   \gamma },\\
    &   y=\frac{(3
   \gamma -4) (4 \bar{\Sigma} +4
   \bar{\Omega}_k-5)}{8 (2
   \gamma -3)},\\
   &    z=\bar{\Omega},
   \end{align}
we obtain  new equations \begin{align}
   \label{NewsystemforDa}
      & x'=A x+f_1(x,y,z,\gamma),\\
        \label{NewsystemforDb}
      & y'=B y+f_2(x,y,z,\gamma),\\
        \label{NewsystemforDc}
      & z'=C z +f_3(x,y,z,\gamma),
   \end{align}
   with 
   \begin{align}
   & A= - \frac{3}{2}, \;  B= 3(1-\gamma), \;  C= 0,
\\
       & f_1(x,y,z,\gamma)=\frac{3 (5 \gamma -8) (\gamma -2) x^3}{12-8 \gamma } \nonumber \\
       & +x^2
   \left(\frac{18-13 \gamma }{8 \gamma -12}+\frac{3 (7 \gamma -12)
   (\gamma -2)^2 y}{4 (2 \gamma -3) (3 \gamma -4)}\right) \nonumber \\
   & +x
   \Bigg(\frac{3 \gamma  (\gamma -2)^3 y^2}{4 (4-3 \gamma )^2 (2
   \gamma -3)} \nonumber \\
   & +\frac{1}{4} \left(-10 \gamma +\frac{3}{3-2 \gamma
   }+\frac{8}{3 \gamma -4}+11\right) y\Bigg) \nonumber \\
   & +z^2 \Bigg(\frac{3
   (\gamma -1)}{8 \gamma -12}+\frac{3 (5 \gamma -8) (\gamma -1)
   x}{12-8 \gamma } \nonumber \\
   & +\frac{3 (\gamma -2) (\gamma -1) y}{12-8 \gamma
   }\Bigg)  -\frac{3 (\gamma -2)^4 y^3}{4 (4-3 \gamma )^2 (2 \gamma
   -3)} \nonumber \\
   & -\frac{(\gamma  (3 \gamma  (12 \gamma -43)+152)-60) (\gamma
   -2) y^2}{4 (4-3 \gamma )^2 (2 \gamma -3)}, 
\end{align}
\begin{align}
       & f_2(x,y,z,\gamma)= \frac{3 (\gamma -2)
   (3 \gamma -4) x^3}{12-8 \gamma } \nonumber \\
   & +x^2 \Bigg(\frac{3}{8} \left(-6
   \gamma +\frac{1}{2 \gamma -3}+11\right)  +\frac{3 (\gamma -2) (5
   \gamma -6) y}{12-8 \gamma }\Bigg) \nonumber \\
   & +x \left(\frac{3 (13 \gamma
   -18) (\gamma -2)^2 y^2}{4 (2 \gamma -3) (3 \gamma
   -4)}+\frac{(30-19 \gamma ) y}{4 \gamma -6}\right) \nonumber \\
   & +z^2
   \Bigg(\frac{3 (\gamma -1) (3 \gamma -4)}{6-4 \gamma }+\frac{3
   (\gamma -1) (3 \gamma -4) x}{12-8 \gamma } \nonumber \\
   & +\frac{3 (\gamma -1)
   (7 \gamma -10) y}{12-8 \gamma }\Bigg) -\frac{3 (\gamma -2)^3 (7
   \gamma -10) y^3}{4 (4-3 \gamma )^2 (2 \gamma -3)} \nonumber \\
   & +\frac{1}{8}
   \left(-30 \gamma +\frac{16}{4-3 \gamma }+\frac{3}{2 \gamma
   -3}+33\right) y^2-6 \gamma  y, 
\end{align}
\begin{align}
   & f_3(x,y,z,\gamma)=z \Bigg(\left(3-\frac{3 \gamma
   }{2}\right) x^2 \nonumber \\
   &+x \left(\frac{3 (\gamma -2)^2 y}{3 \gamma
   -4}-2\right) -\frac{3 (\gamma -2)^3 y^2}{2 (4-3 \gamma
   )^2} \nonumber \\
   & +\left(-2 \gamma +\frac{2}{12-9 \gamma }+\frac{7}{3}\right)
   y\Bigg)  -\frac{3}{2} (\gamma -1) z^3.
   \end{align}
   Note that if $J_1$ is the linearization matrix of system \eqref{NewsystemforDa},  \eqref{NewsystemforDb}, \eqref{NewsystemforDc} then the eigensystem of  $J_1(0,0,0)$ is $$\left(
\begin{array}{ccc}
 -\frac{3}{2} & 3-3 \gamma & 0 
    \\
 (1,0,0) & (0,1,0) & (0,0,1) \\
\end{array}
\right).$$
This implies that the local center manifold of the origin for \eqref{NewsystemforDa},  \eqref{NewsystemforDb}, \eqref{NewsystemforDc} is given by the graph 
\begin{small}
\begin{align}
\label{graphmanifoldD}
 W^{c}_{loc}(\mathbf{0})=  &  \Bigg\{(x,y,z)\in \mathbb{R}^3: x=h_1(z), y= h_2(z), \nonumber \\
 & h_1(0)=h_2(0)=0, \nonumber \\
 & h_1'(0)=h_2'(0)=0, |z|<\delta\Bigg\} 
\end{align}
\end{small}
for some $\delta>0$. 

\begin{figure}[t!]
    \centering
    \includegraphics[width=0.4\textwidth]{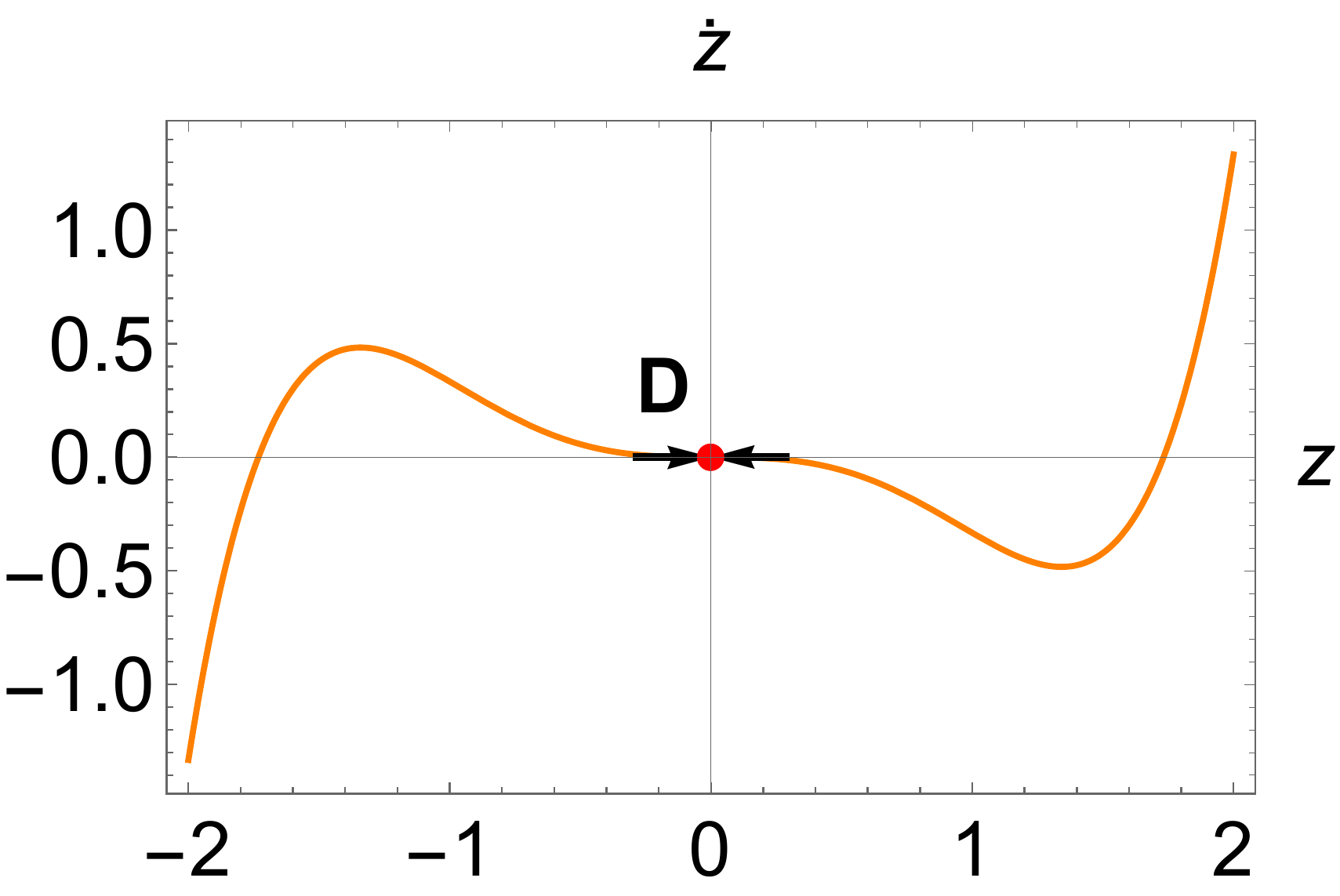}
    \caption{One dimensional flow for \eqref{centermanifoldD} for $z\in [-2,2]$. The origin is stable.}
    \label{fig:my_label0x0x0}
\end{figure}

\noindent
Therefore, we can use Taylor series to define 
\begin{align}
\label{expansion1}
   & h_1(z)=a_1 z^2+a_2
   z^3+a_3 z^4+a_4
   z^5+a_5 z^6+O(z^7),\\
   \label{expansion2}
  & h_2(z)=b_1 z^2+b_2
   z^3+b_3 b^4+b_4
   z^5+b_5 z^6+O(z^7).
\end{align} 
The following quasilinear differential equations 
\begin{align}
    \mathcal{N}(h_1(z))\equiv h_1'(z)(C z+f_3(h_1(z),h_2(z),z,\gamma))& \nonumber \\-A h_1(z)-f_1(h_1(z),h_2(z),z,\gamma),& \end{align} 
    \begin{align}
    \mathcal{N}(h_2(z))\equiv h_2'(z)(C z+f_3(h_1(z),h_2(z),z,\gamma)) \nonumber &\\-B h_2(z)-f_2(h_1(z),h_2(z),z,\gamma),
\end{align}
must be solved for   $a_i,b_i$ up to order six to approximate the center manifold. We notice that the even terms are zero and $a_1= \frac{\gamma -1}{4 \gamma
   -6}, \; a_3= \frac{5
   \gamma -6}{48 \gamma
   -72}, \; a_5=\frac{5
   \gamma -6}{72 \gamma
   -108},    b_1=\frac{4-3 \gamma }{2 (2
   \gamma -3)}, \; b_3= \frac{4-3
   \gamma }{8 (2 \gamma
   -3)}, \; b_5=\frac{4-3
   \gamma }{12 (2 \gamma
   -3)}$. 
Hence,  we obtain
   \begin{align}
& h_1(z)=\frac{(5 \gamma -6) z^6}{72
   \gamma -108}+\frac{(5
   \gamma -6) z^4}{48 \gamma
   -72}+\frac{(\gamma -1)
   z^2}{4 \gamma -6}, \\
&   h_2(z)=\frac{(4-3 \gamma ) z^6}{12 (2
   \gamma -3)}+\frac{(4-3
   \gamma ) z^4}{8 (2 \gamma
   -3)}+\frac{(4-3 \gamma )
   z^2}{2 (2 \gamma -3)}.
   \end{align} 
Letting $\gamma=\frac{4}{3}$ and defining new variables 
   \begin{align}
    &   x=\bar{\Omega}_k-\frac{3}{4}, \;    y=\bar{\Sigma}
   +\bar{\Omega
   _k}-\frac{5}{4}, \;    z=\bar{\Omega},
   \end{align}
we obtain new equations
 \begin{align}
      & x'=A x+f_1(x,y,z),      \label{NewsystemforDrada}\\
      & y'=B y+f_2(x,y,z),   \label{NewsystemforDradb}\\
      &  z'=C z +f_3(x,y,z), \label{NewsystemforDradc}
   \end{align}
   where \begin{align}
   & A= - \frac{3}{2}, \;  B= -1, \;  C= 0,
\\
    &    f_1(x,y,z)=2 x^3-4 x^2 y-\frac{x^2}{2}+2 x y^2-3
   x y \nonumber\\ &-x z^2+\frac{3 y^2}{2}-\frac{3
   z^2}{4},
\\
   &     f_2(x,y,z)=x^3-x^2 y+2 x^2-x y^2-7 x y \nonumber \\ &-\frac{x
   z^2}{2}+y^3+3 y^2-\frac{y
   z^2}{2}-z^2 ,
\\
  &          f_3(x,y,z)= x^2 z-2 x y z-2 x z+y^2 z+y
   z-\frac{z^3}{2}.
        \end{align}
        Note that if $J_1$ is the linearization matrix of system \eqref{NewsystemforDrada},  \eqref{NewsystemforDradb}, \eqref{NewsystemforDradc} then the eigensystem of  $J_1(0,0,0)$ is $$\left(
\begin{array}{ccc}
 -\frac{3}{2} & -1 & 0 \\
 \{1,0,0\} & \{0,1,0\} & \{0,0,1\} \\
\end{array}
\right).$$
Then, using Taylor series, the local center manifold of the origin for \eqref{NewsystemforDrada}, \eqref{NewsystemforDradb}, \eqref{NewsystemforDradc} is given by the graph \eqref{graphmanifoldD} where \begin{align}
    & h_1(z)= -\frac{z^6}{18}-\frac{z^4}{12}-\frac{z^2}{2}, \\
    & h_2(z)=-\frac{z^6}{6}-\frac{z^4}{4}-z^2. 
\end{align}
Letting $\gamma=\frac{3}{2}$  and defining  new variables 
   \begin{align}
    &   x=\bar{\Omega} _k-\frac{3}{4},\\
    &   y=-\frac{3}{32} (4 \bar{\Sigma} +4 \bar{\Omega}_k-5),\\
   &    z=\bar{\Omega},
   \end{align}
we obtain  new equations
 \begin{align}
   \label{NewsystemforDwa}
     &  x'=A_1 x + A_2 y+f_1(x,y,z),\\
        \label{NewsystemforDwb}
     &  y'=B y+f_2(x,y,z),\\
        \label{NewsystemforDwc}
     &  z'=C z +f_3(x,y,z),
   \end{align}
   where \begin{align}
   & A_1= - \frac{3}{2}, \; A_2=1, \; B= - \frac{3}{2}, \;  C= 0,
\\
   &     f_1(x,y,z)=\frac{3 x^3}{2}+8 x^2 y-\frac{7 x^2}{8}+\frac{32 x y^2}{3}\nonumber \\ &+\frac{22 x y}{3}-\frac{3 x
   z^2}{2}+8 y^2-\frac{9 z^2}{8},
\end{align}
\begin{align}
       & f_2(x,y,z)=-\frac{9 x^3}{32}-\frac{3 x^2 y}{4}-\frac{9 x^2}{16}+2 x y^2-\frac{25 x y}{4}\nonumber \\ &+\frac{9
   x z^2}{32}+\frac{16 y^3}{3}-6 y^2-\frac{3 y z^2}{4}+\frac{9 z^2}{16} ,
\\
 & f_3(x,y,z)=\frac{3 x^2 z}{4}+4 x y z-2 x z+\frac{16 y^2 z}{3}-2 y z-\frac{3 z^3}{4}.
        \end{align}
        Note that if $J_1$ is the linearization matrix of system \eqref{NewsystemforDwa},  \eqref{NewsystemforDwb}, \eqref{NewsystemforDwc} then the eigensystem of  $J_1(0,0,0)$ is $$\left(
\begin{array}{ccc}
 -\frac{3}{2} & -\frac{3}{2} & 0 \\
 \{1,0,0\} & \{0,0,0\} & \{0,0,1\} \\
\end{array}
\right).$$
Then, using Taylor series, the local center manifold of the origin is\eqref{NewsystemforDwa}, \eqref{NewsystemforDwb}, \eqref{NewsystemforDwc} is given by the graph \eqref{graphmanifoldD} where 
\begin{align}
    & h_1(z)= -\frac{z^6}{18}-\frac{z^4}{12}-\frac{z^2}{2}, \\
    & h_2(z)=\frac{z^6}{16}+\frac{3 z^4}{32}+\frac{3 z^2}{8}. 
\end{align}
Finally, in these three cases (which corresponds to $\gamma>1$), the dynamics on the center manifold is given by the following equation 
   \begin{equation}
   \label{centermanifoldD}
      z'=-\frac{z^3}{2}+\frac{z^5}{6}+O(z^7).
   \end{equation}
This is a gradient-like equation
   \begin{equation}
       z'=-\nabla U(z), \quad U(z)= \frac{z^4}{8} - \frac{z^6}{36},  
   \end{equation}
   for which the origin is a degenerate minimum of second order, i.e., $U'(0)=U''(0)=U'''(0)=0, U^{(iv)}(0)=3>0$. Therefore, the origin is locally asymptotically stable as shown in Figure \ref{fig:my_label0x0x0}.
   
\noindent   
Letting $\gamma=1$, we deduce that the center manifold of $D$ is 2-dimensional. On the other hand, the system admits the lines of equilibrium points  $(\bar{\Omega}, \bar{\Sigma}, \bar{\Omega}_k)= (\bar{\Omega}^*, 0,  0)$ and $(\bar{\Omega}, \bar{\Sigma}, \bar{\Omega}_k)= (\bar{\Omega}^*, \frac{1}{2}, \frac{3}{4})$ where $\bar{\Omega}^*$  is an arbitrary number which satisfies $\bar{\Omega}^*\in[0,1]$. 
Therefore, $D$ as well as $F_0$ are not isolated fixed points anymore. 

\noindent    
To analyze the stability of $D: (\bar{\Omega}, \bar{\Sigma}, \bar{\Omega}_k)=(0, \frac{1}{2}, \frac{3}{4})$ we define  new variables
   \begin{align}
    &   x=\frac{1}{8} (4 \bar{\Sigma} -4 \bar{\Omega}_{k}+1), \;   y=\frac{1}{8} (4 \bar{\Sigma} +4 \bar{\Omega}_{k}-5), \nonumber \\
    & z= \bar{\Omega},
   \end{align}
to obtain new equations
 \begin{align}
   \label{NewsystemforDa1a}
     &  x'=A x +f_1(x,y,z),\\
        \label{NewsystemforDa1b}
     &  y'=C_1 y +f_2(x,y,z),\\
        \label{NewsystemforDa1c}
     &  z'=C_2 z +f_3(x,y,z),
   \end{align} 
     where 
     \begingroup\makeatletter\def\f@size{9}\check@mathfonts
     \begin{align}
   & A= - \frac{3}{2}, \;  C_1= 0, \;  C_2= 0,
\\
   &     f_1(x,y,z)=\frac{9 x^3}{4}+x^2 \left(\frac{15
   y}{4}+\frac{5}{4}\right) \nonumber \\
   & +x \left(\frac{3
   y^2}{4}-y\right)-\frac{3 y^3}{4}-\frac{y^2}{4},
\\
       & f_2(x,y,z)=-\frac{3
   x^3}{4}+x^2 \left(\frac{3 y}{4}+\frac{3}{2}\right) \nonumber \\
   & +x
   \left(\frac{15 y^2}{4}+\frac{11 y}{2}\right)+\frac{9 y^3}{4}+2
   y^2,
\\
            & f_3(x,y,z)=z \left(\frac{3 x^2}{2}+x (3 y+2)+\frac{3
   y^2}{2}+y\right).
        \end{align}
        \endgroup
The local center manifold of the origin for \eqref{NewsystemforDa1a},  \eqref{NewsystemforDa1b}, \eqref{NewsystemforDa1c} is given by the graph 
\begin{small}
\begin{align}
\label{graphmanifoldaD*}
 W^{c}_{loc}(\mathbf{0})=  &  \Bigg\{(x,y,z)\in \mathbb{R}^3: x=h(y, z), \; h(0,0)=0, \nonumber \\
 & \frac{\partial h}{\partial y}(0,0)=\frac{\partial h}{\partial z}(0,0)=0, |(y,z)^T|<\delta\Bigg\},
\end{align}
\end{small}
for some $\delta>0$. 

\begin{figure}[t!]
    \centering 
    \includegraphics[width=0.4\textwidth]{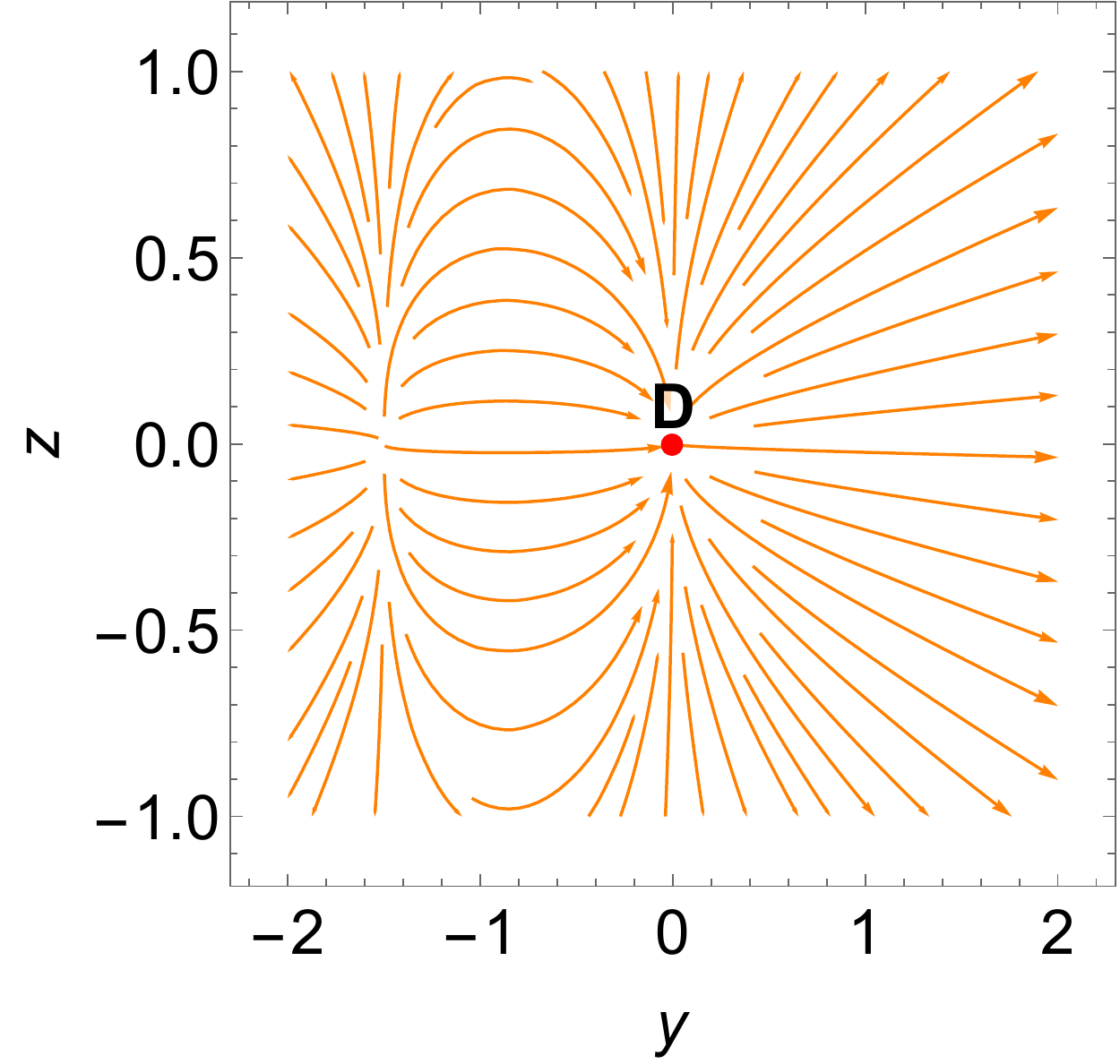}
    \caption{Two dimensional flow for \eqref{A225}, \eqref{A226}. The physical region of the phase space is $y:=\frac{1}{8} (4 \bar{\Sigma} +4 \bar{\Omega}_{k}-5)<0 $, $z:= \bar{\Omega}>0$.}
    \label{fig:my_labelA225-226}
\end{figure}  
\noindent 
$h(y,z)$ satisfies the quasilinear partial differential eq. 
\begin{small}
\begin{align}
\label{quasilinear17}
    & \Bigg(\frac{3}{4} h^3-\frac{3}{4} (y+2) h^2-\frac{1}{4}
   y (15 y+22) h \nonumber \\
   & -\frac{1}{4} y^2 (9 y+8)\Bigg)
  \frac{\partial h}{\partial y} \nonumber \\
   & +\left(-\frac{3}{2} z h^2-(3 y+2) z
   h-\frac{1}{2} y (3 y+2) z\right)
   \frac{\partial h}{\partial z} \nonumber \\
   & +\frac{9}{4} h^3+\frac{5}{4} (3 y+1)
   h^2+\frac{1}{4} (y (3 y-4)-6) h \nonumber \\
   & -\frac{1}{4} y^2 (3
   y+1)=0.
\end{align}  
\end{small}
\noindent
 We propose the Taylor expansion 
\begin{align}
&  h(y,z)=   {a_1} y^2+{a_2} y z+{a_3} z^2 \nonumber \\
& +{b_1} y^3+{b_2}
   y^2 z+{b_3} y z^2+{b_4} z^3 + \mathcal{O}(4),
\end{align}
where $\mathcal{O}(4)$ denotes terms of fourth order in the vector norm. 
Therefore, equation \eqref{quasilinear17} can be expressed as
\begin{small}
\begin{align}
 & \frac{1}{4} (-6 a_1-1) y^2+y^2 z \left(-4
   a_2-\frac{3 b_2}{2}\right) \nonumber \\
   & -\frac{3 a_2 y
   z}{2}-\frac{3}{2} y z^2 (2 a_3+b_3)   -\frac{3
   a_3 z^2}{2}-\frac{3 b_4 z^3}{2}  = \mathcal{O}(4).
\end{align}
\end{small}
Equating the terms of the same power in $y,z$ we have a solution  $a_1= -\frac{1}{6}, \; a_2= 0, \; a_3=
   0,   b_1\; \text{arbitrary}, \; b_2= 0, \; b_3=0, \; b_4=0$.
Then, we obtain 
\begin{equation}
    h(y, z)= -\frac{y^2}{6}+ b_1 y^3 + \mathcal{O}(4). 
\end{equation}
The dynamics at the center manifold is given by 
\begin{align}
   & y'=2 y^2 +\frac{4 y^3}{3}, \label{A225}\\
   & z'=y z + \frac{7 y^2 z}{6}. \label{A226}
\end{align}
\noindent 
In Figure  \ref{fig:my_labelA225-226}  a two dimensional flow for \eqref{A225} and \eqref{A226}  where   is shown that the origin is unstable (saddle type) for $y\neq 0$ is presented.  However, if we restrict the analysis to $y<0$, $D$ is asymptotically stable and behaves as a local attractor.  

\noindent
To analyze the stability of an arbitrary point $D^*: (\bar{\Omega}, \bar{\Sigma}, \bar{\Omega}_k)=(\bar{\Omega}^*, \frac{1}{2}, \frac{3}{4})$ with $\bar{\Omega}^*\neq 0$ we define new variables 
   \begin{align}
    &   x=\frac{1}{8} (4 \bar{\Sigma} -4 \bar{\Omega}_k+1)\\
    &   y=\frac{1}{6} (\bar{\Omega}^* (-4 \bar{\Sigma} +4
   \bar{\Omega}_k-7)+6 \bar{\Omega})\\
   &    z=\frac{1}{8} \bar{\Omega}^*
   (4 \bar{\Sigma} +4 \bar{\Omega}_k-5),
   \end{align}
   \begin{figure}[t!]
    \centering 
    \includegraphics[width=0.4\textwidth]{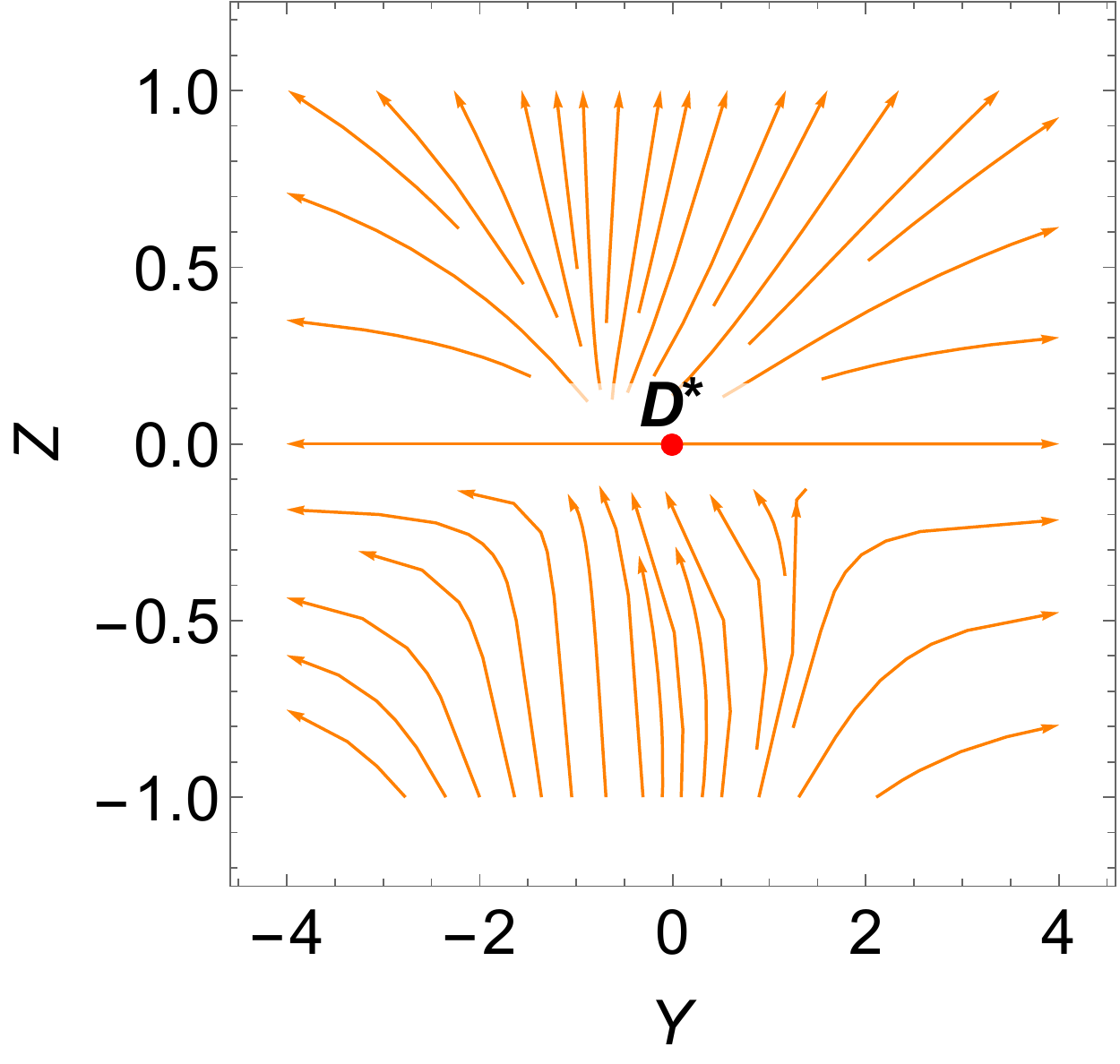}
    \caption{Two dimensional flow for \eqref{225}, \eqref{226}. The origin is unstable.}
    \label{fig:my_label225-226}
\end{figure}  
\noindent 
to obtain  new equations
 \begin{align}
   \label{NewsystemforD1a}
     &  x'=A x +f_1(x,y,z),\\
        \label{NewsystemforD1b}
     &  y'=C_1 y + B_1 x+ B_2 z+f_2(x,y,z),\\
        \label{NewsystemforD1c}
     &  z'=C_2 z +f_3(x,y,z),
   \end{align}
    where 
     \begingroup\makeatletter\def\f@size{9}\check@mathfonts
    \begin{align}
   & A= - \frac{3}{2}, \; B_1=4 \bar{\Omega}^*, \; B_2=1, \;  C_1= 0, \;  C_2= 0,   
\\
   &     f_1(x,y,z)=\frac{9 x^3}{4}+z \left(\frac{15 x^2}{4 {\bar{\Omega}^*}}-\frac{x}{{\bar{\Omega}^*}}\right)+\frac{5 x^2}{4}\nonumber \\
   & +z^2
   \left(\frac{3 x}{4 \bar{\Omega}^*{}^2}-\frac{1}{4
   \bar{\Omega}^*{}^2}\right)-\frac{3 z^3}{4 \bar{\Omega}^*{}^3},
\\
       & f_2(x,y,z)=-x^3 \bar{\Omega}^*+x^2 \left(\frac{3 y}{2}+\frac{5
   \bar{\Omega}^*}{2}\right) \nonumber \\ &  +z \left(-x^2+x \left(\frac{3
   y}{\bar{\Omega}^*}+\frac{17}{3}\right)+\frac{y}{\bar{\Omega}^*}\right)\nonumber \\
   & +z^2 \left(\frac{x}{\bar{\Omega}^*}+\frac{3 y}{2 \bar{\Omega}^*{}^2}+\frac{11}{6 \bar{\Omega}^*}\right)+2 x
   y+\frac{z^3}{\bar{\Omega}^*{}^2},
\\
            & f_3(x,y,z)=-\frac{3 x^3 \bar{\Omega}^*{}}{4}+\frac{3 x^2 \bar{\Omega}^*}{2} \nonumber \\
            & +\left(\frac{3
   x^2}{4}+\frac{11 x}{2}\right) z+z^2 \left(\frac{15 x}{4
   \bar{\Omega}^*}+\frac{2}{\bar{\Omega}^*}\right)+\frac{9
   z^3}{4 \bar{\Omega}^*{}^2}.
        \end{align}
        \endgroup
The local center manifold of the origin for \eqref{NewsystemforD1a},  \eqref{NewsystemforD1b}, \eqref{NewsystemforD1c} is given by the graph 
\begin{small}
\begin{align}
\label{graphmanifoldD*}
 W^{c}_{loc}(\mathbf{0})=  &  \Bigg\{(x,y,z)\in \mathbb{R}^3: x=h(y, z), h(0,0)=0, \nonumber \\
 & \frac{\partial h}{\partial y}(0,0)=\frac{\partial h}{\partial z}(0,0)=0, |(y,z)^T|<\delta\Bigg\},
\end{align}
\end{small}
\noindent 
for some $\delta>0$. 

\noindent
$h(y,z)$ satisfies the quasilinear partial differential equation 
\begingroup\makeatletter\def\f@size{9}\check@mathfonts
\begin{align}
   & \frac{\partial h}{\partial z} \Bigg(\frac{3}{4} \bar{\Omega}^*{}
   h^3-\frac{3}{4} \left(2 \bar{\Omega}^*{}+z\right)
   h^2 \nonumber \\
   & -\frac{z \left(22 \bar{\Omega}^*{}+15 z\right) h}{4
   \bar{\Omega}^*}-\frac{z^2 \left(8 \bar{\Omega}^*{}+9
   z\right)}{4 \bar{\Omega}^*{}^2}\Bigg) \nonumber \\
   & +\frac{\partial h}{\partial y}
   \Bigg(\bar{\Omega}^*h^3+\left(-\frac{5 \tilde{\Omega
   }}{2}-\frac{3 y}{2}+z\right) h^2 \nonumber \\
   & -\frac{\left(6 y \bar{\Omega}^*+12 \bar{\Omega}^*{}^2+17
   z \bar{\Omega}^*{}+9 y z+3 z^2\right) h}{3 \tilde{\Omega
   }} \nonumber \\
   & -\frac{z \left(6 y
   \bar{\Omega}^*+6 \bar{\Omega}^*{}^2+11 z \bar{\Omega}^*{}+9 y
   z+6 z^2\right)}{6 \bar{\Omega}^*{}^2}\Bigg) \nonumber \\
   & +\frac{5
   \left(\bar{\Omega}^*+3 z\right) h^2}{4 \bar{\Omega}^*}+\frac{1}{4} \left(\frac{z \left(3 z-4 \tilde{\Omega
   }\right)}{\bar{\Omega}^*{}^2}-6\right) h \nonumber \\
   & -\frac{z^2
   \left(\bar{\Omega}^*+3 z\right)}{4 \bar{\Omega}^*{}^3}+\frac{9}{4} h^3=0. \label{quasilinear}
\end{align}
\endgroup
\noindent 
We propose the Taylor expansion 
\begin{align}
&  h(y,z)=   {a_1} y^2+{a_2} y z+{a_3} z^2 \nonumber \\
& +{b_1} y^3+{b_2}
   y^2 z+{b_3} y z^2+{b_4} z^3 + \mathcal{O}(4),
\end{align}
where $\mathcal{O}(4)$ denotes terms of fourth order in the vector norm. 
Therefore, equation \eqref{quasilinear} can be expressed as 
\begin{small}
\begin{align}
   & y z^2 \left(-4 \bar{\Omega}^*{} \left(2 {a_1}
   {a_3}+{a_2}^2\right)-\frac{11 {a_1}+12 {a_2}}{3
   \bar{\Omega}^*{}}-2 {b_2}-\frac{3 {b_3}}{2}\right) \nonumber \\
   & +y^2 z \left(-12 {a_1} {a_2} \bar{\Omega}^*{}-\frac{3
   {a_1}}{\bar{\Omega}^*{}}-3 {b_1}-\frac{3
   {b_2}}{2}\right) \nonumber \\
   & +y z \left(-2 {a_1}-\frac{3
   {a_2}}{2}\right)-\frac{3 {a_1} y^2}{2} \nonumber \\
   & -\frac{z^3
   \left(2 \bar{\Omega}^*{}^2 \left({a_2} \left(24 {a_3}
   \bar{\Omega}^*{}^2+11\right)+30 {a_3}+6 {b_3}
   \bar{\Omega}^*{}+9 {b_4} \bar{\Omega}^*{}\right)+9\right)}{12 \bar{\Omega}^*{}^3} \nonumber \\
   & +z^2
   \left(-{a_2}-\frac{3 {a_3}}{2}-\frac{1}{4 \text{$\Omega
   $1}^2}\right)= \mathcal{O}(4).
\end{align}
\end{small}
Equating the terms of the same power in $y,z$ we have a solution $ a_1= 0, \; a_2= 0, \; a_3=
   -\frac{1}{6 \bar{\Omega}^*{}^2},   b_1= \frac{3
   b_3}{8}, \; b_2= -\frac{3 b_3}{4}, \; b_4=
   \frac{1}{18} \left(\frac{1}{\bar{\Omega}^*{}^3}-12
   b_3\right)$. For simplicity, we set 
$b_3=\frac{1}{12\bar{\Omega}^*{}^3}, b_4=0$. Then, we obtain 
\begin{equation}
    h(y, z)= \frac{y^3}{32 \bar{\Omega}^*{}^3}-\frac{y^2 z}{16 \bar{\Omega}^*{}^3}+\frac{y z^2}{12 \bar{\Omega}^*{}^3}-\frac{z^2}{6
   \bar{\Omega}^*{}^2} + \mathcal{O}(4). 
\end{equation}
The dynamics at the center manifold is given by 
\begin{align}
   & y'= z +\frac{y z}{\bar{\Omega}^*{}}+\frac{7 z^2}{6 \bar{\Omega}^*{}}+ \frac{y^3}{8 \bar{\Omega}^*{}^2}-\frac{y^2 z}{4
   \bar{\Omega}^*{}^2} \nonumber \\
   & +\frac{3 y z^2}{2 \bar{\Omega}^*{}^2} +\frac{z^3}{18
   \bar{\Omega}^*{}^2},\\
   & z'= \frac{4 z^3}{3 \bar{\Omega}^*{}^2}+\frac{2
   z^2}{\bar{\Omega}^*{}}.
\end{align}
 Using a re-scaling  $(T, Y, Z)=  ( \bar{\Omega}^* \tau, \frac{y}{\bar{\Omega}^*}, \frac{z}{\bar{\Omega}^*}), \; \bar{\Omega}^*>0$, 
we obtain a topologically equivalent system 
\begin{align}
\label{225}
   & \frac{d Y}{d T}= Z + Y Z +\frac{7 Z^2}{6} + \frac{Y^3}{8}-\frac{Y^2 Z}{4}+\frac{3 Y Z^2}{2} +\frac{Z^3}{18}, \\
   \label{226}
   & \frac{d Z}{d T}=2 Z^2 + \frac{4 Z^3}{3}.
\end{align}
In Figure  \ref{fig:my_label225-226}  a two dimensional flow for \eqref{225}, \eqref{226}, where it is shown that the origin is unstable (saddle type) for $Z\neq 0$ is presented.

\section{Numerical simulation}
\label{numerics}

 In this section we present numerical evidence that support the main Theorem of section \ref{SECT:II} by solving numerically  full and time-averaged systems obtained for each metric, namely LRS Bianchi III and open  FLRW. For this purpose   an algorithm in the programming language \textit{Python} was implemented. The systems of differential equations were solved using the \textit{solve\_ivp} code provided by the \textit{SciPy} open-source \textit{Python}-based ecosystem. The integration method used was \textit{Radau} that is an implicit Runge-Kutta method of the Radau IIa family of order $5$  with a relative and absolute tolerances of $10^{-4}$ and $10^{-7}$, respectively. All systems of differential equations were integrated with respect to  $\tau$, instead of $t$, with the range of integration $-40\leq\tau\leq 10$ for   original systems and $-40\leq\tau\leq 100$  for  averaged systems. All of them partitioned in $10000$ data points. Furthermore, each full and time-averaged systems were solved considering only one matter component, these are cosmological constant ($\gamma=0$), non relativistic matter or dust ($\gamma=1$), radiation ($\gamma=4/3$) and stiff fluid ($\gamma=2$). Thereby the vacuum solutions corresponds to those where $\Omega=\Omega_m\equiv 0$ and the solutions without matter component corresponds to $\Omega_m\equiv 0$. Finally we have considered fixed constants $\mu=\sqrt{2}/2$, $b=\sqrt{2}/5$ and $\omega=\sqrt{2}$, that lead to a value of $f=\frac{b \mu ^3}{\omega ^2-2 \mu ^2}=1/10$, that fulfills condition $f\geq 0$. With this values a generalized harmonic potential of the form 
 \begin{equation}
     V(\phi)=\frac{\phi ^2}{2}+\frac{1}{100}(1-\cos(10\phi))
 \end{equation}
is obtained. 

\subsection{LRS Bianchi III}

For the LRS Bianchi III metric we integrate: 
\begin{enumerate}
    \item The full system  given by \eqref{BIIIunperturbed1}. 
\item The time-averaged system \eqref{BIIIavrgsyst}.
\end{enumerate} 
\begin{table}[H]
\caption{\label{Tab1} Seven initial data sets for  simulation of  full system  \eqref{BIIIunperturbed1} and time-averaged system \eqref{BIIIavrgsyst}. All initial conditions are chosen in order to fulfill  equality $\bar{\Sigma}^2(0)+\bar{\Omega}^2(0)+\bar{\Omega}_{k}(0)+\bar{\Omega}_{m}(0)=1$.}
\footnotesize\setlength{\tabcolsep}{5pt}
    \begin{tabular}{lccccccc}\hline
Sol.  & \multicolumn{1}{c}{$H(0)$} & \multicolumn{1}{c}{$\bar{\Sigma}(0)$} & \multicolumn{1}{c}{$\bar{\Omega}^2(0)$} & \multicolumn{1}{c}{$\bar{\Omega}_{k}(0)$}  & \multicolumn{1}{c}{$\bar{\Omega}_m(0)$}  & \multicolumn{1}{c}{$\bar{\Phi}(0)$}  & \multicolumn{1}{c}{$t(0)$}  \\\hline
        i &  $0.1$ & $0.1$ & $0.9$ & $0.09$ & $0$ & $0$ & $0$\\
        ii &  $0.1$ & $0.4$ & $0.1$ & $0.74$ & $0$ & $0$ & $0$\\
        iii &  $0.1$ & $0.6$ & $0.1$ & $0.54$ & $0$ & $0$ & $0$\\
        iv &  $0.02$ & $0.48$ & $0.02$ & $0.7496$ & $0$ & $0$ & $0$\\
        v &  $0.1$ & $0.48$ & $0.02$ & $0.7496$ & $0$ & $0$ & $0$\\
        vi &  $0.1$ & $0.5$ & $0.01$ & $0.74$ & $0$ & $0$ & $0$\\
        vii &  $0.1$ & $0$ & $0.684$ & $0.001$ & $0.315$ & $0$ & $0$\\\hline
    \end{tabular}
\end{table}
As initial conditions we use seven data set presented in the Table \ref{Tab1}  as initial conditions for a better comparison of both systems. For data set $vii$ current values of $\Omega_{m}(0)=0.315$ and $\Omega_{k}(0)=0.001$ according to \cite{Planck2018} were considered. 

\begin{figure*}[h!]
    \centering
    \subfigure[\label{fig:BIIIParticular3D} Projections in the space $(\Sigma, H, \Omega^2)$. The surface is given by the constraint $\Omega^{2}=1-\Sigma^{2}$.]{\includegraphics[scale = 0.40]{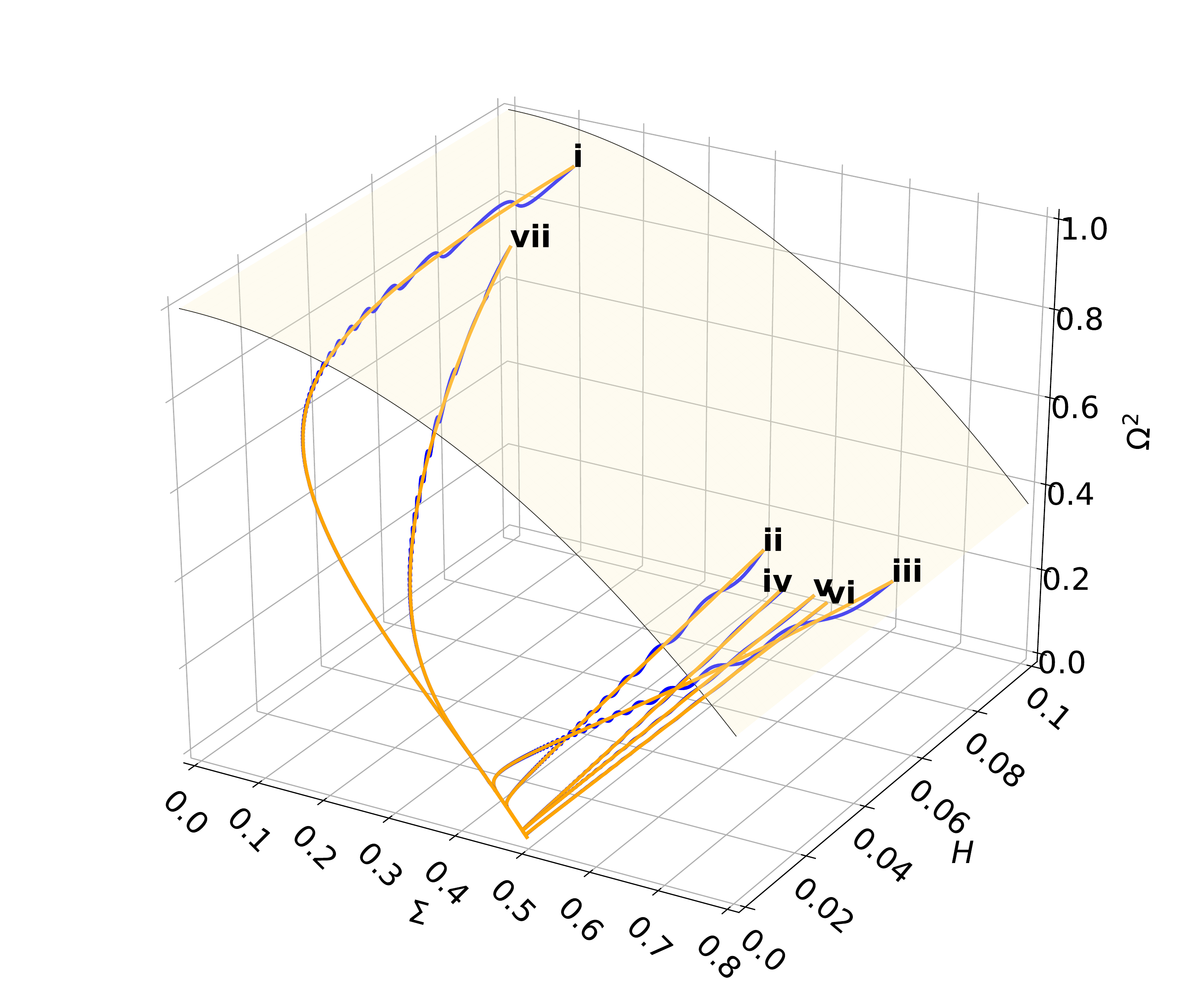}}
    \subfigure[\label{fig:BIIIParticular2D} Projection in the space $(\Sigma, \Omega^2)$. The black line represent the constraint $\Omega^{2}=1-\Sigma^{2}$.]{\includegraphics[scale = 0.53]{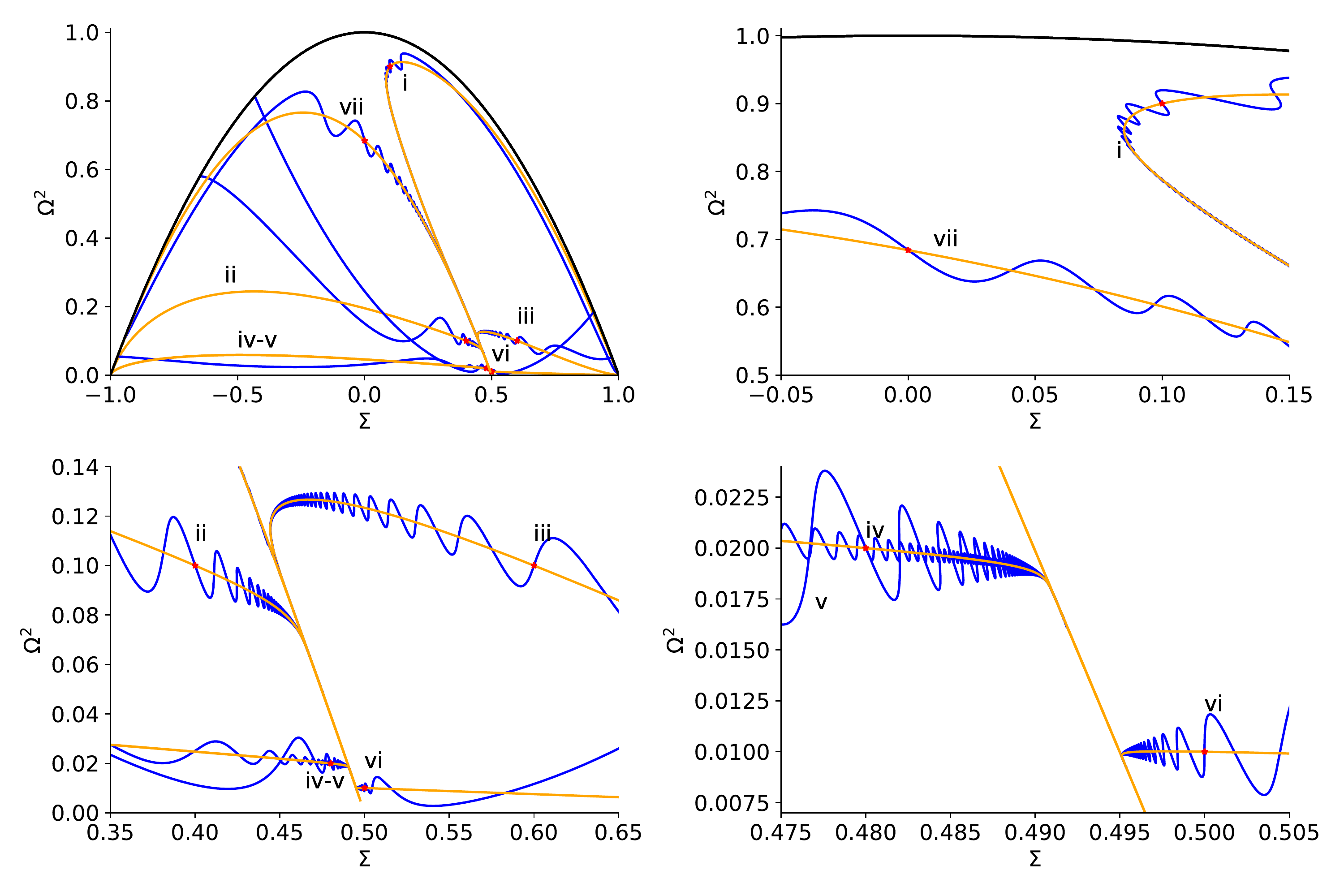}}
    \caption{Some solutions of the full system \eqref{BIIIunperturbed1} (blue) and time-averaged system \eqref{BIIIavrgsyst} (orange), corresponding to LRS Bianchi III metric, when $\gamma=b=f=0$, $\omega^2=2$, $\mu=1$, $\Omega_m=0$ and $\Omega_k=1-\Sigma^2-\Omega^2$, with the identification $\Omega^2 \mapsto \Omega $, for which the results of \cite{Fajman:2020yjb} are recovered. We have used initial data sets presented in the Table \ref{Tab1}.}
\end{figure*}

It is important to mention that the first six initial conditions correspond to initial conditions presented in  Table 2 of \cite{Fajman:2020yjb} and the additional data set $vii$ is obtained considering  current values of $\Omega_m(0)=0.315$ and $\Omega_k(0)=0.001$ according to \cite{Planck2018}. 
Even more, the model presented in \cite{Fajman:2020yjb} is contained in our model when $\gamma=b=f=0$, $\omega^2=2$, $\mu=1$, $\Omega_m=0$ and $\Omega_k=1-\Sigma^2-\Omega^2$ (then $\gamma$ does not appears in the model presented in \cite{Fajman:2020yjb}) with the identification $\Omega^2\mapsto\Omega$. As can be seen in Figures \ref{fig:BIIIParticular3D} and \ref{fig:BIIIParticular2D} where some solutions of the full system \eqref{BIIIunperturbed1} and time-averaged system \eqref{BIIIavrgsyst} are presented; showing that our results are in complete agreement with results presented in \cite{Fajman:2020yjb} for the limiting case.

\begin{figure*}
    \centering
    \subfigure[\label{fig:BIIICC3DS} Projections in the space $(\Sigma, H, \Omega^2)$. The surface is given by the constraint $\Omega^{2}=1-\Sigma^{2}$.]{\includegraphics[scale = 0.40]{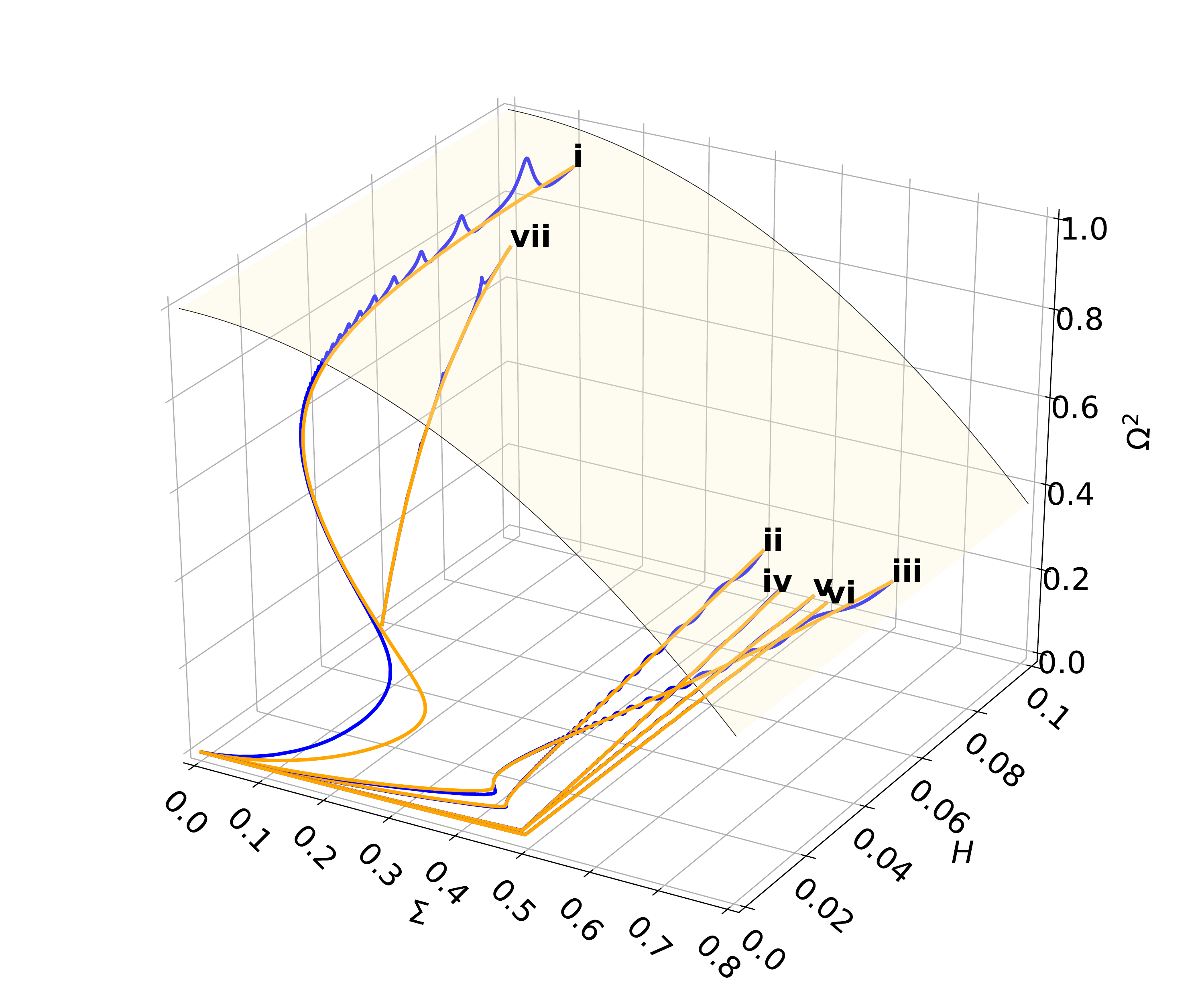}}
    \subfigure[\label{fig:BIIICC2DS} Projection in the space $(\Sigma, \Omega^2)$. The black line represent the constraint $\Omega^{2}=1-\Sigma^{2}$.]{\includegraphics[scale = 0.53]{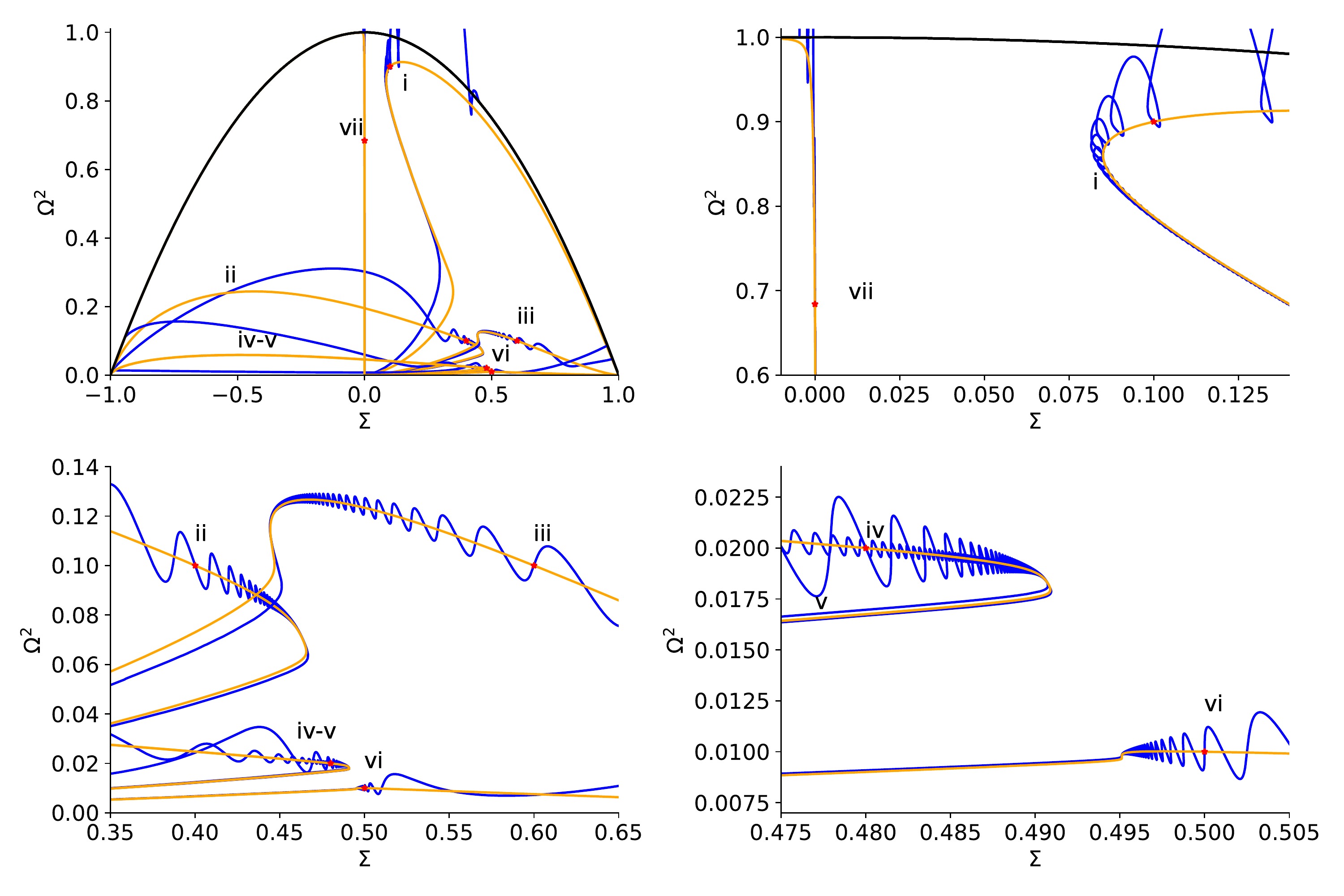}}
    \caption{Some solutions of the full system \eqref{BIIIunperturbed1} (blue) and time-averaged system \eqref{BIIIavrgsyst} (orange) for the LRS Bianchi III metric when $\gamma=0$, in the projection $\Omega_{k}=0$. We have used for both systems initial data sets presented in the Table \ref{Tab1}.}
\end{figure*}

\begin{figure*}
    \centering
    \subfigure[\label{fig:BIIICC3DK} Projections in the space $(\Omega_{k}, H, \Omega^2)$. The surface is given by the constraint $\Omega^{2}=1-\Omega_{k}$.]{\includegraphics[scale = 0.40]{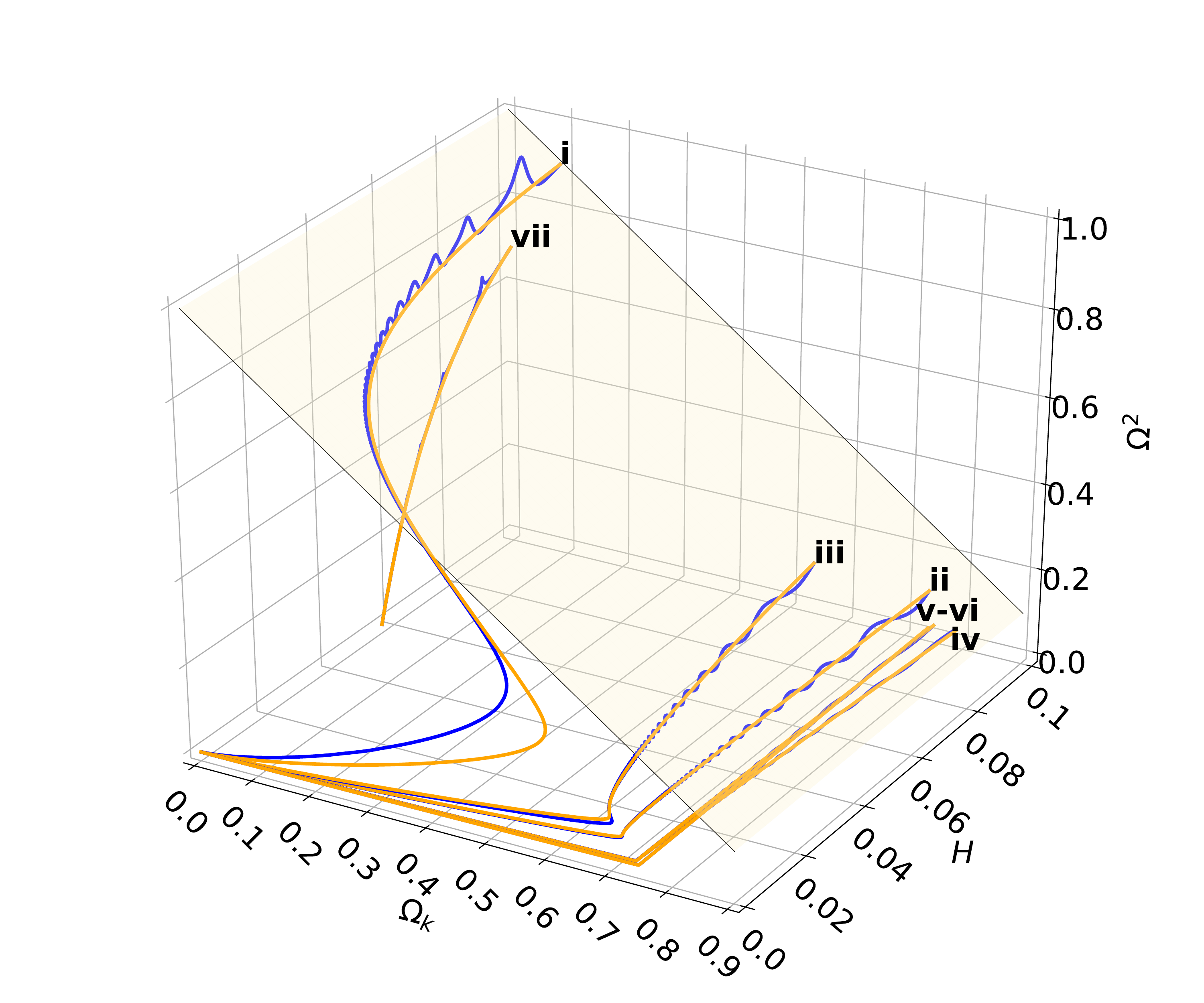}}
    \subfigure[\label{fig:BIIICC2DK} Projection in the space $(\Omega_{k}, \Omega^2)$. The black line represent the constraint $\Omega^{2}=1-\Omega_{k}$.]{\includegraphics[scale = 0.53]{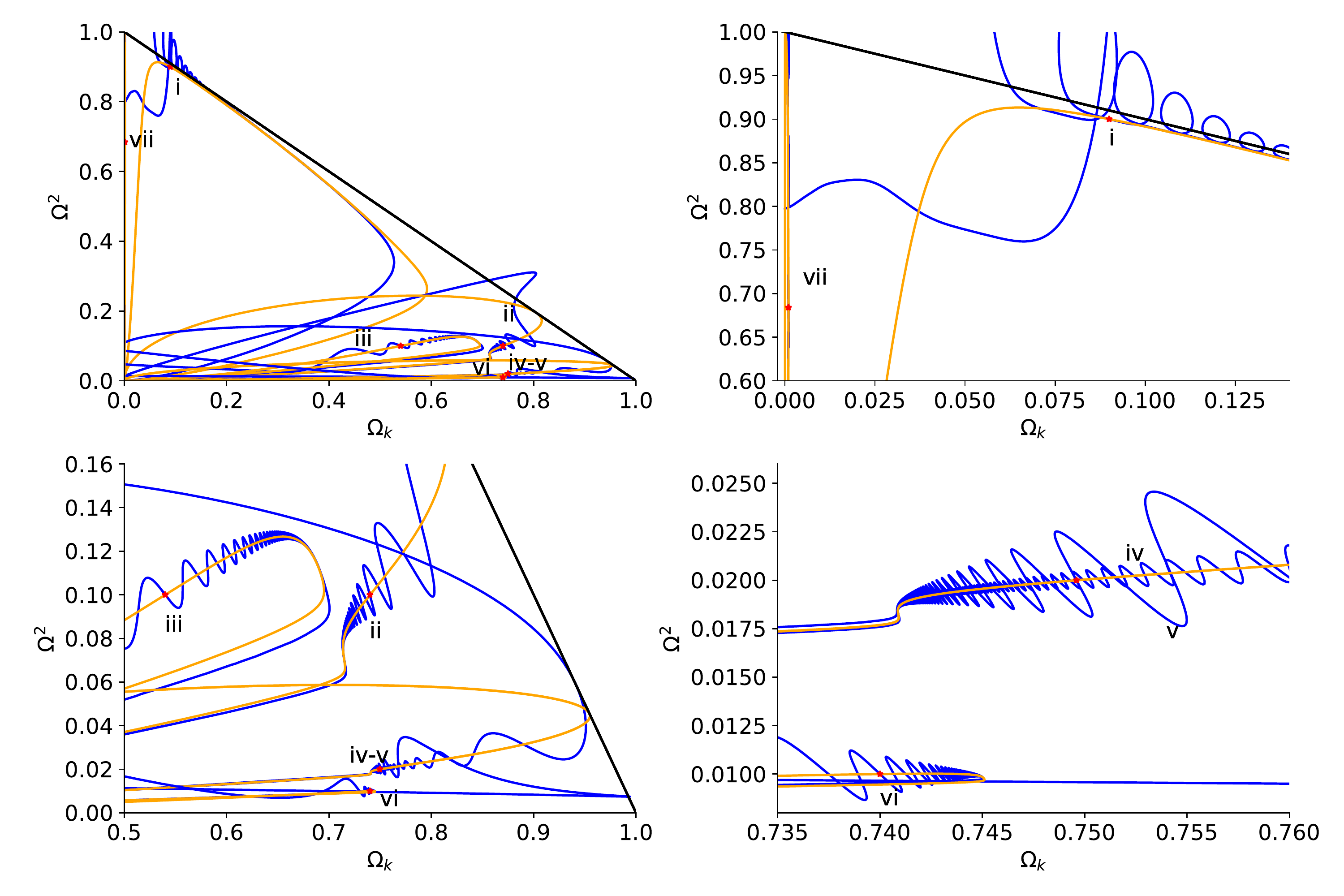}}
    \caption{Some solutions of the full system \eqref{BIIIunperturbed1} (blue) and time-averaged system \eqref{BIIIavrgsyst} (orange) for the LRS Bianchi III metric when $\gamma=0$, in the projection $\Sigma=0$. We have used for both systems initial data sets presented in the Table \ref{Tab1}.}
\end{figure*}

\begin{figure*}
    \centering
    \subfigure[\label{fig:BIIIDust3DS} Projections in the space $(\Sigma, H, \Omega^2)$. The surface is given by the constraint $\Omega^{2}=1-\Sigma^{2}$.]{\includegraphics[scale = 0.40]{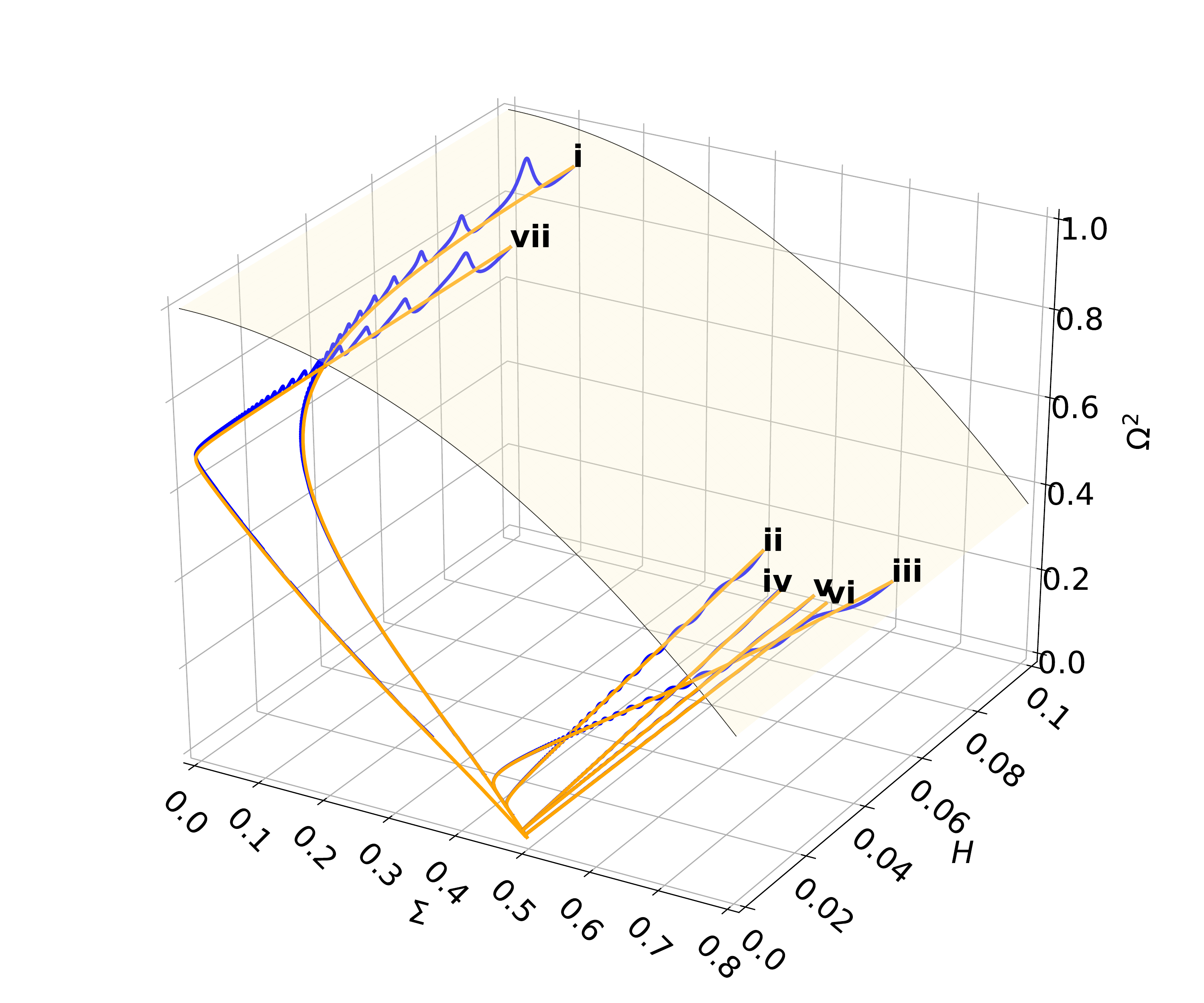}}
    \subfigure[\label{fig:BIIIDust2DS} Projection in the space $(\Sigma, \Omega^2)$. The black line represent the constraint $\Omega^{2}=1-\Sigma^{2}$.]{\includegraphics[scale = 0.53]{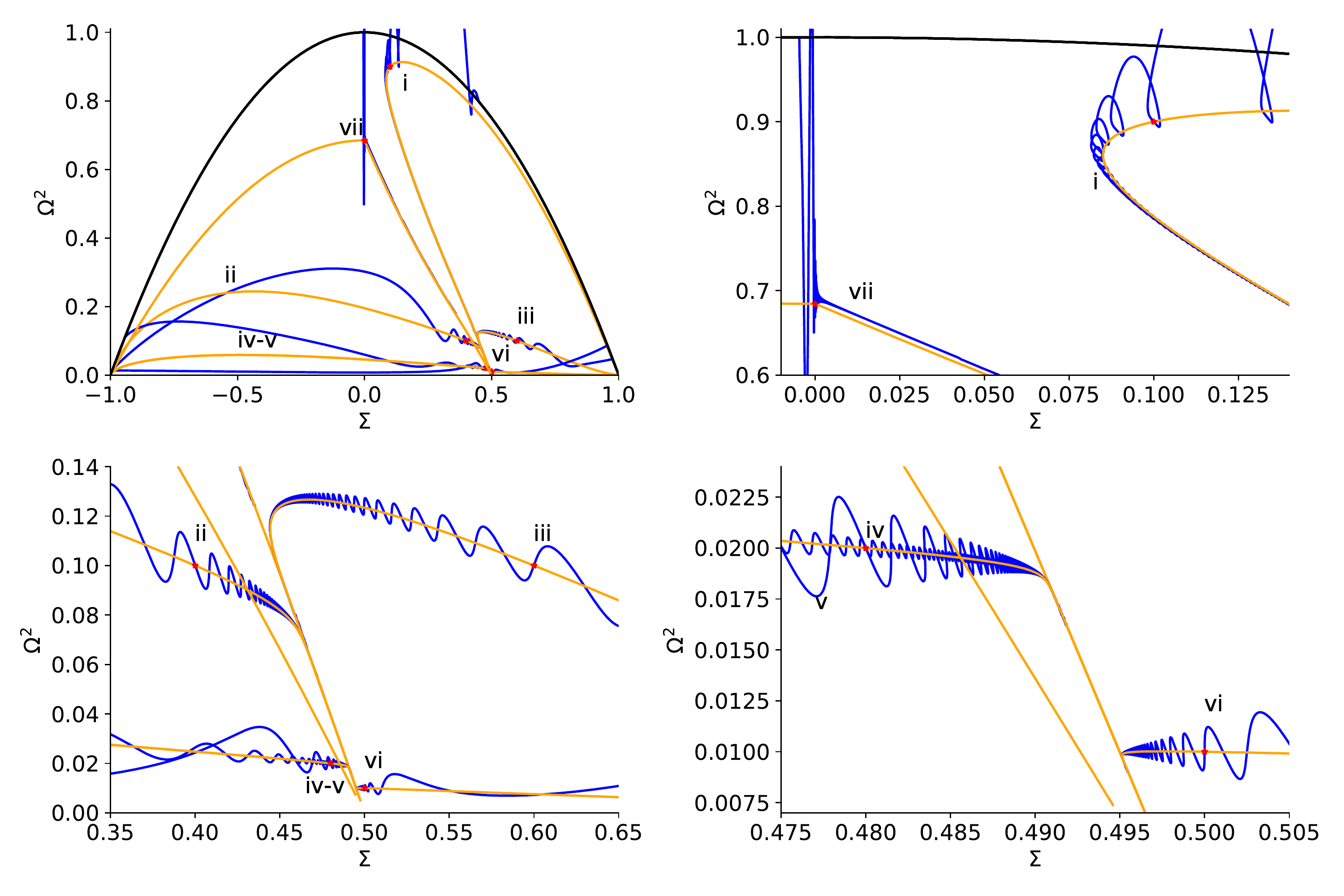}}
    \caption{Some solutions of the full system \eqref{BIIIunperturbed1} (blue) and time-averaged system \eqref{BIIIavrgsyst} (orange) for the LRS Bianchi III metric when $\gamma=1$, in the projection $\Omega_{k}=0$. We have used for both systems initial data sets presented in the Table \ref{Tab1}.}
\end{figure*}

\begin{figure*}
    \centering
    \subfigure[\label{fig:BIIIDust3DK} Projections in the space $(\Omega_{k}, H, \Omega^2)$. The surface is given by the constraint $\Omega^{2}=1-\Omega_{k}$.]{\includegraphics[scale = 0.40]{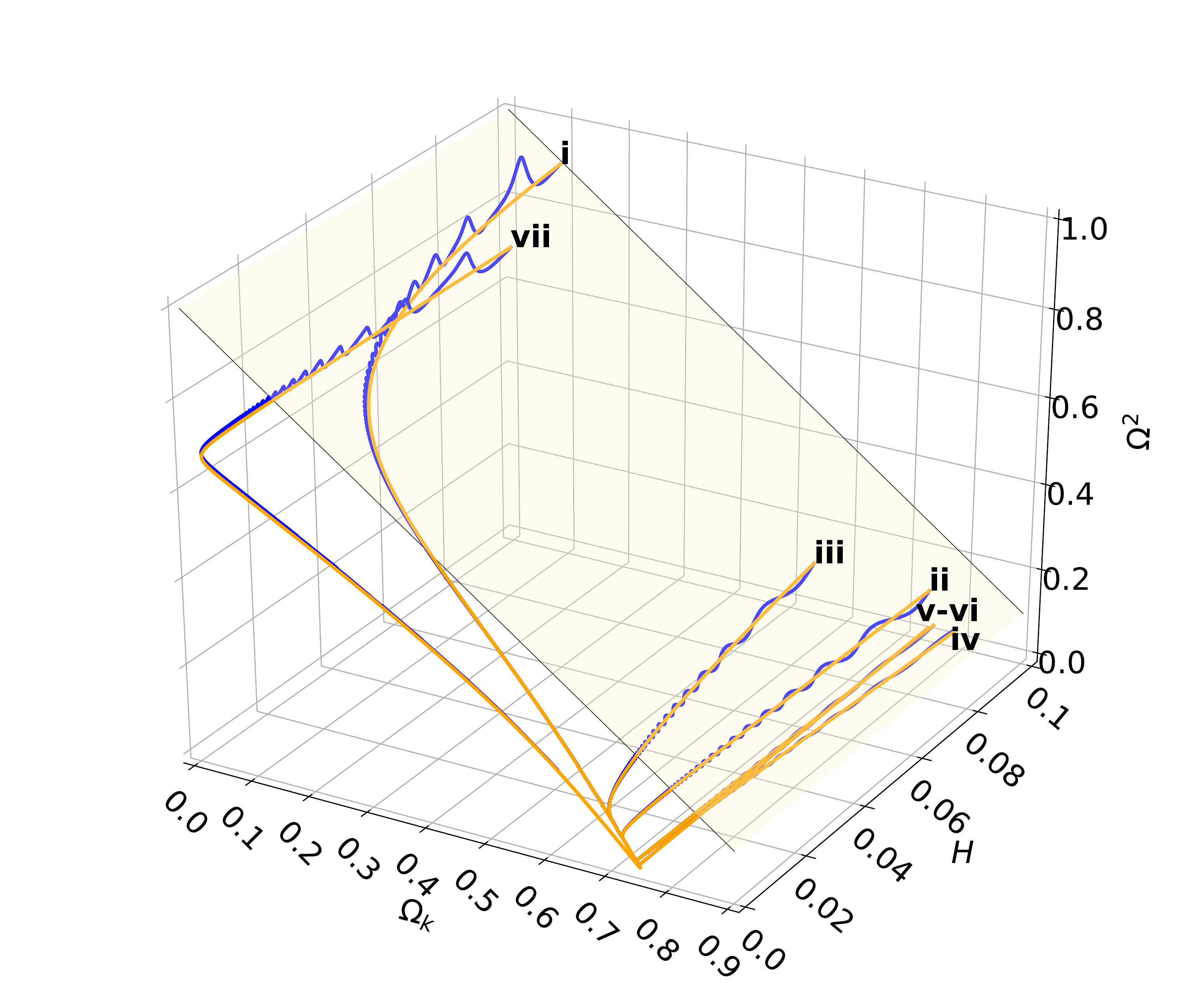}}
    \subfigure[\label{fig:BIIIDust2DK} Projection in the space $(\Omega_{k}, \Omega^2)$. The black line represent the constraint $\Omega^{2}=1-\Omega_{k}$.]{\includegraphics[scale = 0.53]{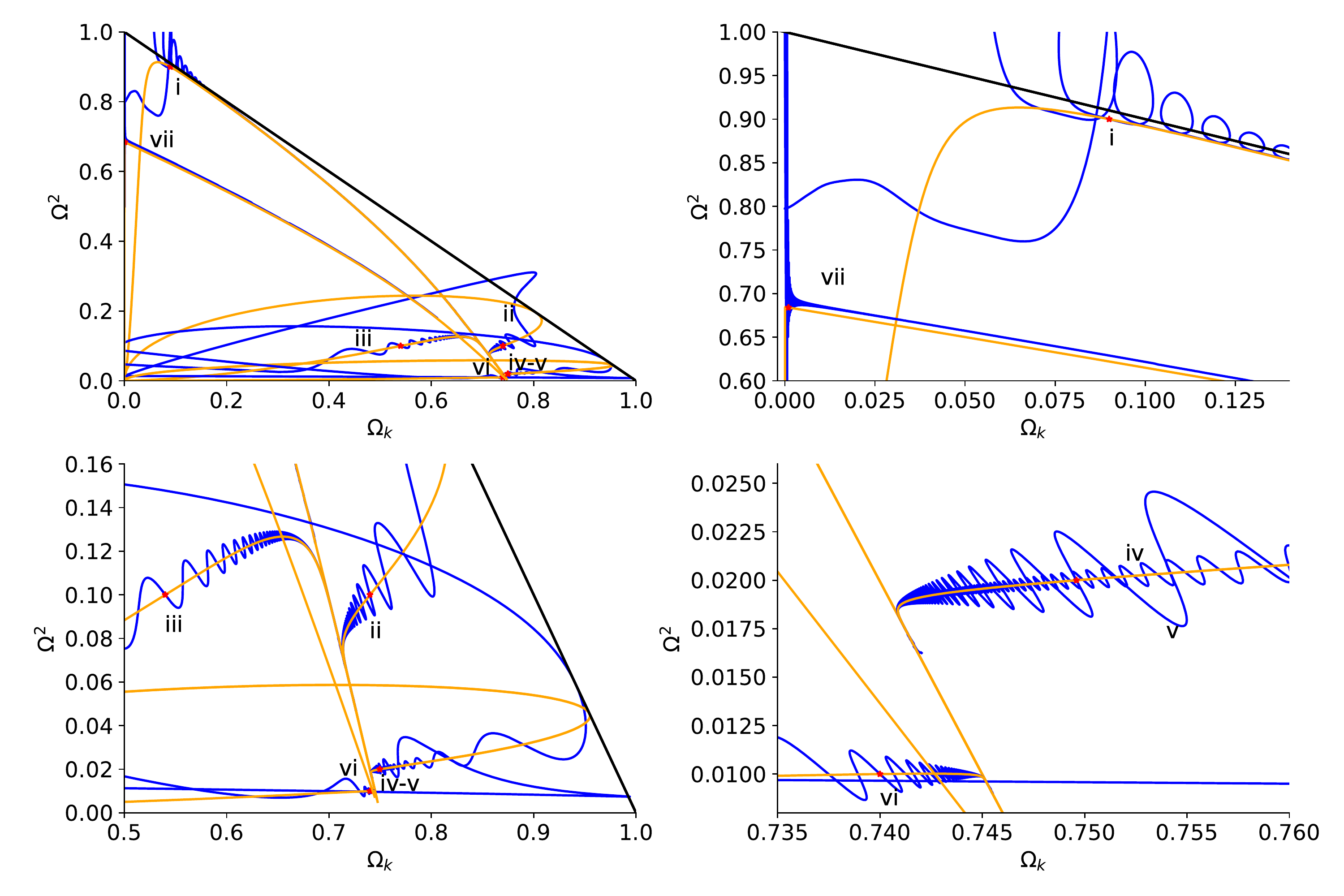}}
    \caption{Some solutions of the full system \eqref{BIIIunperturbed1} (blue) and time-averaged system \eqref{BIIIavrgsyst} (orange) for the LRS Bianchi III metric when $\gamma=1$, in the projection $\Sigma=0$. We have used for both systems initial data sets presented in the Table \ref{Tab1}.}
\end{figure*}

\begin{figure*}
    \centering
    \subfigure[\label{fig:BIIIRad3DS} Projections in the space $(\Sigma, H, \Omega^2)$. The surface is given by the constraint $\Omega^{2}=1-\Sigma^{2}$.]{\includegraphics[scale = 0.40]{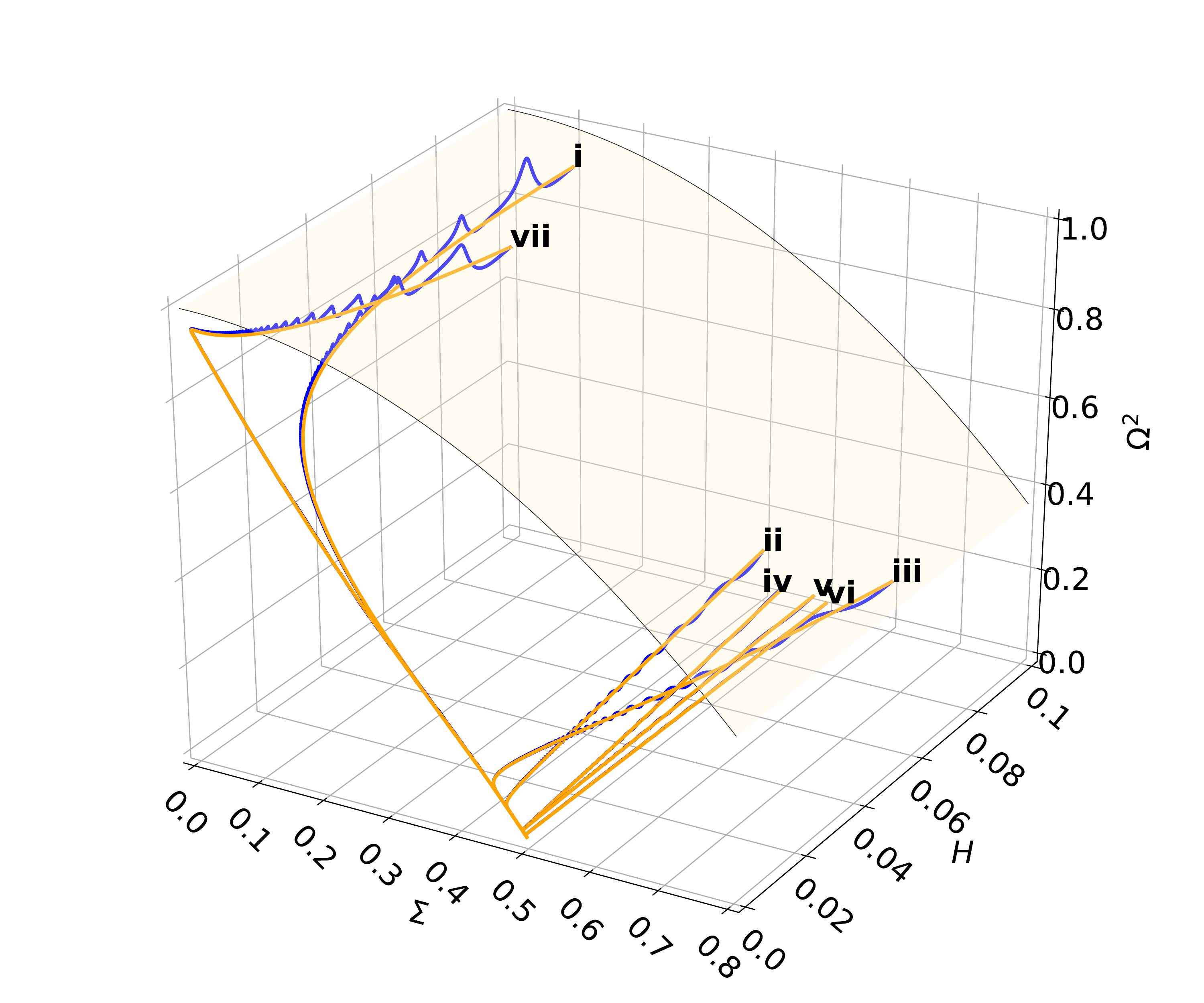}}
    \subfigure[\label{fig:BIIIRad2DS} Projection in the space $(\Sigma, \Omega^2)$. The black line represent the constraint $\Omega^{2}=1-\Sigma^{2}$.]{\includegraphics[scale = 0.53]{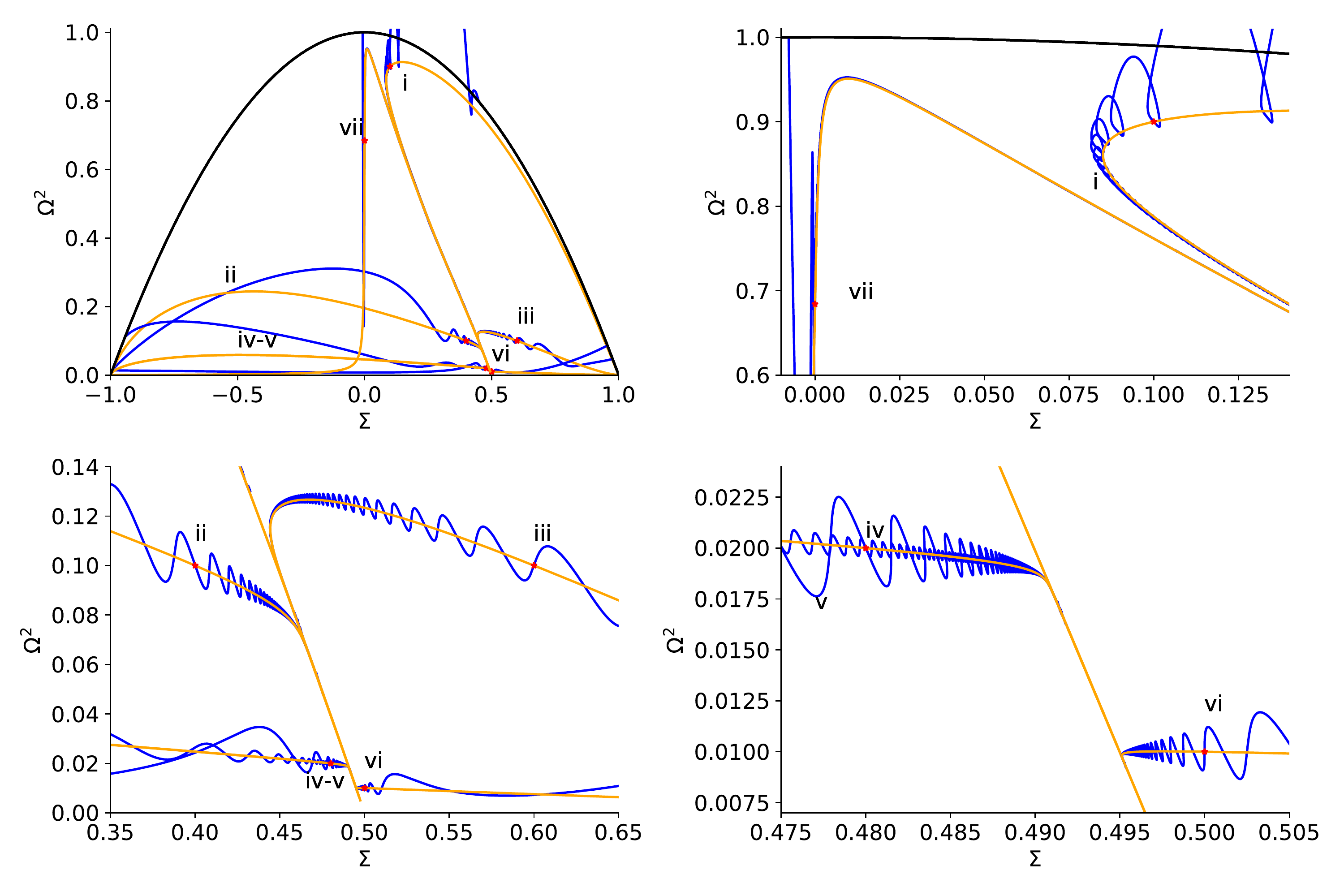}}
    \caption{Some solutions of the full system \eqref{BIIIunperturbed1} (blue) and time-averaged system \eqref{BIIIavrgsyst} (orange) for the LRS Bianchi III metric when $\gamma=4/3$, in the projection $\Omega_{k}=0$. We have used for both systems initial data sets presented in the Table \ref{Tab1}.}
\end{figure*}

\begin{figure*}
    \centering
    \subfigure[\label{fig:BIIIRad3DK} Projections in the space $(\Omega_{k}, H, \Omega^2)$. The surface is given by the constraint $\Omega^{2}=1-\Omega_{k}$.]{\includegraphics[scale = 0.40]{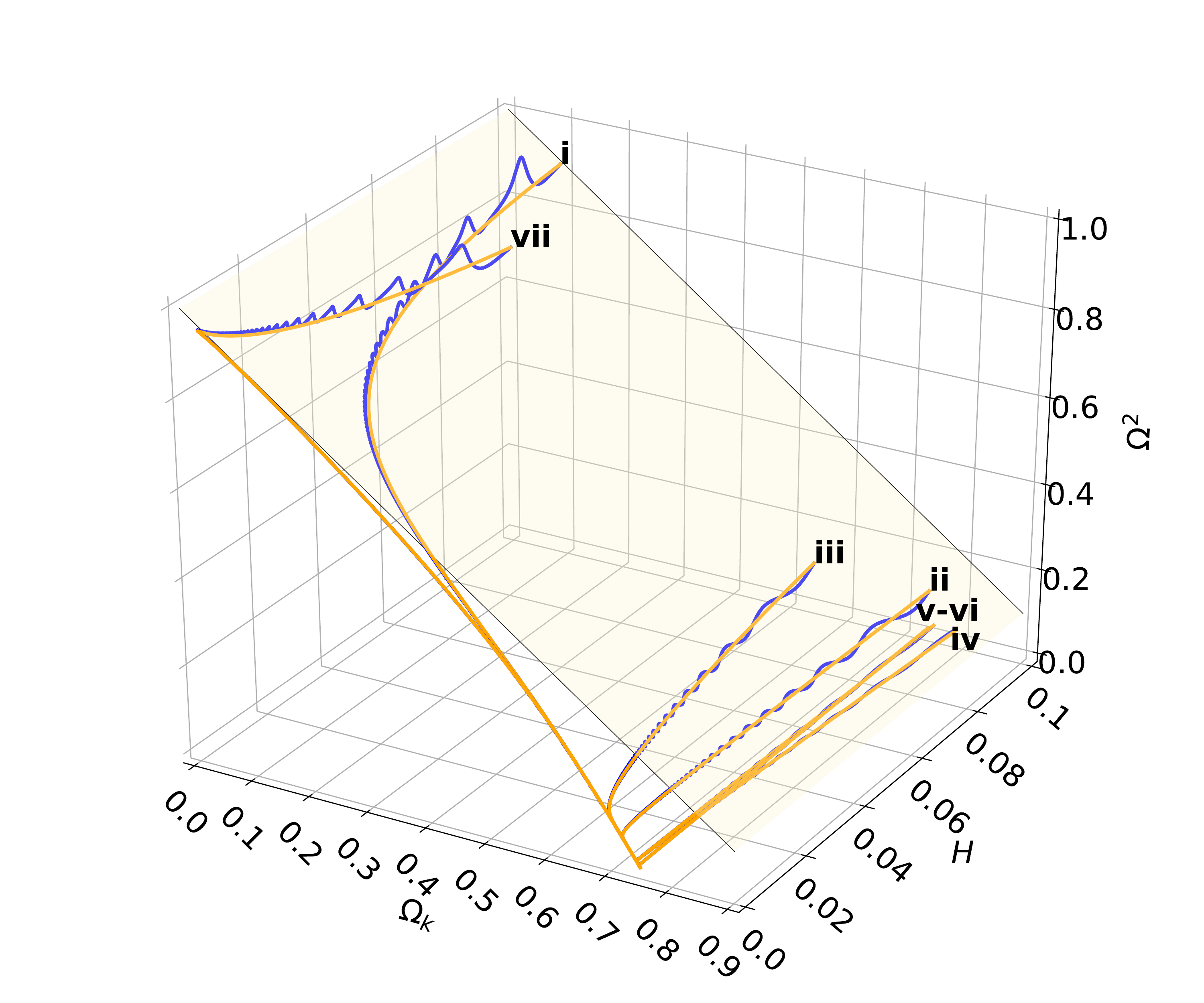}}
    \subfigure[\label{fig:BIIIRad2DK} Projection in the space $(\Omega_{k}, \Omega^2)$. The black line represent the constraint $\Omega^{2}=1-\Omega_{k}$.]{\includegraphics[scale = 0.53]{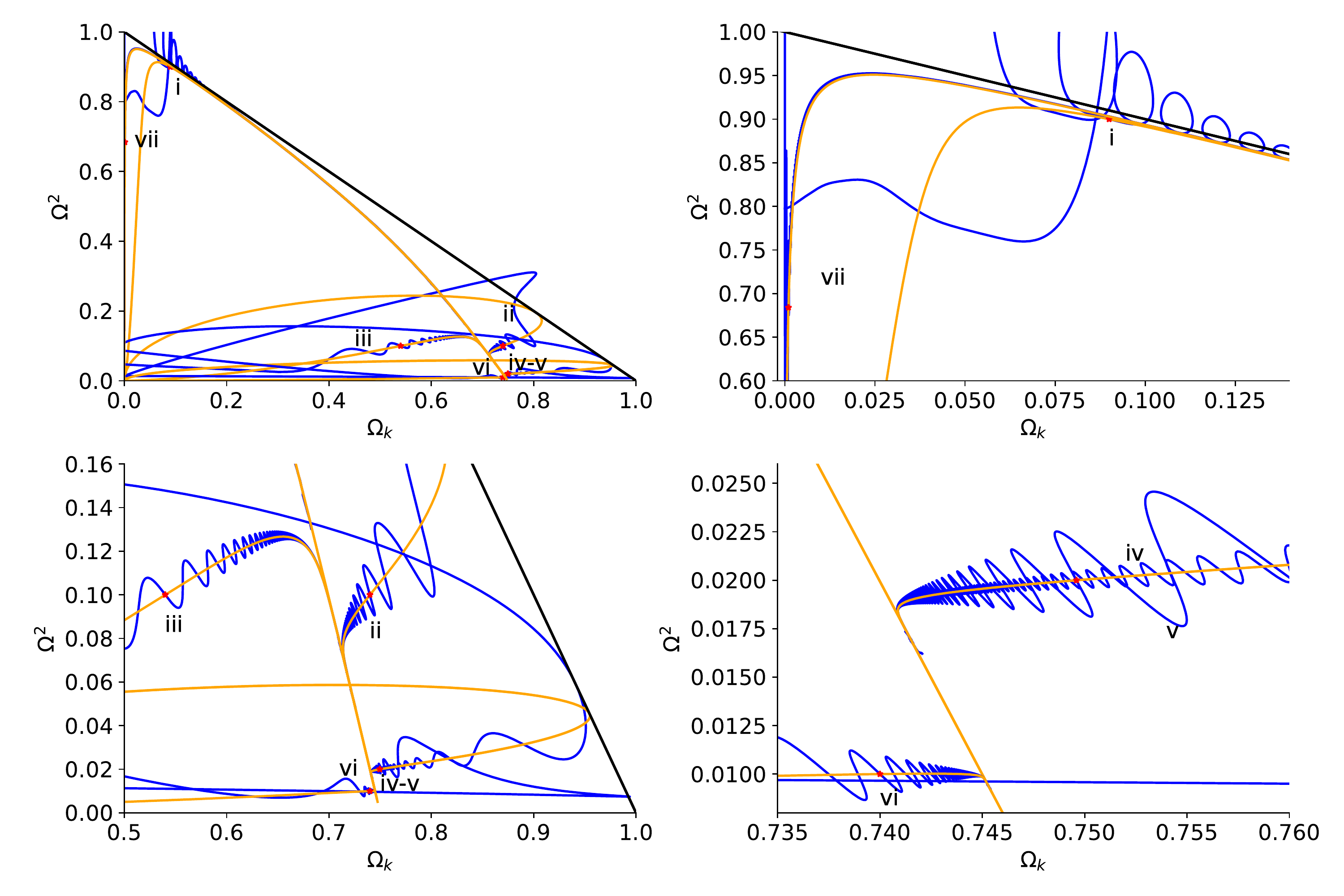}}
    \caption{Some solutions of the full system \eqref{BIIIunperturbed1} (blue) and time-averaged system \eqref{BIIIavrgsyst} (orange) for the LRS Bianchi III metric when $\gamma=4/3$, in the projection $\Sigma=0$. We have used for both systems initial data sets presented in the Table \ref{Tab1}.}
\end{figure*}

\begin{figure*}
    \centering
    \subfigure[\label{fig:BIIIStiff3DS} Projections in the space $(\Sigma, H, \Omega^2)$. The surface is given by the constraint $\Omega^{2}=1-\Sigma^{2}$.]{\includegraphics[scale = 0.40]{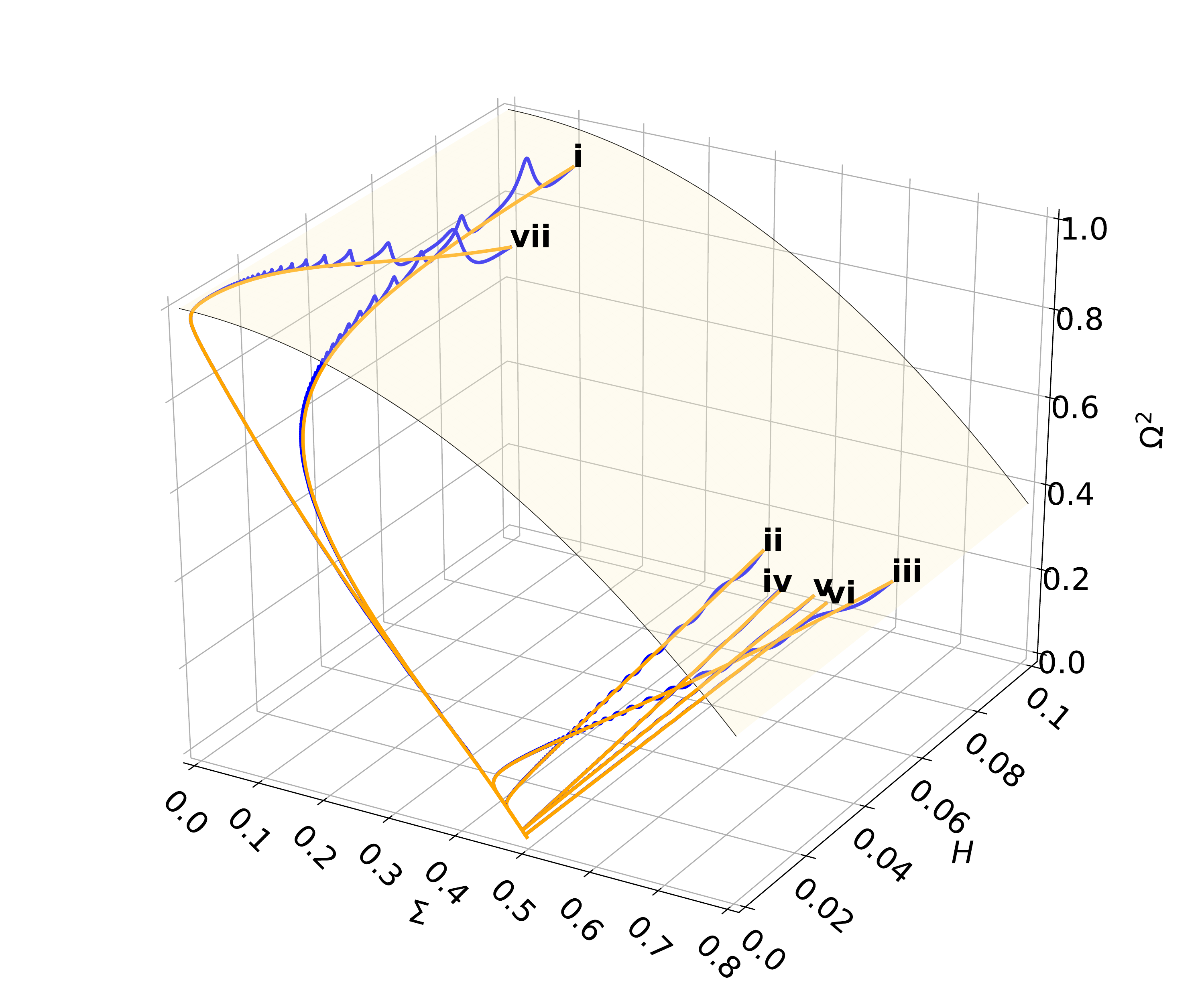}}
    \subfigure[\label{fig:BIIIStiff2DS} Projection in the space $(\Sigma, \Omega^2)$. The black line represent the constraint $\Omega^{2}=1-\Sigma^{2}$.]{\includegraphics[scale = 0.53]{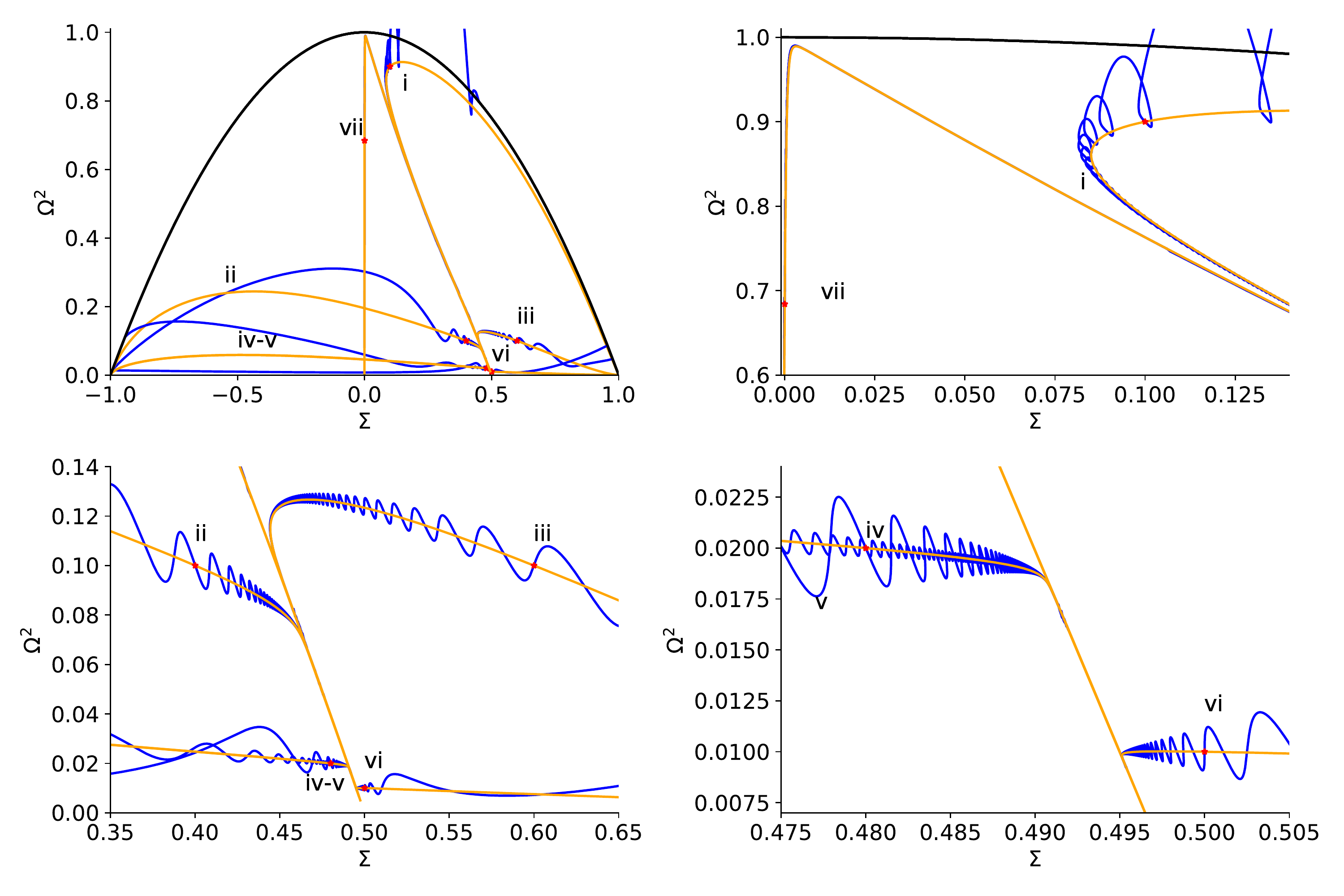}}
    \caption{Some solutions of the full system \eqref{BIIIunperturbed1} (blue) and time-averaged system \eqref{BIIIavrgsyst} (orange) for the LRS Bianchi III metric when $\gamma=2$, in the projection $\Omega_{k}=0$. We have used for both systems initial data sets presented in the Table \ref{Tab1}.}
\end{figure*}

\begin{figure*}
    \centering
    \subfigure[\label{fig:BIIIStiff3DK} Projections in the space $(\Omega_{k}, H, \Omega^2)$. The surface is given by the constraint $\Omega^{2}=1-\Omega_{k}$.]{\includegraphics[scale = 0.40]{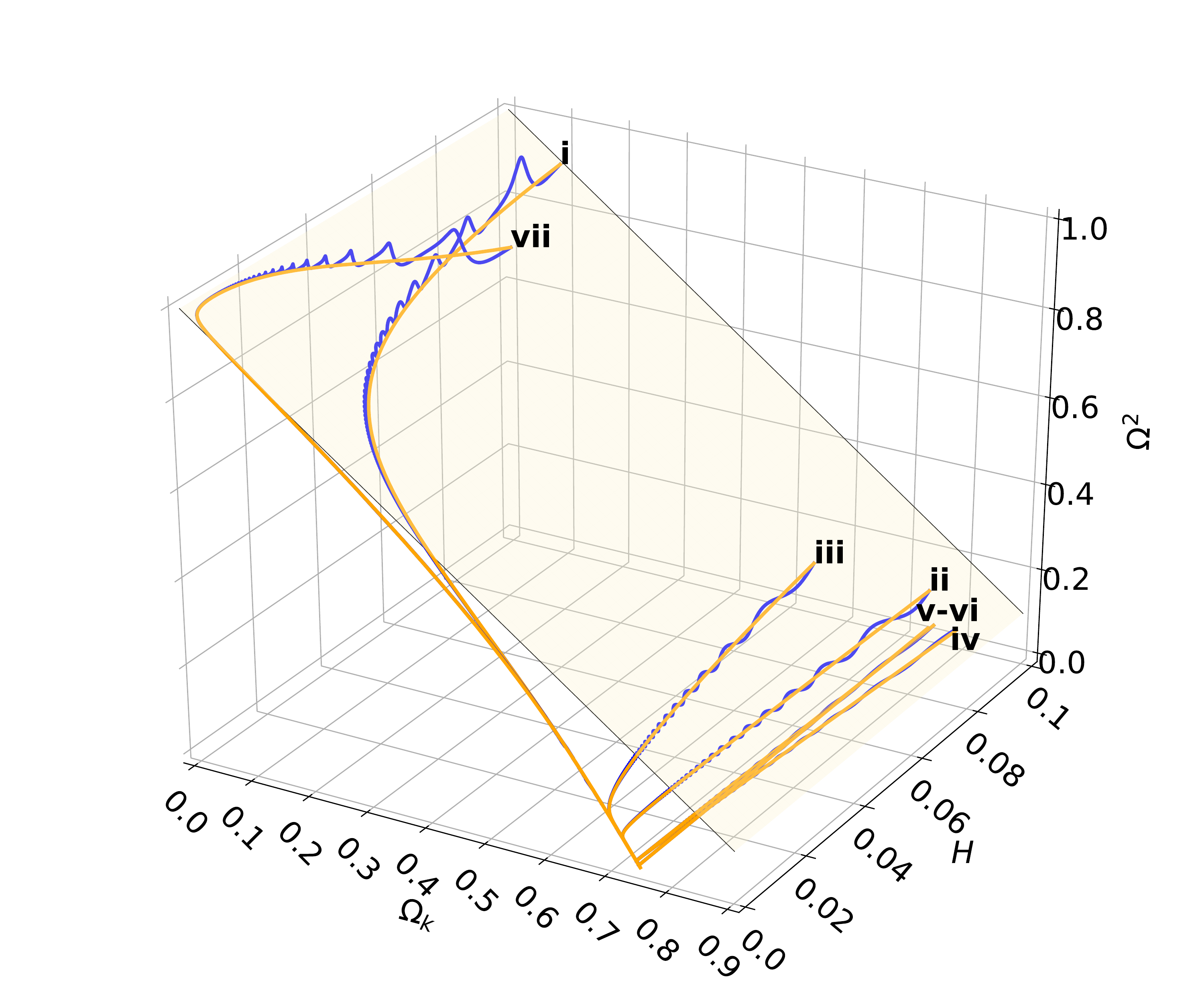}}
    \subfigure[\label{fig:BIIIStiff2DK} Projection in the space $(\Omega_{k}, \Omega^2)$. The black line represent the constraint $\Omega^{2}=1-\Omega_{k}$.]{\includegraphics[scale = 0.53]{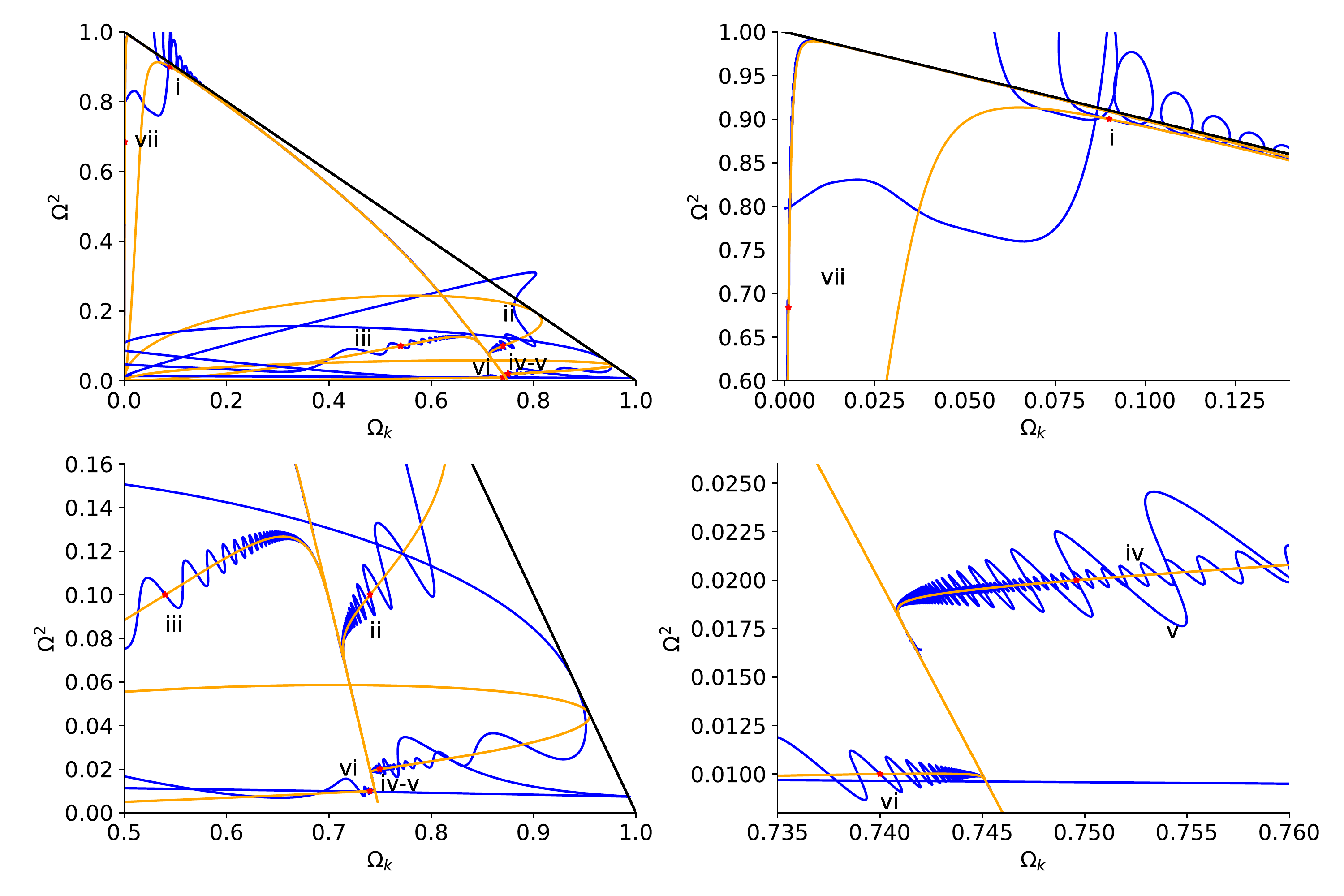}}
    \caption{Some solutions of the full system \eqref{BIIIunperturbed1} (blue) and time-averaged system \eqref{BIIIavrgsyst} (orange) for the LRS Bianchi III metric when $\gamma=2$, in the projection $\Sigma=0$. We have used for both systems initial data sets presented in the Table \ref{Tab1}.}
\end{figure*}

In Figures \ref{fig:BIIICC3DS}-\ref{fig:BIIIStiff2DK} projections of some solutions of the full system \eqref{BIIIunperturbed1} and time-averaged system \eqref{BIIIavrgsyst} in the $(\Sigma, H, \Omega^{2})$ and $(\Omega_{k}, H, \Omega^{2})$ space with their respective projection when $H=0$ considering for both systems the same initial data sets from Table \ref{Tab1} are presented. Figures \ref{fig:BIIICC3DS}-\ref{fig:BIIICC2DK} show solutions for matter fluid   corresponding to cosmological constant ($\gamma=0$). Figures \ref{fig:BIIIDust3DS}-\ref{fig:BIIIDust2DK} show solutions for matter fluid   corresponding to dust ($\gamma=1$). Figures \ref{fig:BIIIRad3DS}-\ref{fig:BIIIRad2DK} show solutions for matter fluid   corresponding to radiation ($\gamma=\frac{4}{3}$). Figures \ref{fig:BIIIStiff3DS}-\ref{fig:BIIIStiff2DK} show solutions for matter fluid   corresponding to a stiff fluid ($\gamma=2$). These figures are evidence that the main Theorem presented in section \ref{SECT:II} is fulfilled for  LRS Bianchi III metric.

\subsection{FLRW  $k=-1$}

For the FLRW metric with negative curvature ($k=-1$, $\Omega_{k}>0$) we integrate:
\begin{enumerate}
    \item The full system given by \eqref{unperturbed1FLRW}. 
\item The time-averaged system \eqref{avrgsystFLRW}.
\end{enumerate}
\begin{table}[H]
\caption{\label{Tab3a} Seven initial data sets for  simulation of  full system  \eqref{unperturbed1FLRW} and time-averaged system \eqref{avrgsystFLRW}. All the conditions are chosen in order to fulfill equality $\bar{\Omega}^{2}(0)+\bar{\Omega}_{k}(0)+\bar{\Omega}_{m}(0)=1$.}
\footnotesize\setlength{\tabcolsep}{5pt}
    \begin{tabular}{lcccccc}\hline
Sol.  & \multicolumn{1}{c}{$H(0)$}  & \multicolumn{1}{c}{$\bar{\Omega}^2(0)$} & \multicolumn{1}{c}{$\bar{\Omega}_{k}(0)$}  & \multicolumn{1}{c}{$\bar{\Omega}_m(0)$}  & \multicolumn{1}{c}{$\bar{\Phi}(0)$}  & \multicolumn{1}{c}{$t(0)$}  \\\hline
        i &  $0.1$ & $0.9$ & $0.09$ & $0.01$ & $0$ & $0$\\
        ii &  $0.1$  & $0.1$ & $0.74$ & $0.16$ & $0$ & $0$\\
        iii &  $0.1$  & $0.1$ & $0.54$ & $0.36$ & $0$ & $0$\\
        iv &  $0.02$  & $0.02$ & $0.7496$ & $0.2304$  & $0$ & $0$\\
        v &  $0.1$ &  $0.02$  & $0.7496$ & $0.2304$ & $0$ & $0$\\
        vi &  $0.1$  & $0.01$ &  $0.3$ & $0.69$ & $0$ & $0$\\
        vii &  $0.1$  & $0.684$ & $0.001$ & $0.315$ & $0$ & $0$\\\hline
    \end{tabular}
\end{table}
Seven data set presented in Table  \ref{Tab3a} were used as initial conditions. For data set $vii$ current values of $\Omega_{m}(0)=0.315$ and $\Omega_{k}(0)=0.001$ according to \cite{Planck2018} were considered. 

\begin{figure*}
    \centering
    \subfigure[\label{fig:OpenFLRWCC3D} Projections in the space $(\Omega_{k}, H, \Omega^2)$. The surface is given by the constraint $\Omega^{2}=1-\Omega_{k}$.]{\includegraphics[scale = 0.40]{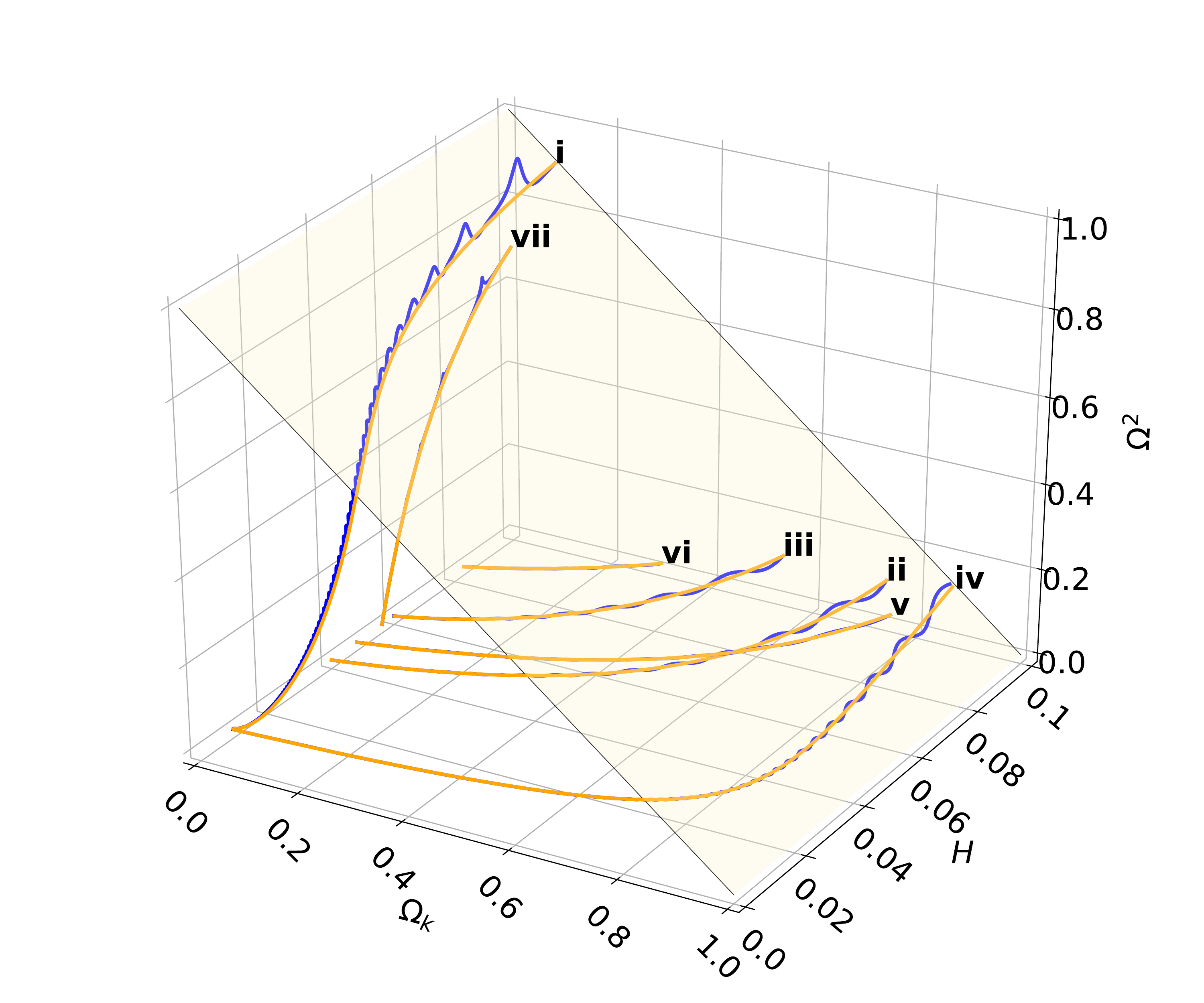}}
    \subfigure[\label{fig:OpenFLRWCC2D} Projection in the space $(\Omega_{k}, \Omega^2)$. The black line represent the constraint $\Omega^{2}=1-\Omega_{k}$.]{\includegraphics[scale = 0.53]{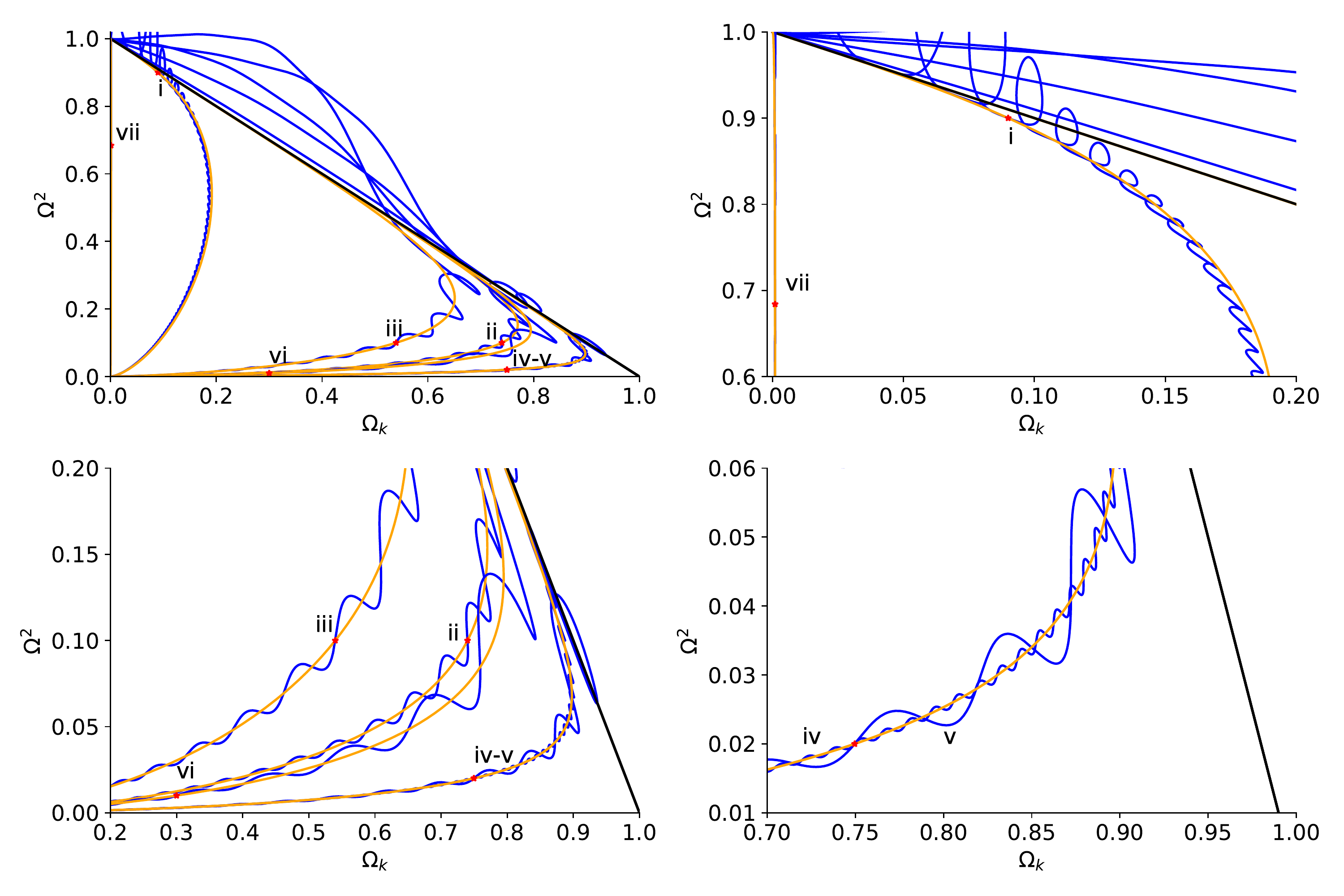}}
    \caption{Some solutions of the full system \eqref{unperturbed1FLRW} (blue) and time-averaged system \eqref{avrgsystFLRW} (orange) for the FLRW metric with negative curvature ($k=-1$) when $\gamma=0$. We have used for both systems initial data sets presented in the Table \ref{Tab3a}.}
\end{figure*}

\begin{figure*}
    \centering
    \subfigure[\label{fig:OpenFLRWDust3D} Projections in the space $(\Omega_{k}, H, \Omega^2)$. The surface is given by the constraint $\Omega^{2}=1-\Omega_{k}$.]{\includegraphics[scale = 0.40]{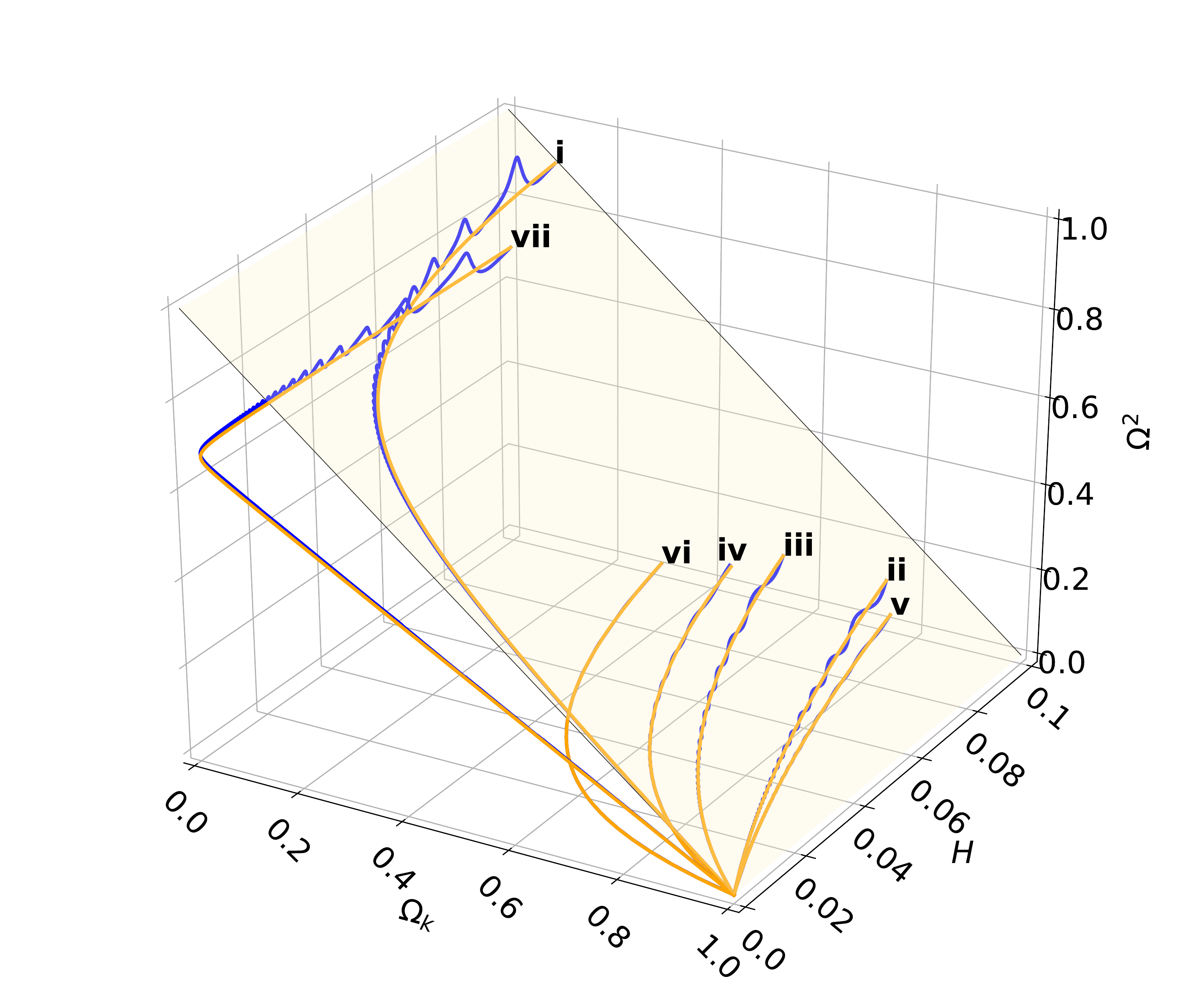}}
    \subfigure[\label{fig:OpenFLRWDust2D} Projection in the space $(\Omega_{k}, \Omega^2)$. The black line represent the constraint $\Omega^{2}=1-\Omega_{k}$.]{\includegraphics[scale = 0.53]{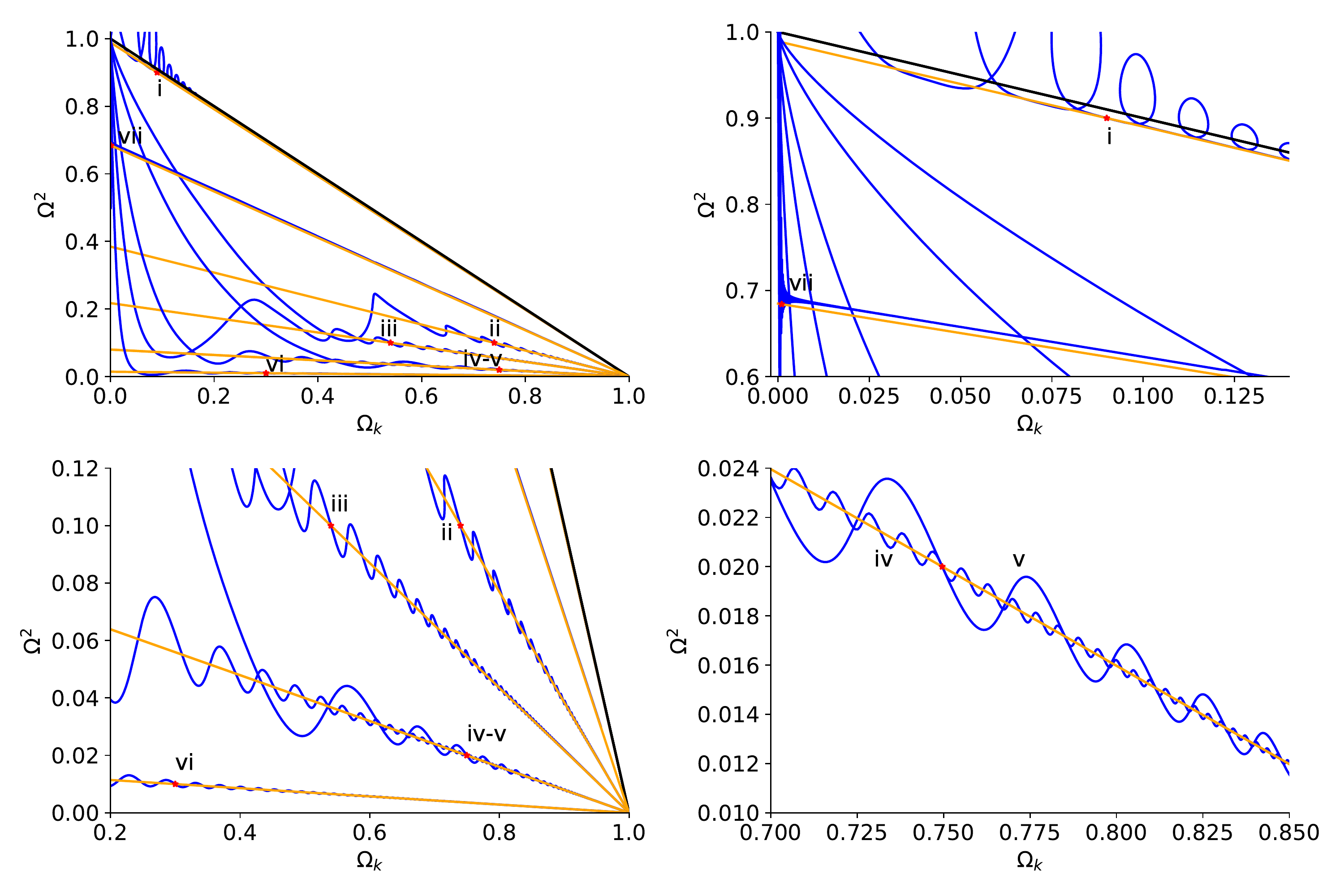}}
    \caption{Some solutions of the full system \eqref{unperturbed1FLRW} (blue) and time-averaged system \eqref{avrgsystFLRW} (orange) for the FLRW metric with negative curvature ($k=-1$) when $\gamma=1$. We have used for both systems initial data sets presented in the Table \ref{Tab3a}.}
\end{figure*}

\begin{figure*}
    \centering
    \subfigure[\label{fig:OpenFLRWRad3D} Projections in the space $(\Omega_{k}, H, \Omega^2)$. The surface is given by the constraint $\Omega^{2}=1-\Omega_{k}$.]{\includegraphics[scale = 0.40]{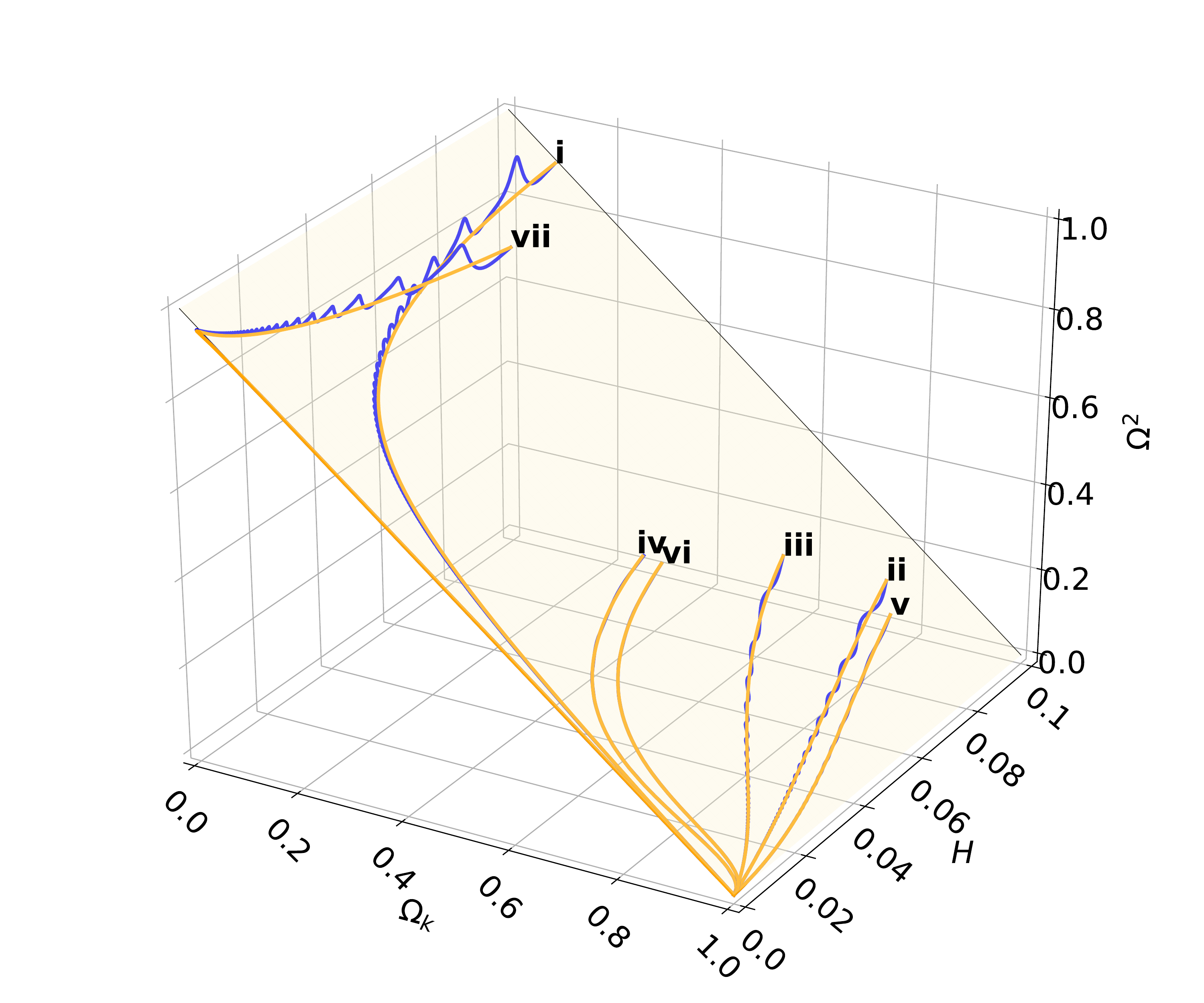}}
    \subfigure[\label{fig:OpenFLRWRad2D} Projection in the space $(\Omega_{k}, \Omega^2)$. The black line represent the constraint $\Omega^{2}=1-\Omega_{k}$.]{\includegraphics[scale = 0.53]{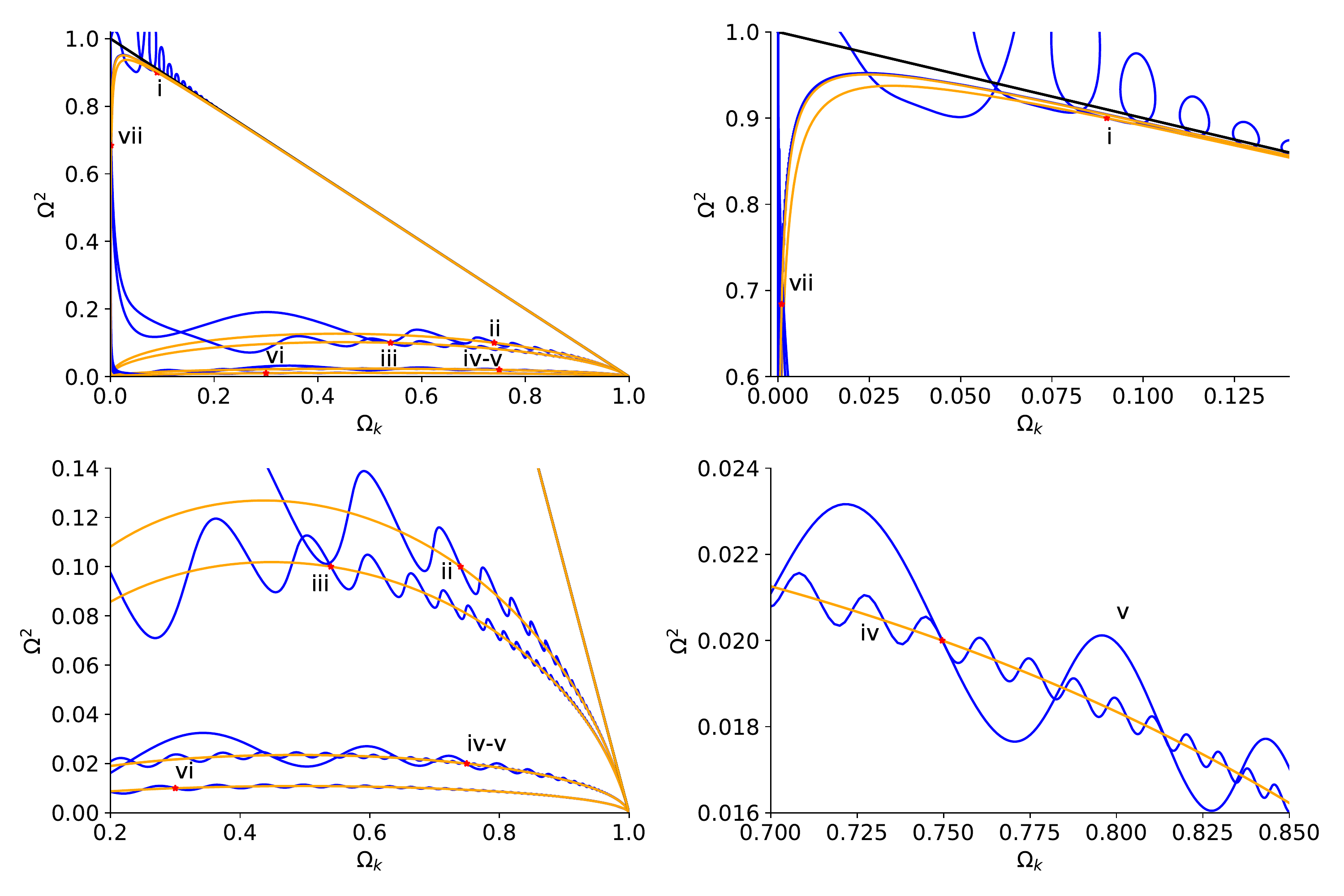}}
    \caption{Some solutions of the full system \eqref{unperturbed1FLRW} (blue) and time-averaged system \eqref{avrgsystFLRW} (orange) for the FLRW metric with negative curvature ($k=-1$) when $\gamma=4/3$. We have used for both systems initial data sets presented in the Table \ref{Tab3a}.}
\end{figure*}

\begin{figure*}
    \centering
    \subfigure[\label{fig:OpenFLRWStiff3D} Projections in the space $(\Omega_{k}, H, \Omega^2)$. The surface is given by the constraint $\Omega^{2}=1-\Omega_{k}$.]{\includegraphics[scale = 0.40]{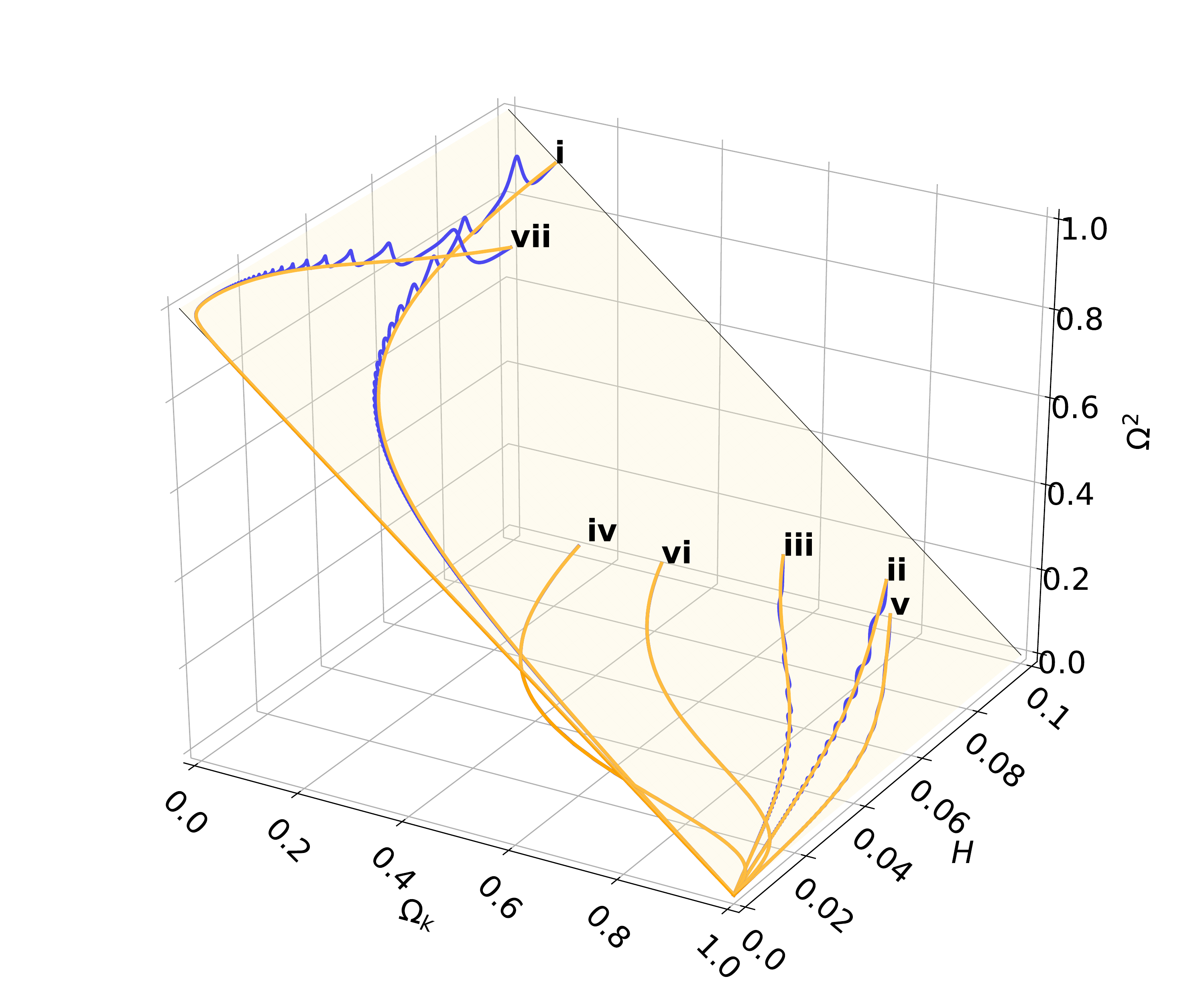}}
    \subfigure[\label{fig:OpenFLRWStiff2D} Projection in the space $(\Omega_{k}, \Omega^2)$. The black line represent the constraint $\Omega^{2}=1-\Omega_{k}$.]{\includegraphics[scale = 0.53]{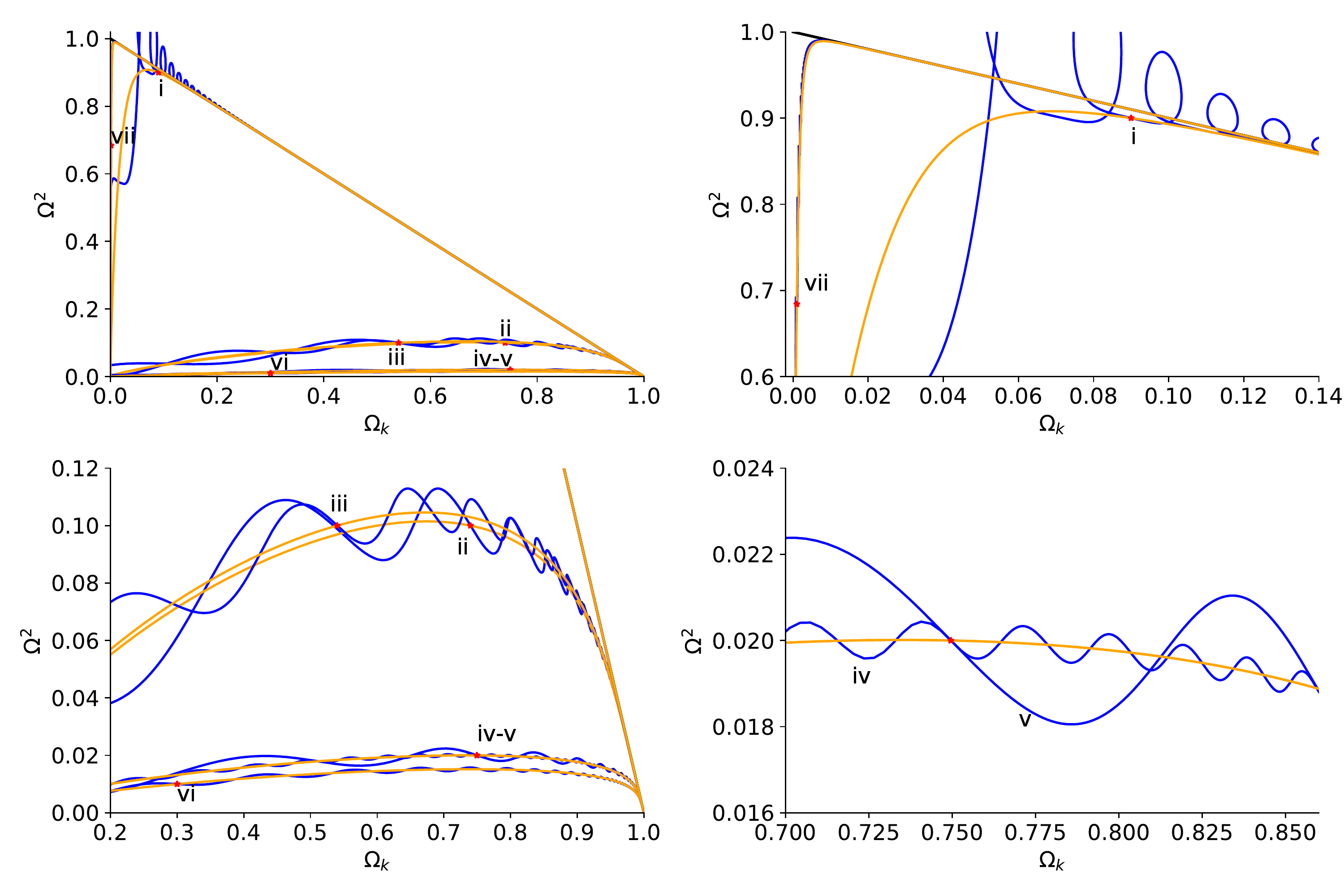}}
    \caption{Some solutions of the full system \eqref{unperturbed1FLRW} (blue) and time-averaged system \eqref{avrgsystFLRW} (orange) for the FLRW metric with negative curvature ($k=-1$) when $\gamma=2$. We have used for both systems initial data sets presented in the Table \ref{Tab3a}.}
\end{figure*}

In figures \ref{fig:OpenFLRWCC3D}-\ref{fig:OpenFLRWStiff2D}  projections of some solutions of the full system \eqref{unperturbed1FLRW} and time-averaged system \eqref{avrgsystFLRW} for the FLRW metric with negative curvature ($k=-1$) in the $(\Omega_{k}, H, \Omega^{2})$ space with their respective projection when $H=0$ and considering in both systems the same initial data sets from table \ref{Tab3a} are presented. Figures \ref{fig:OpenFLRWCC3D}-\ref{fig:OpenFLRWCC2D} show solutions for matter fluid   corresponding to cosmological constant ($\gamma=0$). Figures \ref{fig:OpenFLRWDust3D}-\ref{fig:OpenFLRWDust2D} show solutions for matter fluid   corresponding to dust ($\gamma=1$). Figures \ref{fig:OpenFLRWRad2D}-\ref{fig:OpenFLRWRad3D} show solutions for matter fluid   corresponding to radiation ($\gamma=4/3$). Figures \ref{fig:OpenFLRWStiff3D}-\ref{fig:OpenFLRWStiff2D} show solutions for matter fluid   corresponding to stiff fluid ($\gamma=2$). 
It is interesting to note that in the FLRW with negative curvature case, when the matter fluid  corresponds to a cosmological constant, $H$ tends asymptotically to constant values depending on the initial conditions which is consistent to de Sitter expansion. In  addition, for any $\gamma<\frac{2}{3}$ and $\Omega_k>0$, $\Omega_k \rightarrow 0$. On the other hand, when $\gamma>\frac{2}{3}$ and $\Omega_k>0$ the universe becomes curvature dominated asymptotically ($\Omega_k \rightarrow 1$). These figures are evidence that the main Theorem presented in  section \ref{SECT:II} is fulfilled for  FLRW metric with negative curvature. 

\bibliographystyle{alpha}

\end{document}